\documentclass{article}

\usepackage{PRIMEarxiv}

\usepackage[utf8]{inputenc} 
\usepackage[T1]{fontenc}    
\usepackage{url}            
\usepackage{booktabs}       
\usepackage{amsfonts}       
\usepackage{nicefrac}       
\usepackage{microtype}      
\usepackage{lipsum}
\usepackage{fancyhdr}       
\usepackage{graphicx}       
\graphicspath{{media/}}     

\usepackage{array}
\usepackage{mathtools}
\usepackage{color}
\usepackage{cite}
\usepackage{amsmath}
\usepackage[indent,bf]{caption}
\usepackage{rotating}
\usepackage{setspace}
\usepackage{ragged2e}
\usepackage{longtable}
\usepackage[colorlinks=true,pdftex,citecolor=blue,bookmarks=true,bookmarksnumbered=true,breaklinks=true]{hyperref}
\usepackage{pdfsync}
\usepackage{enumerate}
\usepackage{titlesec}
\usepackage{etoolbox}
\usepackage{longtable}
\usepackage{scrextend}
\usepackage[T1]{fontenc}
\usepackage{amssymb}
\usepackage{bm}
\usepackage{subcaption}
\usepackage{caption}
\usepackage{float}
\usepackage{epstopdf}
\usepackage{pdfpages}
\allowdisplaybreaks[1]
\usepackage{setspace}
\usepackage{xcolor}
\usepackage[capitalise]{cleveref}

\usepackage{multirow}

\definecolor{ss}{rgb}{0,0,0}

\newcommand{\specialcell}[2][c]{%
  \begin{tabular}[#1]{@{}c@{}}#2\end{tabular}}

\title{Multilevel Scalable Solvers for Stochastic Linear and Nonlinear Problems
}

\author{
  Sudhi Sharma \\
  Department of Civil and Environmental Engineering \\
  Carleton University \\
  Ottawa\\
   \And
  Pierre Jolivet \\
  Sorbonne Université \\
  CNRS, LIP6 \\
  Paris\\
   \AND
   Victorita Dolean \\
   Department of Mathematics and Statistics \\
   University of Strathclyde \\
   Glasgow \\
   and \\
   Laboratoire J.A. Dieudonne, CNRS \\
   Universite Cote d Azur \\
   Nice \\
   \And
   Abhijit Sarkar \\
   Department of Civil and Environmental Engineering \\
   Carleton University \\
  Ottawa\\
}

\begin{document}
\maketitle

\begin{abstract}
This article discusses the uncertainty quantification (UQ) for time-independent linear and nonlinear partial differential equation (PDE)-based systems with random model parameters carried out using sampling-free intrusive stochastic Galerkin method leveraging multilevel scalable solvers constructed combining two-grid Schwarz method and AMG. High-resolution spatial meshes along with a large number of stochastic expansion terms increase the system size leading to significant memory consumption and computational costs. Domain decomposition (DD)-based parallel scalable solvers are developed to this end for linear and nonlinear stochastic PDEs. A generalized minimum residual (GMRES) iterative solver equipped with a multilevel preconditioner consisting of restricted additive Schwarz (RAS) for the fine grid and algebraic multigrid (AMG) for the coarse grid is constructed to improve scalability. Numerical experiments illustrate the scalabilities of the proposed solver for stochastic linear and nonlinear Poisson problems.

\end{abstract}


\section{Introduction}
For realistic numerical prediction, uncertainty in the model parameters (including initial or boundary conditions), model error, sparsity and noise in the observational data are generally accounted for using the probabilistic or non-probabilistic framework. In the probabilistic UQ framework, the forward propagation of uncertainty in model parameters poses computational challenges due to the nonlinear transformation of random input to output even for a linear model. The Monte Carlo simulations (MCS) provide a robust way to quantify these uncertainties by propagating a number of independent samples of the input through a computational model. The corresponding output realizations (samples) of the model are used to construct the joint probability density function of the quantities of interest to a desired accuracy. However, the slow convergence rate of the method necessitates a large number of input samples demanding repeated solutions of the computational model \cite{book_raulphsmith,xiu_review,xiu_collocation}. In the MCS framework, the computational cost associated with these simulations for high-resolution PDE models becomes prohibitive. The polynomial chaos expansion (PCE)-based approaches have gained considerable interest for UQ by expanding the random quantities (input and output) as a truncated series of polynomials of random variables \cite{PC_challenges,UQ_habibnajim,PCKF_ghanem,inverse_najm1,inverse_najm2}. Among these, non-intrusive spectral projection and collocation approaches are used widely \cite{xiu_collocation,inverse_xiu,motamed2013stochastic,eldred_nisp_colloc}. However, the substantial increase in the number of quadrature or collocation points with respect to the increasing number of random variables incurs high computational costs for these methods \cite{compare_eldred_NISP,knio_book}.

In this investigation, the sampling-free intrusive stochastic Galerkin method is used which transforms a stochastic PDE to a coupled set of deterministic PDEs \cite{knio_book,book_ghanem}. In this methodology, the input and output are expanded using PCE; and then the application of Galerkin projection converts the stochastic PDE into a system of a coupled set of deterministic PDEs solved numerically using iterative solvers. The number of unknowns arising from the numerical discretization of coupled PDEs to be solved rapidly overwhelms the typical memory for a desktop computer with increasing random parameters and the resolution of computational grids. Solutions to these linear or nonlinear sets of equations are typically carried out using iterative solvers equipped with efficient preconditioners. Several strategies to expedite the convergence of these iterative solvers are proposed in the literature as discussed below.

The block sparse structure and diagonal dominance of the linear (or linearized) system matrices arising from the numerical discretization of the coupled PDEs in the intrusive Galerkin framework permit the use of block-diagonal or mean-based preconditioners \cite{SG_pellisetti}. However, these preconditioners lose efficiency for large uncertainties with significant contributions from off-diagonal terms in the system matrices. Other variants utilizing a Kronecker product representation of the system matrices are presented in \cite{SG_kroneckermean, SG_matthieskeese, SG_kroneckerPC}. A hierarchical structure of the global system matrix in an intrusive PCE scheme is exploited to develop a preconditioner based on the Schur complement of the blocks in \cite{sousedik,sousedik_2}. Even though the preconditioners performed well in numerical experiments with a uniform random variable, the condition number bound for the preconditioned system shows a multiplicative growth which leads to larger computational costs for other random variables \cite{sousedik_2}. Application of an approximate Gauss-Seidel algorithm as a preconditioner to Krylov solvers is used to solve advection-diffusion equations with random input parameters in \cite{SG_GaussSeidel_Roger}. However, these solvers were only tested for small-scale system sizes running sequentially. \textcolor{ss}{Multigrid-based methods for stochastic problems are also applied in \cite{knio_book,SG_AMG,SG_MG,mg_diffusion}. However, these methods require efficient fine-tuning of several parameters for different class of problems. Moreover, their convergence can deteriorate for large polynomial chaos order in case of Hermite chaos as shown in \cite{SG_AMG}.}

DD-based approaches provide a means to decompose the domain into several subdomains which can be solved independently and combined to obtain the final solution \cite{book_DD_barryFsmith,book_victorita}. This reduces the memory requirement per core for the problem allowing to solve PDEs in high-resolution computational grids. They also offer an inherently parallelizable algorithm and associated preconditioners for the iterative solver. The theoretical formulation for applying the non-overlapping DD approach to the intrusive stochastic Galerkin method is proposed in \cite{SG_DD_sarkar}. The scalabilities of these solvers are extensively studied for diffusion equation and linear elasticity with many random variables using hundreds of cores  \cite{subberJCP,subberCMAM,subber_thesis,ajit_CMAM,ajit_thesis}. Similarly, an overlapping two-level Schwarz preconditioner is constructed for the stochastic linear symmetric coefficient matrices in \cite{subber_schwarz}. 

Multilevel domain decomposition methods are available in literature \cite{book_DD_barryFsmith, DD_compare, comparisondefmultigriddd, multilevel_pc1, multilevel_pc2, multilevel_pc3} where the hierarchy of coarser levels are used to construct the preconditioner at each level. However, this article utilizes only two grids with an AMG at the coarse level providing multiple levels of error reduction. Even though the use of multigrid with domain decomposition is not new, the specific architecture of the preconditioner and its application to deterministic non-symmetric coefficient matrices is novel. This article extensively investigates the scalabilities of this proposed two-grid Schwarz solver for linear and nonlinear problems in deterministic (see Appendix) and stochastic settings.

The main contributions of this artilce are as follows. A multilevel overlapping DD-based solver for the symmetric and non-symmetric systems arising from the stochastic Galerkin method is proposed. A GMRES iterative solver equipped with a two-grid restricted additive Schwarz (RAS) preconditioner is proposed. The fine grid uses a one-level RAS smoother while the coarse grid is preconditioned using AMG. These multiple levels of error reduction achieved using AMG provide excellent strong and weak scalability for high-resolution computational grids. The proposed solver is also compared with a coarse grid solved using a one-level RAS preconditioner (similar to \cite{subber_schwarz}) which even though performs well for strong scaling, shows poor performance for weak scalability. 

This article is organized as follows. An introduction to the stochastic Galerkin method is provided by formulating linear and nonlinear 
time-independent stochastic PDEs (i.e. Poisson problems) with a non-Gaussian random field (log-normal process) model of diffusion coefficient. The details of the multilevel overlapping DD-based solver are provided next. Further, the numerical section illustrates the scalabilities of the proposed solver for stochastic linear and nonlinear Poisson problems. A verification of the linear and nonlinear Poisson problem using MCS and a numerical experiment to demonstrate the optimal size of the coarse grid is also provided in this section.

\section{Stochastic Galerkin Method for Time-Independent PDEs}
In this section, the mathematical formulation of the sampling-free (intrusive) stochastic Galerkin method is provided for both time-independent linear and nonlinear PDEs.
For simplicity, linear and nonlinear Poisson problems are used for an elementary exposition of the method.

\subsection{Stochastic Linear Poisson Problem}\label{sec:LP}
The stochastic linear Poisson problem with a spatially varying random diffusion coefficient can be written as \cite{keese_galerkin}:
\begin{align}\label{Eq.LP_StrongPDE}
  - \nabla \cdot \Big (c(\mathbf{x},\theta) \nabla u(\mathbf{x},\theta) \Big) &= f(\mathbf{x}) \;\; \rm{on} \;\;  \Omega \times \Phi \\
     u(\mathbf{x},\theta) &= g_d \quad \mathrm{on} \quad   \Gamma_d \times \Phi \\
    \frac{\partial u(\mathbf{x},\theta)}{\partial n } &= g_n \quad \mathrm{on} \quad \Gamma_n \times \Phi
\end{align}
where $c(\mathbf{x},\theta)$ is the diffusion coefficient modelled as a log-normal random field (non-Gaussian; enforcing positivity of diffusion coefficient), $u(\mathbf{x},\theta)$ is the random (non-Gaussian) output and $f(\mathbf{x})$ is the deterministic external forcing. $\Gamma_d$ and $\Gamma_n$ are the Dirichlet and Neumann boundaries with values of $g_d$ and $g_n$ respectively. The boundary conditions are assumed to be deterministic for the current problem even though random boundaries can also be treated in this framework. \textcolor{ss}{Here, $\Phi$ represents the set of all possible outcomes associated with the probability space $(\Phi, \Sigma, P)$ and random variable or $\theta$ represents the random aspect. For the above stochastic linear Poisson problem with random diffusion coefficient, we assume the solution $u(\mathbf{x},\theta)$ lies in the tensor product space formed by the deterministic space of functions and $L_2$ space of second order random variables as $H^1(\Omega) \otimes L_2(\Phi,P)$ \cite{knio_book}.} 
We note that, to ensure well-posedness (and unique solution) to the above stochastic problem, the parameter should be bounded as \cite{motamed2013stochastic,SG_matthieskeese,SG_kroneckerPC}:
\begin{equation}\label{Eq.bounded}
0 < c_{min} < c(\mathbf{x}, \theta) < c_{max} < \infty \quad a.e \quad \mathrm{in} \quad  \Omega \times \Phi.
\end{equation}

The random input log-normal field in Eq.~(\ref{Eq.LP_StrongPDE}) is modelled as the exponential of a Gaussian process $c(\mathbf{x},\theta) = \exp(g(\mathbf{x},\theta))$. The underlying Gaussian process can be represented using the Karhunen-Loeve expansion (KLE) as \cite{lognormal_roger}:
\begin{equation}\label{Eq.KLE}
g(\mathbf{x},\theta) = {g}_0(\mathbf{x})+\sum_{i=1}^{M}{g_i(\mathbf{x})\xi_i(\theta)}
\end{equation}
where $ {g}_0(\mathbf{x})$ is the mean of the random process, $\bm{\xi} = \{ \xi_1, \xi_2, \cdots \xi_M \}$ are the uncorrelated random variables (in this case Gaussian) with zero mean and unit variance and $g_i(\mathbf{x})= \sqrt{\lambda_{i}} \phi_i(\mathbf{x})$ are the coefficients computed from the eigenvalues $\lambda_i$ and eigenfunctions $\phi_i(\mathbf{x})$ of the associated covariance kernel. In this article, we assume the underlying Gaussian process has an exponential covariance kernel as follows \cite{book_ghanem}:
\begin{equation}\label{Eq.CovarKer2DNLP}
\mathcal{K}(\mathbf{x}_1, \mathbf{x}_2) = \sigma^2\ e^{-{\frac{|x_1-x_2|}{b_x}  } - {\frac{|y_1-y_2|}{b_y}}}
\end{equation}
where $\mathbf{x}_i= (x_i, y_i)$, $\sigma$ represents the standard deviation of the covariance kernel and $b_x,b_y$ are the correlation lengths in $x$ and $y$ direction respectively. The eigenvalues and eigenfunctions for the kernel can be computed analytically as \cite{lognormal_roger}:
The number of random variables $M$ required for an accurate expansion depends upon the strength of correlation of the kernel and can be computed based on the relative sum of eigenvalues ($\sum_{i=1} ^{M} \lambda_i / \sum_{i=1} ^{\infty} \lambda_i$) of the kernel \cite{knio_book, book_ghanem}.
The log-normal input random field $c(\mathbf{x},\theta)$ admits a PCE as:
\begin{equation} \label{Eq.inputpc}
c(\mathbf{x},\theta) = \sum_{i=0}^{\infty} \bar{c}_{i}(\mathbf{x}) \Psi_i(\bm{\xi}) \approx \sum_{i=0}^{L} \bar{c}_{i}(\mathbf{x}) \Psi_i(\bm{\xi})
\end{equation}
where $\bar{c}_{i}(\mathbf{x})$ are the deterministic coefficients and $\Psi_i(\bm{\xi})$ are the Hermite polynomial functions of Gaussian random variables known as the chaoses \cite{knio_book,book_ghanem}. The infinite series expansion is curtailed at $L$ terms depending upon the number of random variables ($M$) and order of expansion ($p$) as $L+1 = \frac{(M+p)!}{(M!)(p!)}$ \cite{knio_book,book_ghanem}. The coefficients of the expansion can be computed using sampling/quadrature-based approaches or a sampling-free (intrusive) stochastic Galerkin method which is utilized in this article. The exponential of the Gaussian process expressed using KLE can be analytically evaluated to find the coefficients of the PCE as \cite{lognormal_roger,subber_thesis,ajit_thesis,mohammed_thesis}:
\begin{equation}\label{Eq.inputlogn}
c(\mathbf{x},\theta) = c_0(\mathbf{x}) \sum_{i=0}^{L}  \frac{\langle\Psi_i(\bm{\eta})\rangle}{\langle\Psi_i^2(\bm{\xi})\rangle}\Psi_i(\bm{\xi})
\end{equation} 
where 
\begin{equation}
c_0(\mathbf{x}) = \mathrm{exp}\left[ g_0(\mathbf{x}) + \frac{1}{2}  \sum_{i=1}^{L} g_i^2(\mathbf{x}) \right] 
\end{equation}
is the mean of the log-normal process, $\bm{\eta} = \{ \eta_1, \eta_2, \cdots \eta_L \}$ are the vector of independent random variables with each $\eta_i = \xi_i - g_i$ and \textcolor{ss}{$ \langle . \rangle$ is the expectation operator defined as \cite{book_ghanem, knio_book}:
\begin{equation}
\langle f(\bm{\xi}) \rangle = \int \cdots \int f(\bm{\xi}) p(\bm{\xi}) d \bm{\xi}
\end{equation}
where $p(\bm{\xi})$ represents the joint probability density function of the random variables. The denominator of Eq.~(\ref{Eq.inputlogn}) can be analytically determined  and saved a-priori for all the PCE coefficients.} The PCE coefficients of the input $\bar{c}_{i}(\mathbf{x})$ are completely determined through this analytical treatment of KLE. Note, many details pertaining to KLE and PCE are not provided here for brevity and clarity of the article, but are available in \cite{knio_book,book_ghanem,book_raulphsmith,subber_thesis,ajit_thesis}.

Now, the output $u(\mathbf{x},\theta)$ can also be expanded using PCE as:
\begin{equation}\label{Eq.outpc}
u(\mathbf{x},\theta) = \sum_{j=0}^{N}\bar{u}_{j}(\mathbf{x}) \Psi_j(\bm{\xi}) 
\end{equation}
where $\bar{u}_{j}(\mathbf{x})$ are the deterministic coefficients to be computed and $N$ is the total number of terms in output expansion.
Now, the PCE of the input and output are inserted into the weak form of the stochastic PDE in Eq.~(\ref{Eq.LP_StrongPDE}) for the stochastic Galerkin (Bubnov) projection. The projection is carried out by multiplying the weak form with a test function $\Psi_k(\bm{\xi})$ of the same polynomial chaos basis used for output expansion and applying the expectation operator as \cite{knio_book,book_ghanem}:
\begin{gather}
      \Big \langle \int_{\Omega} \Big( \sum_{i=0}^{L} (\bar{c}_{i}(\mathbf{x}) \Psi_i(\bm{\xi}) ) \nabla(  \sum_{j=0}^{N} \bar{u}_{j}(\mathbf{x}) \Psi_j(\bm{\xi}) ) \cdot  \nabla v(\mathbf{x}) \Big) \; \Psi_k(\bm{\xi})  \; d \mathbf{x}  \Big \rangle \nonumber  \\
       =  \Big \langle \int_{\Omega} f(\mathbf{x}) v(\mathbf{x}) \Psi_k(\bm{\xi})  \; d \mathbf{x} \Big \rangle \quad  k = 0, 1,2, \ldots N\label{Eq.Galerkin_projectL}
\end{gather}
Gathering the chaos terms, we get a coupled system of PDEs as (note, the explicit dependence on $\bm{\xi}$ is dropped for brevity):
\begin{equation}\label{Eq.Galerkin_L_exp}
     \sum_{i=0}^{L} \sum_{j=0}^{N}  \int_{\Omega}  \langle  \Psi_i \Psi_j \Psi_k \rangle  \; ( \bar{c}_{i}(\mathbf{x}) \nabla \bar{u}_{j}(\mathbf{x}) \cdot  \nabla v(\mathbf{x}) ) \; d \mathbf{x}   =  \int_{\Omega} f(\mathbf{x}) v(\mathbf{x}) \langle \Psi_k \rangle \; d \mathbf{x},  \quad \quad k = 0, 1,2, \ldots N
\end{equation}
After rewriting $m_{ijk} = \langle  \Psi_i \Psi_j \Psi_k \rangle$ and $b_k = f(\mathbf{x}) \langle \Psi_k \rangle $, we obtain (note, explicit dependence of $\mathbf{x}$ is dropped for brevity):
\begin{equation}
     \int_{\Omega} \sum_{i=0}^{L} \sum_{j=0}^{N} m_{ijk} \; \bar{c}_{i} \nabla \bar{u}_{j} \cdot \nabla v \; d \mathbf{x}   =   \int_{\Omega} b_k v \; d \mathbf{x}, \quad \quad k = 0, 1,2, \ldots N.
\end{equation}

Note, the above system of equations can be assembled in a Kronecker product format and a preconditioner based on this structure can also be constructed as in \cite{subber_schwarz,SG_kroneckermean,SG_kroneckerPC}. In our implementations, we save and use only the non-zero entries of the multiplication tensors ($m_{ijk}$) and utilize the sparse matrix-vector operations through PETSc \cite{petsc-web-page}. However, the Kronecker product structure can also be accommodated into our implementations without much difficulty. In this article, we use a different approach being convenient for applying DD-based preconditioners as explained afterwards. This system of equations is assembled as a vector-valued PDE (termed as vector PDE from now onward) with the solution vector at each node having $N+1$ elements corresponding to the PCE coefficients. 

We illustrate the construction of a vector PDE for the stochastic Poisson problem by expanding the system for $L = N = 1$ as:
\begin{equation}
    \int_{\Omega} \sum_{i=0}^{1} \sum_{j=0}^{1} m_{ijk} \; \bar{c}_{i} \nabla \bar{u}_{j}\cdot \nabla v_k \; d \mathbf{x}   = \int_{\Omega} b_{k} v_k \; d \mathbf{x},  \quad  \; k = 0, 1.
\end{equation}
Concisely the above equations can be rearranged as follows:

\begin{align}
   \int_{\Omega} \sum_{i=0}^{1} m_{i00} \; \bar{c}_{i} \nabla \bar{u}_{0}\cdot \nabla v_0 \; d \mathbf{x}  + \int_{\Omega} \sum_{i=0}^{1} m_{i10} \; \bar{c}_{i} \nabla \bar{u}_{1}\cdot \nabla v_0 \; d \mathbf{x} =  \int_{\Omega} b_{0} v_0  \; d \mathbf{x}  \\
    \int_{\Omega} \sum_{i=0}^{1} m_{i01} \; \bar{c}_{i} \nabla \bar{u}_{0}\cdot \nabla v_1  \; d \mathbf{x} + \int_{\Omega} \sum_{i=0}^{1} m_{i11} \; \bar{c}_{i} \nabla \bar{u}_{1}\cdot \nabla v_1 \; d \mathbf{x} =  \int_{\Omega} b_{1} v_1  \; d \mathbf{x}
\end{align}
The above equations can be gathered in a vector form as follows:
\begin{equation}\label{Eq.linearVecPDE}
\int_{\Omega} \mathcal{A} \nabla \mathcal{U} : \nabla V \; d \mathbf{x} = \int_{\Omega} \mathcal{B} \cdot V \; d \mathbf{x}
\end{equation}
where,
\begin{equation}\label{Eq.Ajklinear}
\mathcal{A}_{kj} = \sum_{i=0}^{1} m_{ijk}\bar{c}_{i} (\mathbf{x})
\end{equation}

\begin{gather}
\mathcal{U} =
\begin{bmatrix}
 \bar{u}_0\\ \bar{u}_1\\
\end{bmatrix}
\quad
V =
\begin{bmatrix}
v_0 \\ v_1\\
\end{bmatrix}
\quad
\mathcal{B} =
\begin{bmatrix}
b_0\\ b_1\\
\end{bmatrix}
\end{gather}
and the Frobenius inner product of two matrices is defined as $A : B = \sum_{i,j} A_{ij} B_{ij}$ which is a scalar.
For $N+1$ output PCE terms, the vector PDE can be represented same as in Eq.~(\ref{Eq.linearVecPDE}) with coefficients as:
\begin{gather}
\mathcal{A}_{kj} = \sum_{i=0}^{L} m_{ijk}\bar{c}_{i} (\mathbf{x}) ; \quad
\mathcal{U} =
\begin{bmatrix}
 \bar{u}_0\\  \bar{u}_1\\ \vdots \\  \bar{u}_N \\
\end{bmatrix} ;
\quad
V =
\begin{bmatrix}
v_0 \\ v_1 \\ \vdots \\ v_N \\
\end{bmatrix} ;
\quad
\mathcal{B} =
\begin{bmatrix}
b_0\\b_1\\ \vdots \\b_k \\
\end{bmatrix}.
\label{eq:stochatsic_strongform}
\end{gather}

\subsection{Stochastic Nonlinear Poisson Problem}
The stochastic nonlinear Poisson problem with a spatially varying random diffusion coefficient can be written as \cite{keese_galerkin}:
\begin{align}\label{Eq.NLP_StrongPDE}
   - \nabla \cdot \Big ( c(\mathbf{x},\theta) (1 + \alpha u(\mathbf{x},\theta))  \nabla u(\mathbf{x},\theta) \Big) &= f(\mathbf{x}) \;\; \mathrm{on} \;\;  \Omega \times \Phi \\
     u(\mathbf{x},\theta) &= g_d \quad \mathrm{on} \quad   \Gamma_d \times \Phi \\
     \frac{\partial u(\mathbf{x},\theta)}{\partial n} &= g_n \quad \mathrm{on} \quad \Gamma_n \times \Phi
\end{align}
where $c(\mathbf{x},\theta)$ is the random diffusion coefficient, $u(\mathbf{x},\theta)$ is the random response and $\alpha$ is a deterministic constant determining the strength of nonlinearity. $\Gamma_d$ and $\Gamma_n$ are the Dirichlet and Neumann boundaries with values of $g_d$ and $g_n$ respectively. The boundary conditions and $f$ are assumed to be deterministic for the current model. \textcolor{ss}{Similar to the section \ref{sec:LP} on stochastic Linear Poisson, $\Phi$ represents the set of all possible outcomes associated with the probability space $(\Phi, \Sigma, P)$ and random variable or $\theta$ represents the random aspect of the problem. We seek the solution $u(\mathbf{x},\theta)$ in the tensor product space formed by the space of deterministic functions and $L_2$ space of second order random variables as $H^1(\Omega) \otimes L_2(\Phi,P)$ \cite{knio_book}.}
%
Note that, the boundedness and positivity condition for the random field mentioned in Eq.~(\ref{Eq.bounded}) is applicable here too. The following section expands upon the stochastic linear Poisson problem to treat the non-linearities involved along with the stochastic Galerkin projection.

We can expand both $c(\mathbf{x},\theta)$ and $u(\mathbf{x},\theta)$ by PCE similar to Eq.~(\ref{Eq.inputpc}) and Eq.~(\ref{Eq.outpc}). The Galerkin projection of the stochastic PDE can be carried out similar to Eqs.~(\ref{Eq.Galerkin_projectL}) and (\ref{Eq.Galerkin_L_exp}) to obtain:
%
    

\begin{align}
   &  \int_{\Omega} \sum_{i=0}^{L} \sum_{k=0}^{N} \langle  \Psi_i \Psi_k \Psi_l \rangle \;  ( \bar{c}_{i}(\mathbf{x})  \nabla \bar{u}_{k}(\mathbf{x}) \cdot  \nabla v_l(\mathbf{x}) )\; d \mathbf{x} \nonumber  \\
 & + \int_{\Omega} \sum_{i=0}^{L} \sum_{j=0}^{N} \sum_{k=0}^{N} \langle  \Psi_i \Psi_j \Psi_k \Psi_l \rangle \alpha \;
(\bar{c}_{i}(\mathbf{x}) \bar{u}_{j}(\mathbf{x}) \nabla \bar{u}_{k}(\mathbf{x}) \cdot \nabla v_l(\mathbf{x}) )\; d \mathbf{x} \nonumber \\
&=  \int_{\Omega}   f(\mathbf{x}) v_l(\mathbf{x}) \langle \Psi_l \rangle \; d \mathbf{x},  \quad  \; l = 0, 1,2, \ldots N
\end{align}
Using $m_{ijk} = \langle  \Psi_i \Psi_j \Psi_k \rangle$, $t_{ijkl} = \langle  \Psi_i \Psi_j \Psi_k \Psi_l\rangle$, $\alpha=1$ and dropping the explicit dependence on $\mathbf{x}$ for brevity, the coupled nonlinear PDEs can be written as:

\begin{gather}
    \int_{\Omega}  \sum_{i=0}^{L} \sum_{k=0}^{N} m_{ikl} \; \bar{c}_{i} \nabla \bar{u}_{k} \cdot \nabla v_l \; d \mathbf{x}  + \int_{\Omega} \sum_{i=0}^{L} \sum_{j=0}^{N} \sum_{k=0}^{N} t_{ijkl} \; \bar{c}_{i} \bar{u}_{j} \nabla \bar{u}_{k} \cdot \nabla v_l \;  d \mathbf{x} \nonumber \\ =   \int_{\Omega}  b_{l} v_l \; d \mathbf{x}, \quad \; l = 0, 1,2, \ldots N
\end{gather}
where $b_l = f(\mathbf{x}) \langle \Psi_l \rangle $.

The above system of equations contains two parts. The first part is the same as the system of equations derived from the stochastic linear Poisson problem and the second part corresponds to the stochastic nonlinear part of the problem. In this investigation, this nonlinear system of equations is solved using Picard iterations by linearization of the nonlinear coefficient term \cite{book_nonlinear}. By assuming a value from the last known iteration $n$ of the Picard iteration, the above system of equations for the current iteration $n+1$, can be re-written as follows: 
\begin{gather}\label{Eq.stonlp_system}
    \int_{\Omega}  \sum_{i=0}^{L} \sum_{k=0}^{N} m_{ikl} \; \bar{c}_{i} \nabla \bar{u}_{k} ^{n+1} \cdot \nabla v_l \; d \mathbf{x}  + \int_{\Omega} \sum_{i=0}^{L} \sum_{j=0}^{N} \sum_{k=0}^{N} t_{ijkl} \; \bar{c}_{i} \bar{u}_{j} ^{n} \nabla \bar{u}_{k} ^{n+1} \cdot \nabla v_l \; d \mathbf{x}  \nonumber \\ =   \int_{\Omega}  b_{l} v_l \; d \mathbf{x}, \quad  \; l = 0, 1,2, \ldots N
\end{gather}
Note that $\bar{u}_{j} ^{n}$ is used to distinguish the solution vector as already computed from the previous step. Similar to the stochastic linear Poisson problem, these equations are solved as a vector PDE which is illustrated using $L= N = 1$ as:

\begin{gather}
   \int_{\Omega} \sum_{i=0}^{1} m_{i00} \; \bar{c}_{i} \nabla \bar{u}_{0} \cdot \nabla v_0 \; d \mathbf{x} + \int_{\Omega} \sum_{i=0}^{1} m_{i10} \; \bar{c}_{i} \nabla \bar{u}_{1} ^{n+1} \cdot \nabla v_0 \; d \mathbf{x} + \nonumber \\
  \int_{\Omega} \sum_{j=0}^{1} \sum_{i=0}^{1} t_{ij00} \; \bar{c}_{i} \bar{u}_{j} ^{n} \nabla \bar{u}_{0} ^{n+1} \cdot \nabla v_0  \; d \mathbf{x} + \int_{\Omega} \sum_{j=0}^{1} \sum_{i=0}^{1} t_{ij10} \; \bar{c}_{i} \bar{u}_{j} ^{n} \nabla \bar{u}_{1} ^{n+1} \cdot \nabla v_0 \; d \mathbf{x}
  = \int_{\Omega}  b_{0} v_0 \; d \mathbf{x}  \\
\int_{\Omega} \sum_{i=0}^{1} m_{i01} \; \bar{c}_{i} \nabla \bar{u}_{0}  ^{n+1} \cdot \nabla v_1  \; d \mathbf{x} + \int_{\Omega} \sum_{i=0}^{1} m_{i11} \; \bar{c}_{i} \nabla \bar{u}_{1} ^{n+1} \cdot \nabla v_1 \; d \mathbf{x} + \nonumber \\
   \int_{\Omega} \sum_{j=0}^{1} \sum_{i=0}^{1} t_{ij01} \; \bar{c}_{i} \bar{u}_{j} ^{n} \nabla \bar{u}_{0} ^{n+1} \cdot \nabla v_1 \; d \mathbf{x} + \int_{\Omega} \sum_{j=0}^{1} \sum_{i=0}^{1} t_{ij11} \; \bar{c}_{i} \bar{u}_{j} ^{n} \nabla \bar{u}_{1} ^{n+1} \cdot \nabla v_1 \; d \mathbf{x} = \int_{\Omega} b_{1} v_1 \; d \mathbf{x}
\end{gather}
This can be written in the form of a vector PDE as:
\begin{equation}\label{Eq.vectorvalPDENL}
\int_{\Omega} ( \mathcal{A} + \mathcal{T}^{n} ) \nabla \mathcal{U}^{n+1} : \nabla V \; d \mathbf{x}  = \int_{\Omega} \mathcal{B} \cdot V \; d \mathbf{x}
\end{equation}
where 
\begin{align}\label{Eq.AjkTkl1}
     \mathcal{A}_{lk} &= \sum_{i=0}^{1} m_{ikl}\bar{c}_{i} (\mathbf{x})\\ \label{Eq.AjkTk2}
     \mathcal{T}^{n}_{lk} &= \sum_{j=0}^{1} \sum_{i=0}^{1} t_{ijkl} \bar{c}_{i} (\mathbf{x}) \bar{u}_{j} ^{n} (\mathbf{x})
\end{align}
and 
\begin{gather}
\mathcal{U}^{n+1} = 
\begin{bmatrix}
 \bar{u}_0 ^{n+1} \\  \bar{u}_1 ^ {n+1}\\
\end{bmatrix}
\quad
V =
\begin{bmatrix}
v_0\\ v_1\\
\end{bmatrix}
\quad
\mathcal{B} =
\begin{bmatrix}
b_0\\ b_1\\
\end{bmatrix}.
\end{gather}
In general, the whole system of PDEs for stochastic nonlinear Poisson problem in Eq.~(\ref{Eq.stonlp_system}) can be written in vector form as in Eq.~(\ref{Eq.vectorvalPDENL}) with coefficient matrices as:
\begin{align}
     \mathcal{A}_{lk} &= \sum_{i=0}^{L} m_{ikl}\bar{c}_{i} (\mathbf{x})\\
     \mathcal{T}_{lk}^{n} &= \sum_{j=0}^{N} \sum_{i=0}^{L} t_{ijkl} \bar{c}_{i} (\mathbf{x}) \bar{u}_{j} ^{n} (\mathbf{x})
\end{align}
and 
\begin{gather}
\mathcal{U}^{n+1} = 
\begin{bmatrix}
 \bar{u}_0 ^{n+1}\\  \bar{u}_1 ^{n+1} \\ \vdots \\  \bar{u}_N ^{n+1} \\
\end{bmatrix} 
\quad
V =
\begin{bmatrix}
v_0 \\ v_1 \\ \vdots \\ v_N \\
\end{bmatrix} 
\quad
\mathcal{B} =
\begin{bmatrix}
b_0\\b_1\\ \vdots \\b_N \\
\end{bmatrix}.
\label{Eq.UVB}
\end{gather}

This system of coupled PDEs can be discretized in the spatial domain by various numerical methods such as finite element or finite difference methods. For a coarse-resolution computational grid and a small number of random variables, this system can be solved on a typical desktop (depending on the memory available). However, in the case of a large number of random variables and a coarse discretization, the system size increases significantly prohibiting the computations on the desktop. Similarly, even for a small number of random variables and high-resolution computational grids the memory requirement for the problem is overwhelming for a desktop. Furthermore, problems with high-resolution grids and a large number of random variables necessitate a parallel scalable solver. 

Domain decomposition and multigrid methods offer ways to decompose a domain and efficiently distribute the workload to several cores of a cluster. The exhaustive computational requirements of the stochastic PDEs with high resolution and a large number of random variables are handled in this article using a hybrid approach combing concepts from domain decomposition and multigrid. Even though non-overlapping methods such as FETI-DP \cite{nonlinearpc_3,ghosh_feti} and BDDC \cite{bddc_mandel} are developed for linear and nonlinear problems in the deterministic setting, we utilize overlapping methods in this article owing to the robustness and easiness of implementation. A two-grid solver with restricted additive Schwarz in the 
The next section describes the two-grid overlapping Schwarz solver in the stochastic setting.

\section{Two-grid Overlapping Schwarz Solvers}\label{sec.two-grid}
An overlapping DD-based solver is proposed for solving the stochastic system developed in the previous section. Even though many components of a deterministic DD-based solver and stochastic DD-based solver are similar, the size and structure of matrices and vectors differ in these two cases. The linear/linearized system matrices involve $m_{ijk}$ and $t_{ijkl}$ tensors which directly influence their sparse structure \cite{knio_book}. The sparsity of these tensors changes with an increasing number of random variables or order of expansion \cite{knio_book}. The values of these coefficients depend on the number of random variables and the order of expansion used in the input and output PCEs. Moreover, the strong coupling between the solution PCE coefficients may increase the condition number of the system \cite{sousedik_2} (see Appendix \ref{app:condnumbersto}) which becomes more pronounced with an increasing number of random variables and order of expansion as evident from the numerical experiments as explained later.

\begin{figure}
\centering    
\begin{subfigure}[b]{0.475\textwidth}
      		 \centering
        \includegraphics[width=\textwidth]{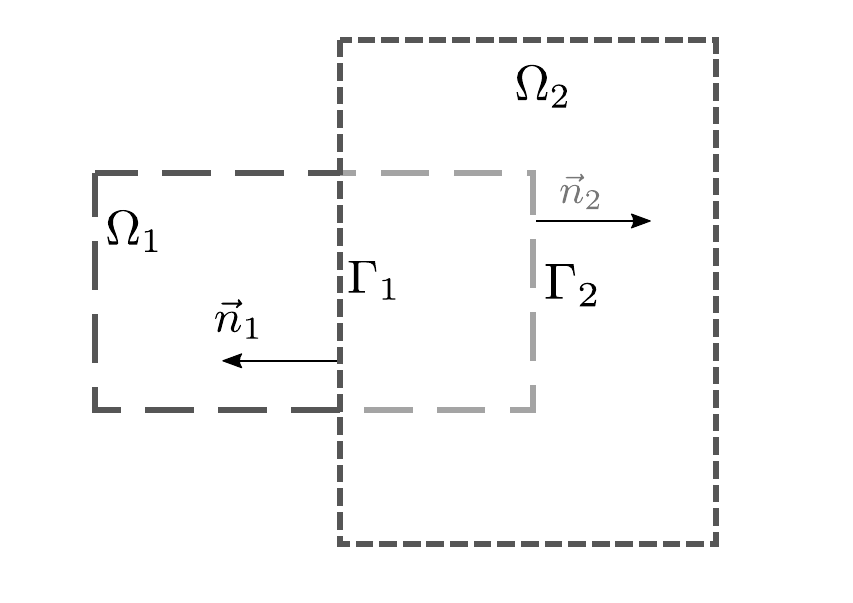} 
        \caption{Overlapping DD}\label{Fig.overlap}
    \end{subfigure} 
        \begin{subfigure}[b]{0.475\textwidth}
       \centering
        \includegraphics[width=\textwidth]{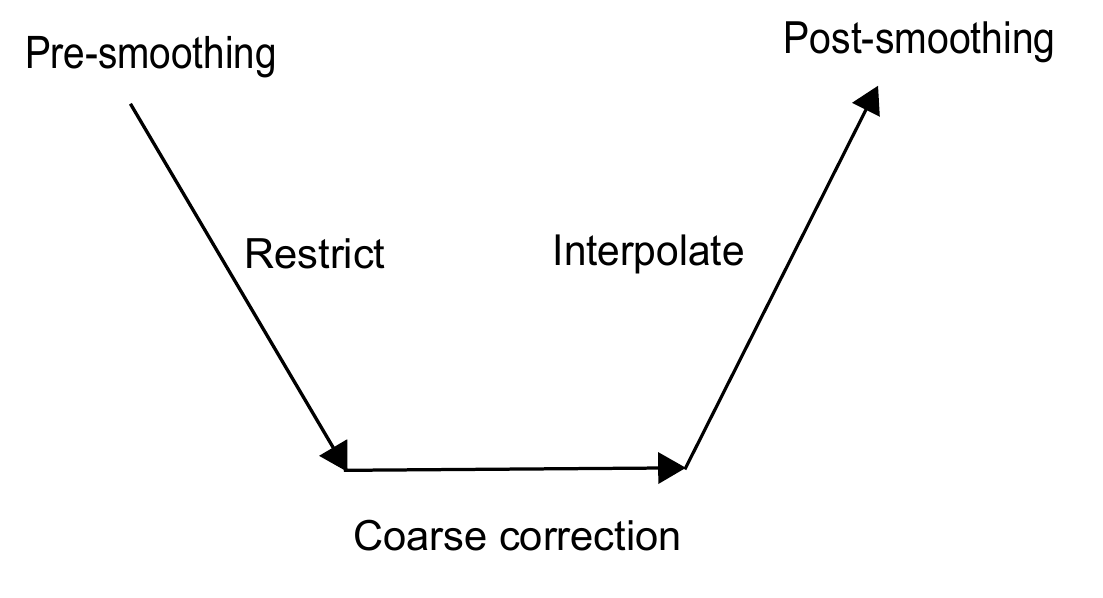} 
        \caption{Architecture of the two-grid Schwarz preconditioner}\label{Fig.arch}
    \end{subfigure}
    \caption{Components of two-grid Schwarz preconditioner}
\end{figure}

The overlapping DD decomposes the problem into several subdomains, which can be solved independently by suitably exchanging information on the overlap. Fig.~\ref{Fig.overlap} shows the two subdomains $\Omega_1$ and $\Omega_2$ and the artificial boundaries $\Gamma_1$ and $\Gamma_2$ created inside each subdomain by the other. Different variants of these methods can be developed by combining the solutions of the subproblems on the overlap \cite{book_victorita}. The restricted additive Schwarz (RAS) iterative method for a linear system $\mathcal{A}\mathcal{U} = \mathcal{F}$ can be written as a fixed-point iteration \cite{book_victorita}:

\begin{align}
		\mathcal{W}^n &= \mathcal{F} - \mathcal{A} \mathcal{U}^{n} \\
        \mathcal{U}^{n+1} &= \mathcal{U}^n + \mathcal{M}^{-1}_{\rm{RAS}} \mathcal{W}^n.
        \label{Eq.RASstofixedpoint}
\end{align}
Here $\mathcal{W}^n$ and $\mathcal{U}^{n}$ are the residual and solution respectively at the $n^{th}$ iteration and
\begin{equation}\label{Eq.RASsto}
\mathcal{M}^{-1}_{\rm{RAS}} = \sum_{i=1}^N \mathcal{R}_i^T \mathcal{D}_i (\mathcal{R}_i \mathcal{A} \mathcal{R}_i^T)^{-1} \mathcal{R}_i
\end{equation}
is the one-level RAS preconditioner, where the matrices $\mathcal{R}_i, \mathcal{R}_i^T, \mathcal{D}_i$ represents the subdomain level restriction operator transferring the solution vector on to the local subdomain, the extension matrix reversing the operation and the Boolean square matrix used for scaling the residuals of each subdomain respectively such that $\mathcal{I} = \sum_{i=1}^N \mathcal{R}_i^T \mathcal{D}_i \mathcal{R}_i $ ($\mathcal{I}$ being the identity matrix). \textcolor{ss}{Even though DD methods can be applied to PDEs as iterative solvers following the fixed-point iteration scheme as in Eq.~(\ref{Eq.RASstofixedpoint})},  they are better suited as preconditioners to Krylov solvers which have better convergence rates \cite{book_victorita}. Thus, we construct solvers with DD-based preconditioners for GMRES iterative method for the stochastic systems. We refer the readers to monographs such as \cite{book_DD_barryFsmith,book_victorita,dd_TFchan,book_DD_widlund,book_quarteroni} for a detailed analysis of various DD-based methods.

One-level DD methods which exchange information only to adjacent subdomains lose scalability with an increasing number of subdomains \cite{book_DD_barryFsmith,book_victorita,dd_TFchan,book_DD_widlund,book_quarteroni}.
A method to propagate the residual in each subdomain globally can be developed by adding a coarse correction to the one-level preconditioners. However, global communication among subdomains incurs an additional parallel overhead which needs to be minimized for the scalable performance of the solvers. Thus, these corrections often become the bottleneck for developing scalable solvers needed for high-resolution (spatial, temporal, and stochastic) computational models. Several variants of coarse corrections are used in the literature to effectively remove the low-frequency error components which remain after the fine grid (one-level) solve \cite{multilevel_pc1,dd_adapted_elastodynamics,deflation>coarsecorrection,NN_BDDC}. This article uses a coarse matrix generated from the coarser triangulation of the domain as follows \cite{book_victorita}:

\begin{equation}
		\mathcal{A}_c = \mathcal{R}_{0} \mathcal{A} \mathcal{R}_{0}^T
        \label{Eq.coarse}
\end{equation}
where $\mathcal{R}_{0}^T$ is the interpolation operator from fine grid to coarse grid. Each solve step in the coarse grid involving $\mathcal{A}_c^{-1}$ could also be solved using GMRES and an associated preconditioner. \textcolor{ss}{Three different solvers are proposed based on a two-grid architecture (see Fig.~\ref{Fig.arch}) with RAS preconditioner for the fine-grid and varying preconditioners for coarse corrections as:}

\begin{itemize}

\item[•] Two-grid RAS: The two-grid RAS preconditioner with the coarse problem solved using an exact LU factorization (parallel MUMPS (\cite{mumps_2}) solver). \textcolor{ss}{Note, numerical experiments using this preconditioner are presented only in the \ref{sec:overlapdd_deter} for the deterministic setting. This is because of the large computational time requirement of the solver compared to the other variants.}

\item[•] Two-grid RAS-V2 (2GV2): The two-grid RAS preconditioner with the coarse problem solved using an inner GMRES equipped with a one-level RAS preconditioner.

\item[•] Two-grid RAS-V3 (2GV3): The two-grid RAS preconditioner with the coarse problem solved using an inner GMRES equipped with AMG preconditioner \cite{boomeramg}.
\end{itemize}

\textcolor{ss}{For the current implementation, the 2GV2 preconditioner has the number of coarse iterations restricted to $100$ and the Krylov solver tolerance (the relative decrease in the residual norm with respect to the norm of the right hand side \cite{petsc}) for the coarse grid (inner GMRES iteration) set to $10^{-2}$ compared to a tolerance of $10^{-5}$ for the fine grid (outer iterations). This is to ensure a faster solution time while using 2GV2, which can have large number of coarse iterations owing to its one-level RAS preconditioner. For other variants of the preconditioner, the default tolerance (as per PETSc \cite{petsc}) of $10^{-5}$ for both inner and outer Krylov solvers.
The coarse grid is distributed to the same number of processes as the fine grid and each sub-block solves for both fine and coarse grid (for 2GV2) are computed using incomplete LU factorization. A variant of GMRES called flexible GMRES (FGMRES) is utilized which permits variable preconditioners to be used in each iteration \cite{FGMRES}. This is because the outer GMRES iteration has a preconditioner which changes at each iteration due to the updated coarse grid corrections. The 2GV3 version uses a parallel implementation of AMG called BoomerAMG \cite{boomeramg}. The number of levels used for the V cycle of AMG is automatically set by BoomerAMG. All implementations leverage FreeFEM++ \cite{FreeFEM} and PETSc \cite{petsc}.}

The architecture of the preconditioner is 
presented in Fig.~\ref{Fig.arch} which resembles a multigrid V-cycle. The pre-smoothing and post-smoothing steps are carried out using a one-level RAS solver which reduces the high-frequency components of the error. Next, the errors are further restricted to the coarse grid to find the appropriate correction. The updated solution is interpolated back to the fine grid and smoothed out again which completes the application of the preconditioner. The final form of this preconditioner can be written as follows \cite{sudhi_MBE}:
\begin{align} \label{Eq.RAS_2L}
\mathcal{M}_2^{-1} = \mathcal{M}_{\rm{RAS}}^{-1} \mathcal{P} + \mathcal{Q}\mathcal{P}\mathcal{Q}^{-1} \mathcal{M}_{\rm{RAS}}^{-1} + \mathcal{Q} - \mathcal{M}_{\rm{RAS}}^{-1} \mathcal{P} \mathcal{A} \mathcal{M}_{\rm{RAS}}^{-1},
\end{align}
where $\mathcal{Q} = \mathcal{R}_{0}^T \mathcal{A}_{c} ^{-1} \mathcal{R}_{0}$ is the coarse correction and  $\mathcal{P} = (\mathcal{I} - \mathcal{AQ})$ is a projection matrix. The preconditioner equation presented is the combined action of fine grid, coarse grid corrections and restriction/interpolation operators exchanging between these grids. The preconditioner action is applied to the residual in a sequence as shown in Fig.~\ref{Fig.arch}. This multiplicative correction approach whereby the solution is updated at each level makes the implementation sequential between levels but provides faster error reduction.
A more detailed derivation and implementational details of the preconditioner are provided in \cite{sudhi_MBE}. The next section presents the numerical experiments comparing the two variants \textcolor{ss}{(comparison using Two-grid RAS is presented only for deterministic setting as shown in \ref{sec:overlapdd_deter})} of the solver for stochastic linear and nonlinear Poisson problems.

\section{Numerical Results}\label{sec.chapt4speedupeffic}
To highlight the benefits of the proposed solvers, numerical results are presented in this section for linear and nonlinear Poisson problems in deterministic and stochastic settings. We compare the performance of the two-grid Schwarz preconditioners for strong and weak scaling to illustrate the benefits of the proposed solver. Strong scaling involves the study of iteration counts or time to solution with an increasing number of processes for a fixed problem. However, weak scaling is the study of iteration counts and time to solution by increasing the global problem size while fixing the problem size per process constant (with a corresponding increase in the number of processes).

The scalability with an increasing number of random variables and order of expansions is also provided for linear and nonlinear Poisson problems for the proposed method. All implementations are carried out in FreeFEM \cite{FreeFEM} and PETSc libraries \cite{petsc}. The outer iterative solver for the two-grid preconditioners is FGMRES (flexible GMRES) which enables variable preconditioners in each iteration \cite{FGMRES}. More details on implementation are available in \cite{sudhi_MBE}. 
The MCS method is used to verify the solution obtained for the stochastic Galerkin method. The underlying Gaussian process has a mean $\mu = 0$ and standard deviation $\sigma = 0.3$. The Gaussian process has an exponential covariance kernel as in Eq.~(\ref{Eq.CovarKer2DNLP}) with the correlation lengths $b_x = b_y = 1$ in each direction. We use $2^{nd}$ order PCE for input and $3^{rd}$ order PCE for output unless stated otherwise. The domain for all numerical experiments is the unit square. 

To compare the two different preconditioners in detail, we present results based on the preconditioner (PC) setup and Krylov subspace solver (KSP) solve time (execution time) generated from PETSc profiling data. \textcolor{ss}{The PC setup time involves the time required for the assembly of the components of the preconditioner. The preconditioner is not constructed explicitly but its action on a vector is computed in parallel at the subdomain level. For the one-level RAS preconditioner as in Eq.~(\ref{Eq.RASsto}), the PC setup time includes the assembly of $\mathcal{R}_i$, $(\mathcal{R}_i \mathcal{A} \mathcal{R}_i^T)$ and $ \mathcal{D}_i$ per process. The factorization of subdomain matrices are considered as PC setup on blocks time in PETSc which is added towards the PC setup time and subtracted from KSP solve time.} The time per process is reported since the time taken for all processes and a single process is almost similar with a very low load imbalance between processes. The KSP solve time includes the total time taken for the Krylov solver from the beginning of the iteration to its end (except the PC setup time). Similar to the PC setup time, we report the KSP solve time also for a single process.

We also present speedup and efficiency plots for strong scaling. Speedup is computed as the ratio of time taken for a sequential solver to the time taken using a parallel solver \cite{book_HPC,book_DD_tarek,book_DD_barryFsmith}. Thus, speedup $S = \frac{T_s}{T_p}$, where $T_s$ and $T_p$ are the time taken for sequential solver and parallel solver respectively. Even though a more accurate analysis of speedup can be established by separating the time required for sequential and parallel parts of a program \cite{book_HPC}, we use the simplified version for convenience. The efficiency for a strong scaling experiment can be computed as $E_s = \frac{S}{np}$, where $np$ is the number of processes used for the parallel solver. On the other hand, the efficiency for a weak scaling experiment is computed as $E_w = \frac{T_s}{T_p}$ where $T_s$ and $T_p$ are the sequential solve time (or equivalent parallel solver time, e.g., $80$ processes used in Fig.~\ref{Fig:weakeff_LP}c)  and parallel solve time respectively \cite{book_HPC,book_victorita}.
Even though we report the total time to solution in the tables of numerical results section along with iteration counts, for speedup and efficiency computations we utilize the PC setup and KSP solve times which essentially characterize the solver without the pre-processing and post-processing parts.

\subsection{Verification for Stochastic Linear Poisson Problem}\label{sec:verify_LP}

\begin{figure}[htbp]
    \centering
    \begin{subfigure}[b]{0.475\textwidth}
       \centering
        \includegraphics[width=\textwidth]{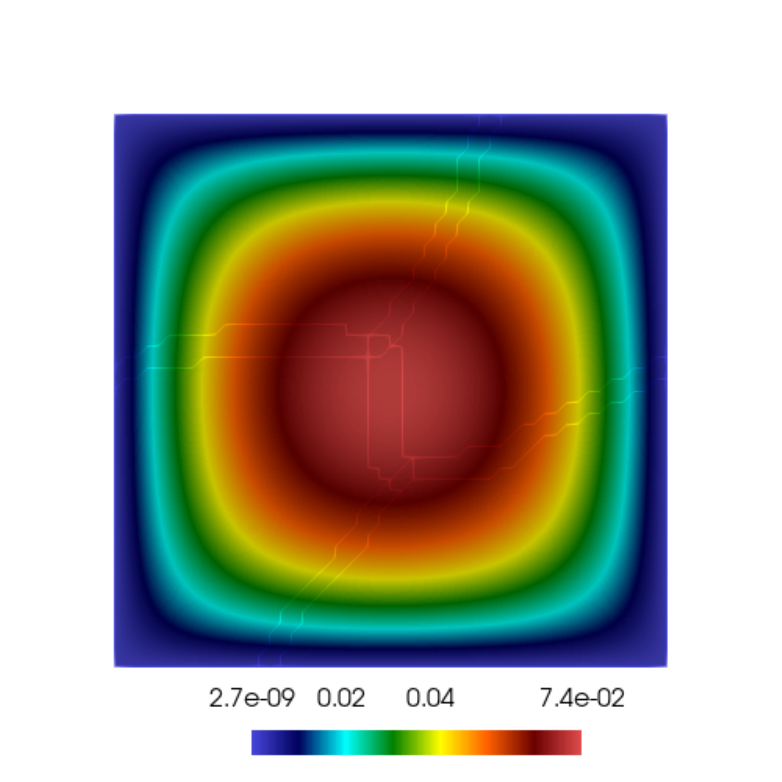} 
        \caption{Mean (SG)}
    \end{subfigure}
    \begin{subfigure}[b]{0.475\textwidth}
      		 \centering
        \includegraphics[width=\textwidth]{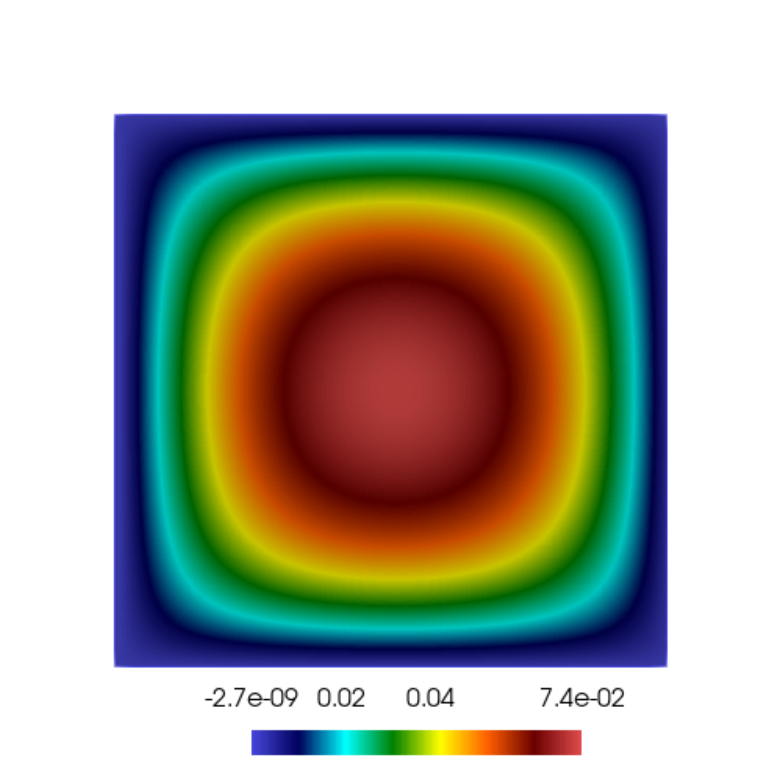} 
        \caption{Mean (MCS)}
    \end{subfigure} 
    \centering
    \begin{subfigure}[b]{0.475\textwidth}
       \centering
        \includegraphics[width=\textwidth]{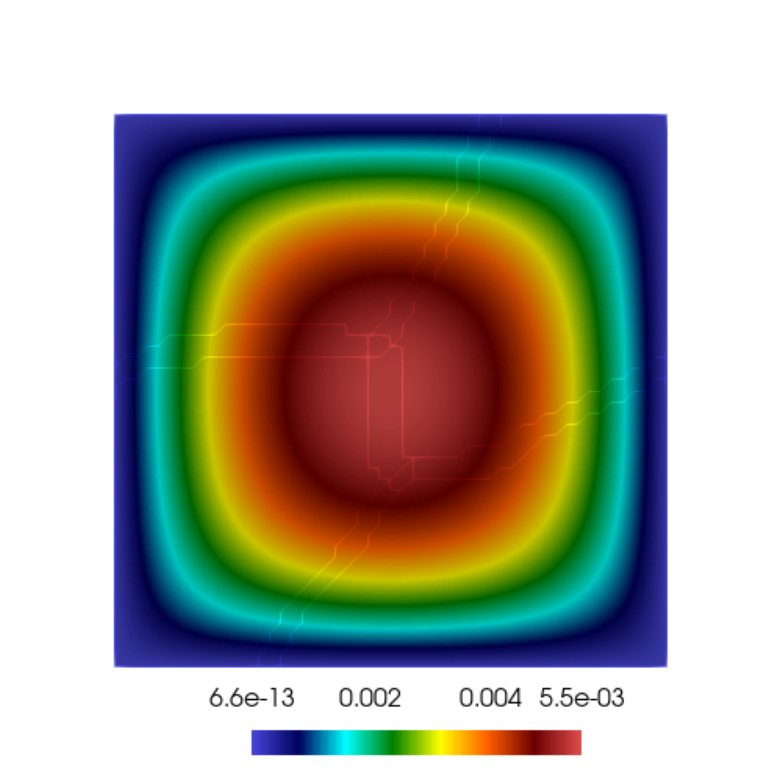} 
        \caption{Standard deviation (SG)}
    \end{subfigure}
    \begin{subfigure}[b]{0.475\textwidth}
      		 \centering
        \includegraphics[width=\textwidth]{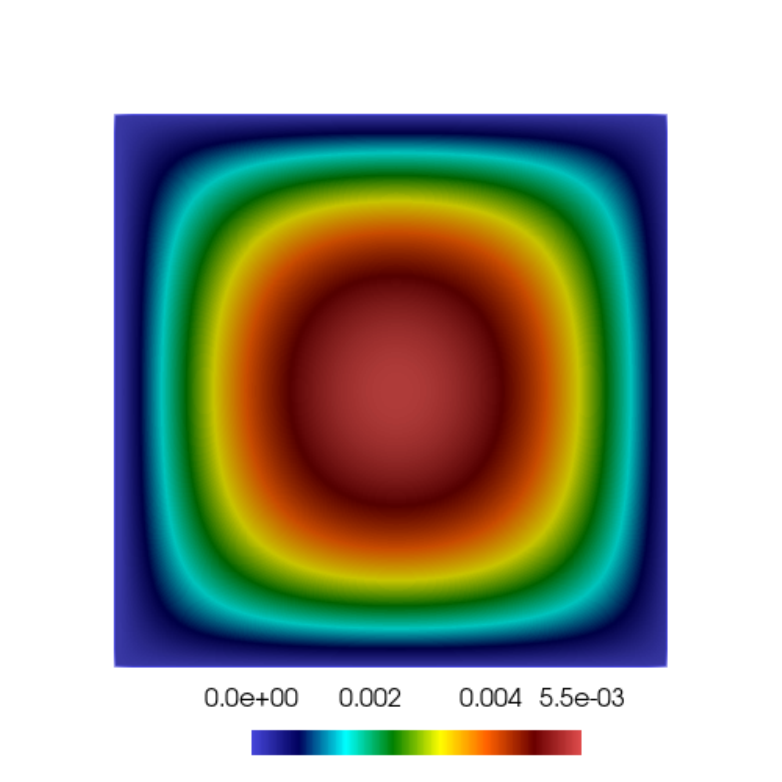} 
        \caption{Standard deviation (MCS)}
    \end{subfigure}
  \caption{Mean and standard deviation of the intrusive stochastic Galerkin (SG) method and MCS for the linear Poisson problem}\label{Fig:mcs_verify_poisson}
\end{figure}

\begin{figure}[htbp]
    \centering
\begin{subfigure}[b]{0.475\textwidth}
       \centering
        \includegraphics[width=\textwidth]{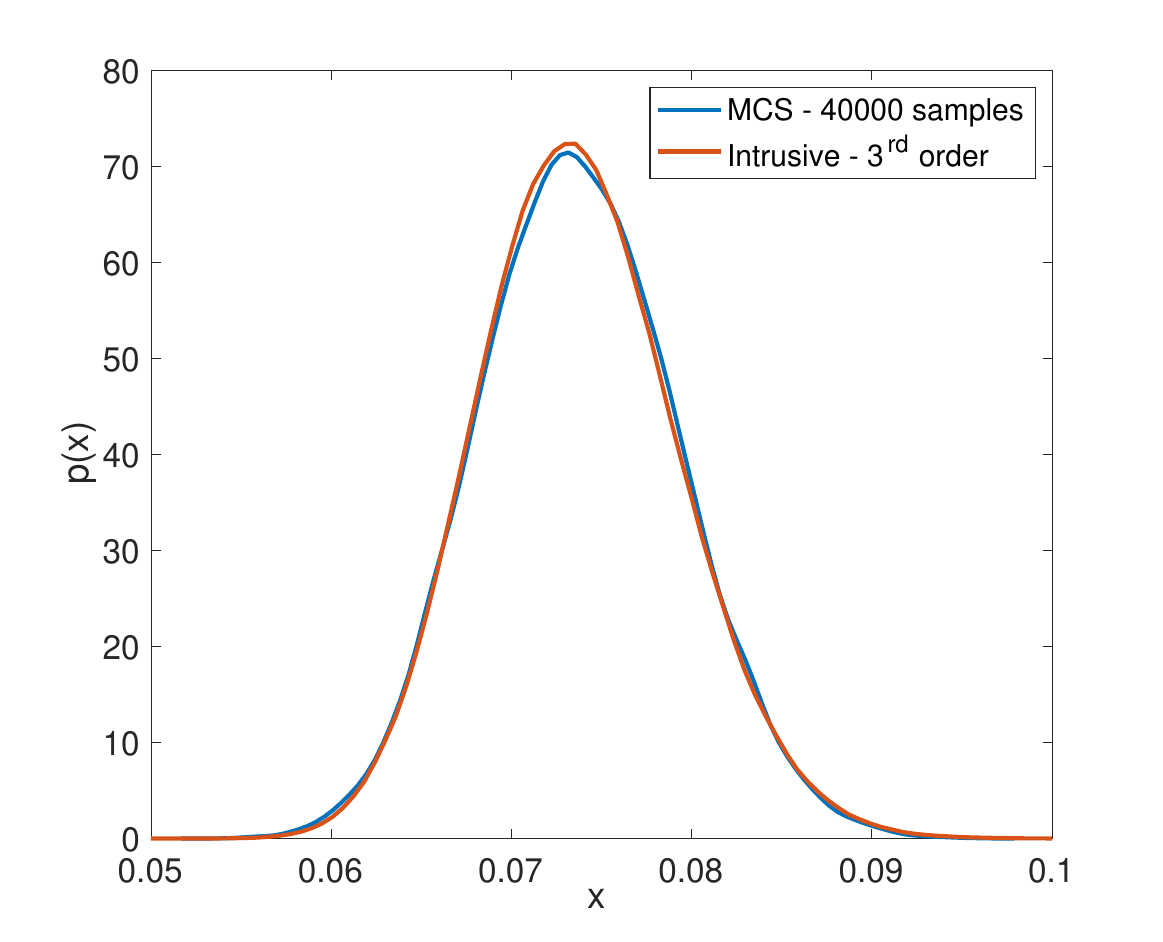} 
        \caption{pdf at the center of domain $(x, y) = (0.5, 0.5)$}
    \end{subfigure}
    \begin{subfigure}[b]{0.475\textwidth}
      		 \centering
        \includegraphics[width=\textwidth]{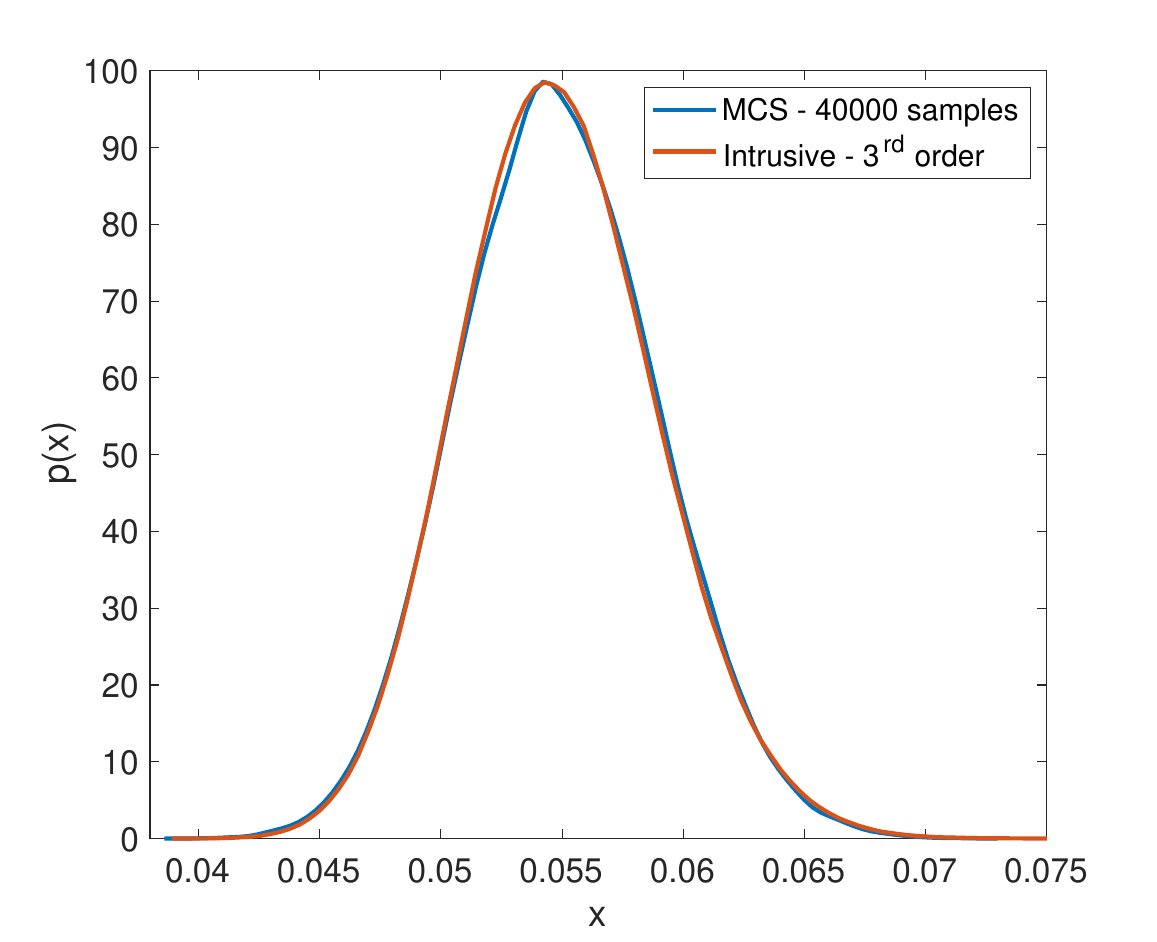} 
        \caption{pdf at $(x, y) = (0.3, 0.7)$}
    \end{subfigure}
\caption{pdf at two points on the domain using MCS and intrusive stochastic Galerkin (SG) method for the linear Poisson problem}\label{Fig:mcs_verify_poisson_pdf}
\end{figure}

The stochastic linear Poisson problem is verified using MCS in this section. An overlapping domain decomposition with four subdomains on a square domain discretized with $10{,}201$ vertices is used for intrusive stochastic Galerkin implementation (referred to as SG or intrusive in captions). The log-normal input field is expanded with $3$ random variables and $2^{nd}$ order expansion while output expansion uses up to $3^{rd}$ order terms. The underlying Gaussian process (used to construct the log-normal input diffusion coefficient) has a mean $\mu = 0$ and standard deviation $\sigma = 0.1$ respectively. An exponential covariance kernel is used to represent the underlying Gaussian process as described in Eq.~(\ref{Eq.CovarKer2DNLP}). The MCS results are generated by computing the statistics of $40{,}000$ samples. The convergence of the MCS is ensured although not shown explicitly. The mean and standard deviation of the solution from the stochastic Galerkin method match with the MCS results as shown in Fig.~\ref{Fig:mcs_verify_poisson}. The pdf of the solution at two different points for both approaches is also shown in Fig.~\ref{Fig:mcs_verify_poisson_pdf}.

\subsection{Scalability for Stochastic Linear Poisson Problem with Mesh Size}

The strong and weak scalabilities of the two-grid RAS solver for the stochastic linear Poisson problem on the unit square are demonstrated in this section. The input has a $2^{nd}$ order expansion with an underlying Gaussian field having the mean of $\mu = 0$ and standard deviation $\sigma = 0.3$. The input Gaussian field is expressed by $3$ random variables and the output has a $3^{rd}$ order expansion having $20$ PCE terms.

The strong scalability with an increasing number of processes from $80$ to $800$ is shown in Table~\ref{tab:ss_lp}. The GMRES iteration counts for the 2GV2 solver are found to be higher than the 2GV3 solver for all processes. However, for both solvers, the iteration counts do not change significantly. The total time to solution for both solvers decreases with an increasing number of processes demonstrating strong parallel scaling. The 2GV3 has a lower time to solution than 2GV2 because of the improved algebraic multigrid coarse preconditioner.

\begin{table}[htbp]
    \centering
    \begin{tabular}{|c|c|c|c|c|c|c|}
    \hline
      \multirow{2}{*}{\specialcell{Number of \\ processes}} &  \multicolumn{2}{c|}{\specialcell{GMRES \\iteration count}} & \multicolumn{2}{c|}{Total time (s)}\\
      \cline{2-5}
      & 2GV2 & 2GV3  & 2GV2 & 2GV3 \\
       \hline
        \hline
      $80$ & $12$ & $7$ & $122.6$ & $71.00 $  \\      
      $160$ & $12$ &$7$  & $61.10$ & $37.35 $  \\
      $320$ & $12$ & $7$ & $33.61$ & $19.32 $   \\
      $400$ & $13$ & $7$  & $29.45$ & $16.95 $   \\
      $640$ & $13$ & $7$  & $18.73$ & $12.07 $   \\
      $800$ & $13$ & $7$  & $17.51$ &  $10.34$   \\
      \hline
    \end{tabular}
    \caption{Strong scalability for the stochastic linear Poisson problem with a total problem size of 12.83 million dof.}
    \label{tab:ss_lp}
\end{table}

\begin{figure}[htbp]
    \centering
    \begin{subfigure}[b]{0.475\textwidth}
       \centering
        \includegraphics[width=\textwidth]{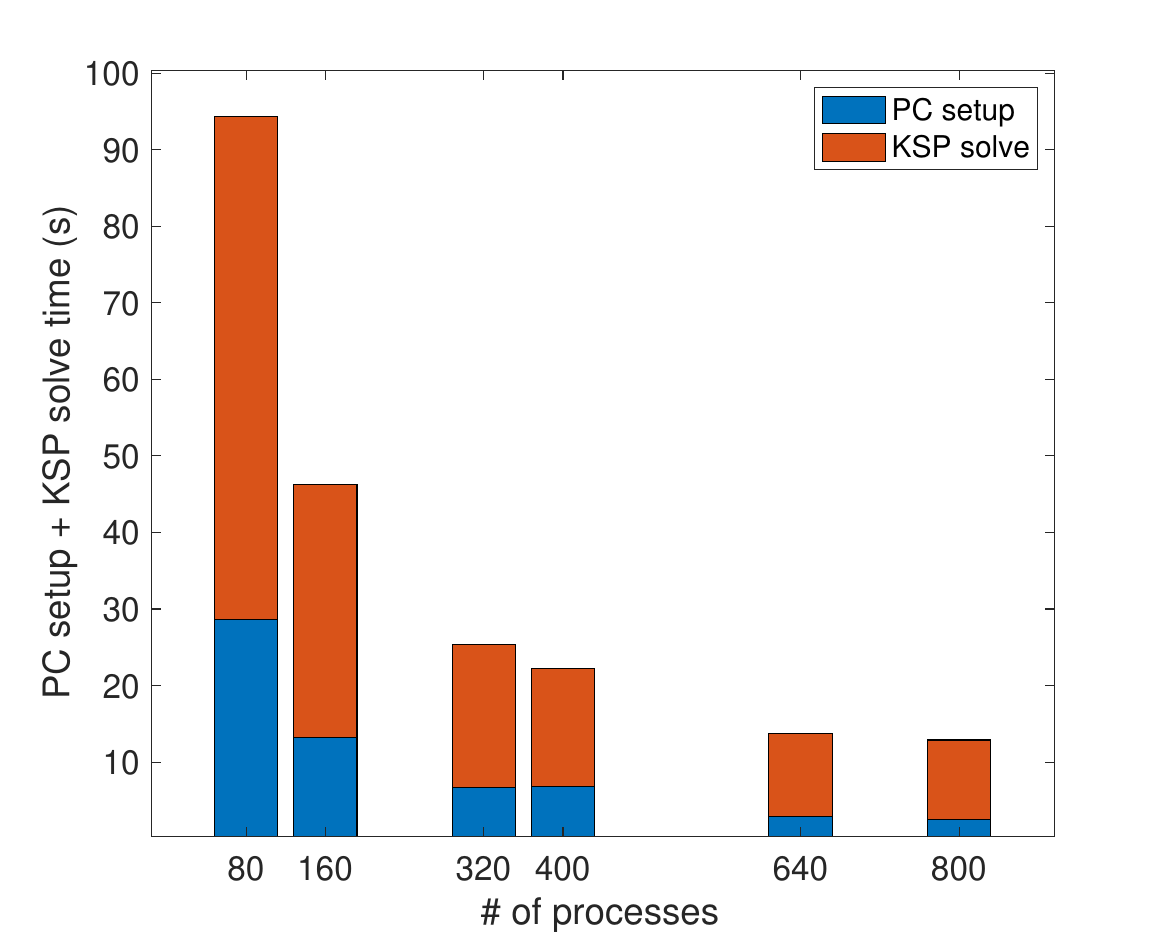} 
        \caption{2GV2}
    \end{subfigure}
        \begin{subfigure}[b]{0.475\textwidth}
       \centering
        \includegraphics[width=\textwidth]{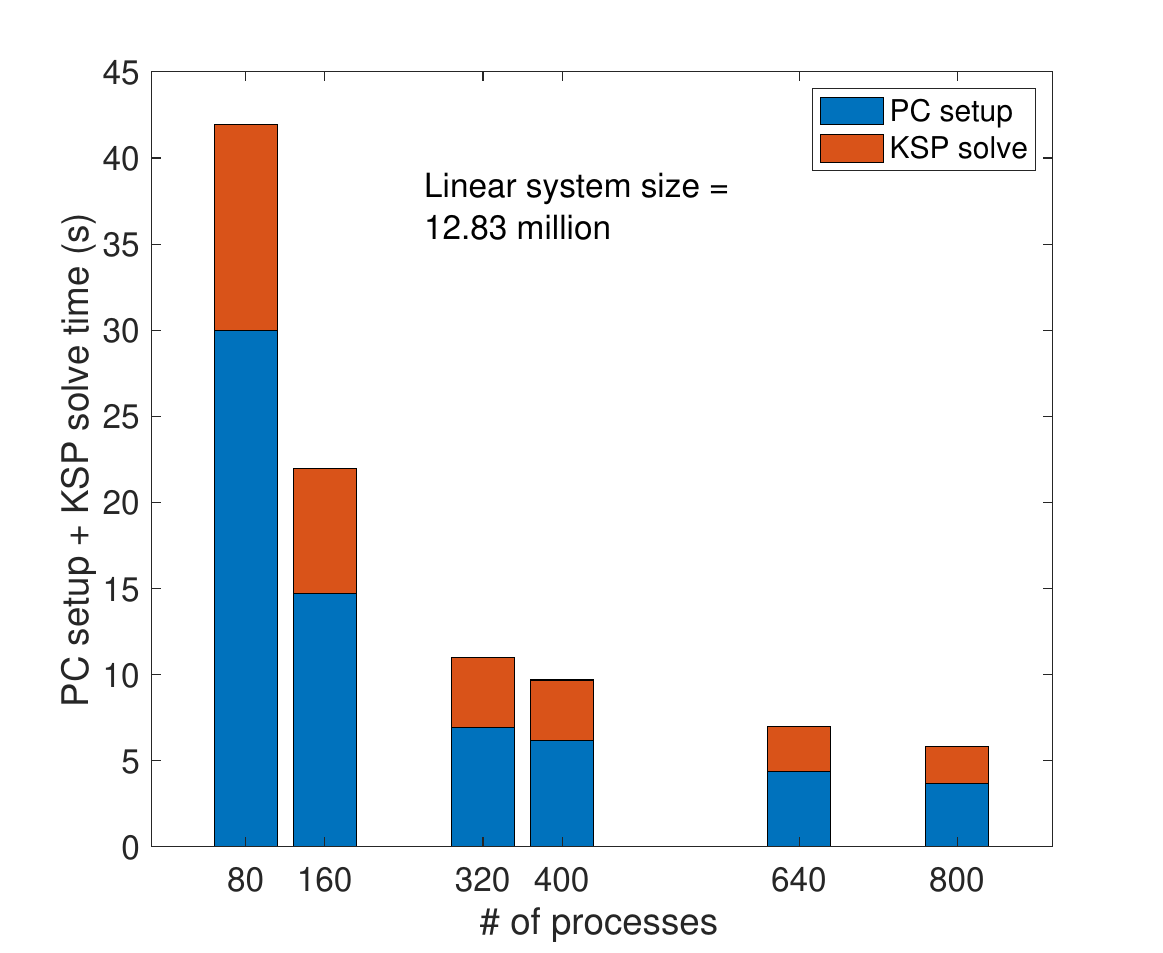} 
        \caption{2GV3}
    \end{subfigure}
    \begin{subfigure}[b]{0.475\textwidth}
      		 \centering
        \includegraphics[width=\textwidth]{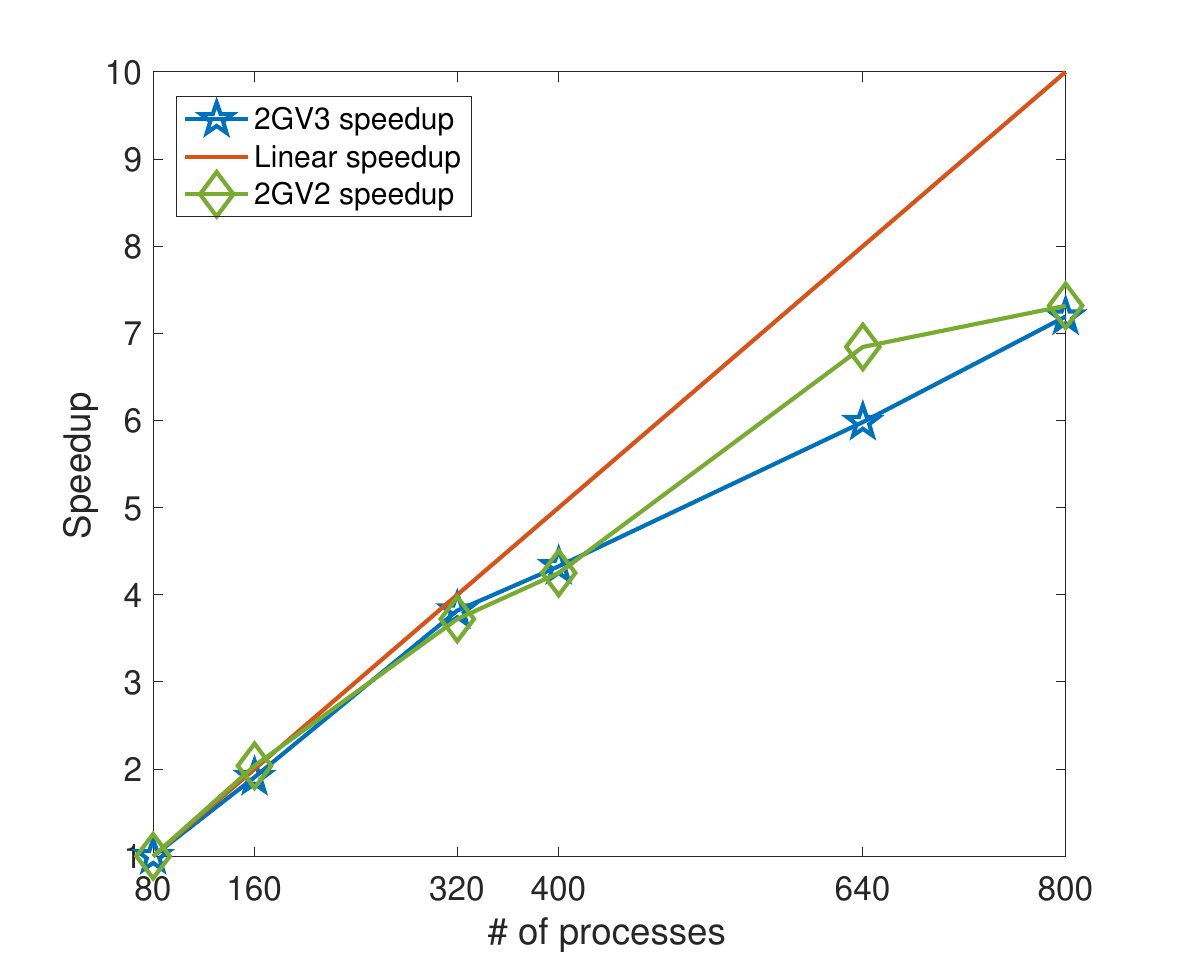} 
        \caption{Speedup}
    \end{subfigure} 
    \centering
    \begin{subfigure}[b]{0.475\textwidth}
       \centering
        \includegraphics[width=\textwidth]{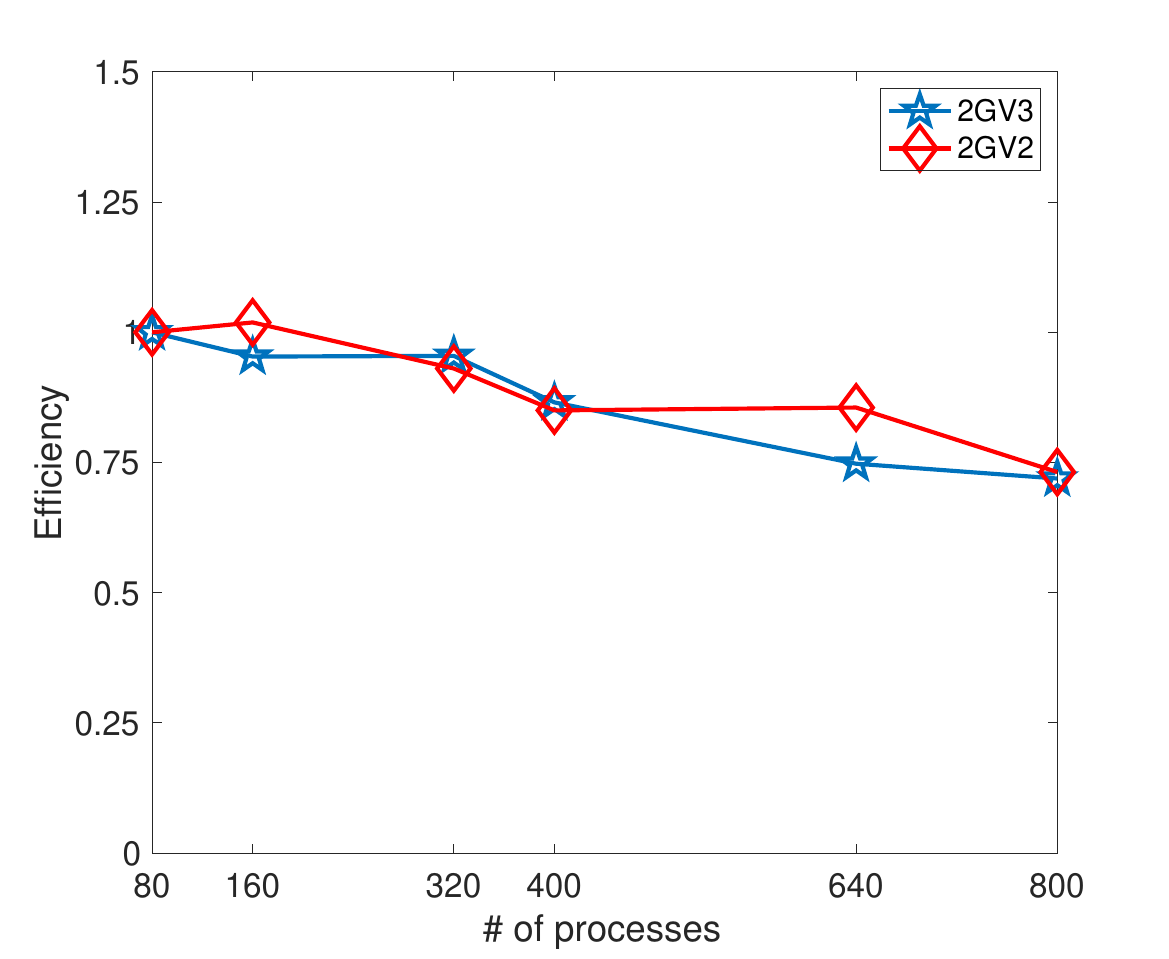} 
        \caption{Efficiency}
    \end{subfigure}
  \caption{Setup time, speedup, and efficiency comparison of preconditioners for the stochastic linear Poisson problem}\label{Fig:speedup_LP}
\end{figure}

Fig.~\ref{Fig:speedup_LP} shows the strong parallel scalability results for the solvers 2GV2 and 2GV3. The time spent per process on PC setup and KSP solve for both 2GV2 and 2GV3 are shown in Fig.~\ref{Fig:speedup_LP}a and Fig.~\ref{Fig:speedup_LP}b. It can be seen that the 2GV3 has a higher setup cost while the KSP solve time is significantly lower than 2GV2. This can be understood from the higher setup time required for the algebraic multigrid for 2GV3 than the one-level Schwarz preconditioner in the coarse grid for 2GV2. A comparison of the speedup and efficiency of both solvers is also shown in Fig.~\ref{Fig:speedup_LP}c and Fig.~\ref{Fig:speedup_LP}d. The speedup and efficiency of 2GV2 and 2GV3 are comparable for strong scaling. A decrease in efficiency by $25 \%$ while increasing the system size by $10$ times for both solvers is observed.

The weak scalability with an increasing global problem size while keeping the problem size per process fixed is provided in Table \ref{tab:ws_lp}. The GMRES iteration counts for the 2GV2 solver increase significantly with the global problem size while 2GV3 has constant iteration counts. The total time to solution for 2GV2 increases rapidly while 2GV3 does not show significant changes with increasing processes. In both strong and weak scaling studies, a significant improvement in iteration counts and total time to the solution can be seen for the 2GV3 preconditioner.

\begin{table}[htbp]
    \centering
    \begin{tabular}{|c|c|c|c|c|c|c|c|}
    \hline
      \multirow{2}{*}{\specialcell{ \small{Number of} processes \\ \small{(System size in million)} } }&  \multicolumn{2}{c|}{\specialcell{GMRES \\iteration count}} & \multicolumn{2}{c|}{Total time (s) }\\
      \cline{2-5}
      & 2GV2 & 2GV3  & 2GV2 & 2GV3 \\
       \hline
       \hline
      $80$ ($12.83$) & $12$ &$7$  & $122.6$ & $71.00 $  \\      
      $160$ ($25.67$) & $15$ &$7$  & $157.7$ & $80.07$ \\
      $320$ ($51.39$) & $21$ & $7$  & $240.3$ &$76.42$  \\
      $400$ ($64.29$) & $23$ &$7$  & $269.9$ & $89.81$  \\
      $640$ ($102.7$) & $29$ &$7$  & $296.2$ & $96.95$ \\
      $800$ ($128.5$) & $34$ &$7$  & $343.4$ & $94.56$  \\
      \hline
    \end{tabular}
    \caption{Weak scalability for the stochastic linear Poisson problem}
    \label{tab:ws_lp}
\end{table}

\begin{figure}[htbp]
     \centering
       \begin{subfigure}[b]{0.475\textwidth}
       \centering
        \includegraphics[width=\textwidth]{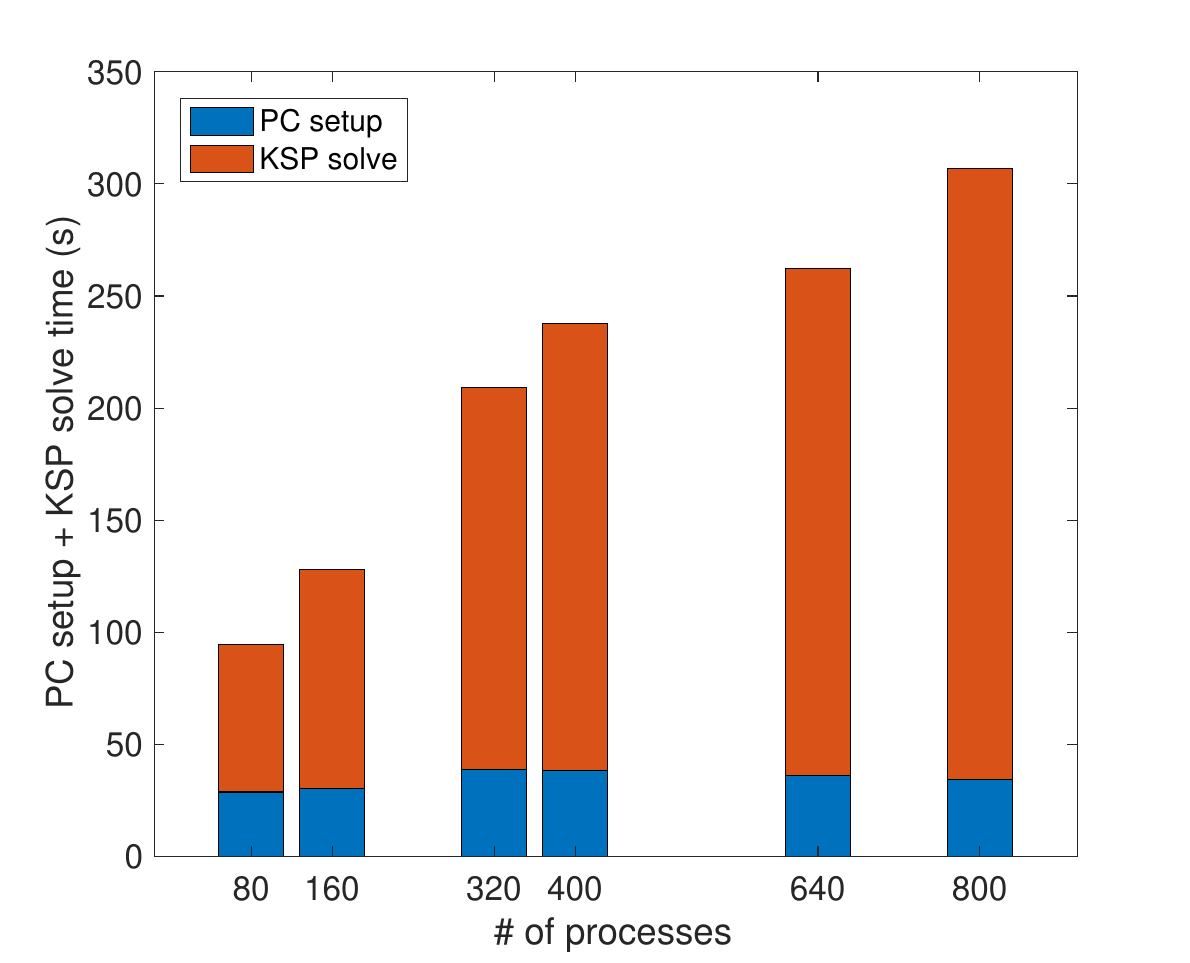} 
        \caption{2GV2}
    \end{subfigure}
    \centering
    \begin{subfigure}[b]{0.475\textwidth}
       \centering
        \includegraphics[width=\textwidth]{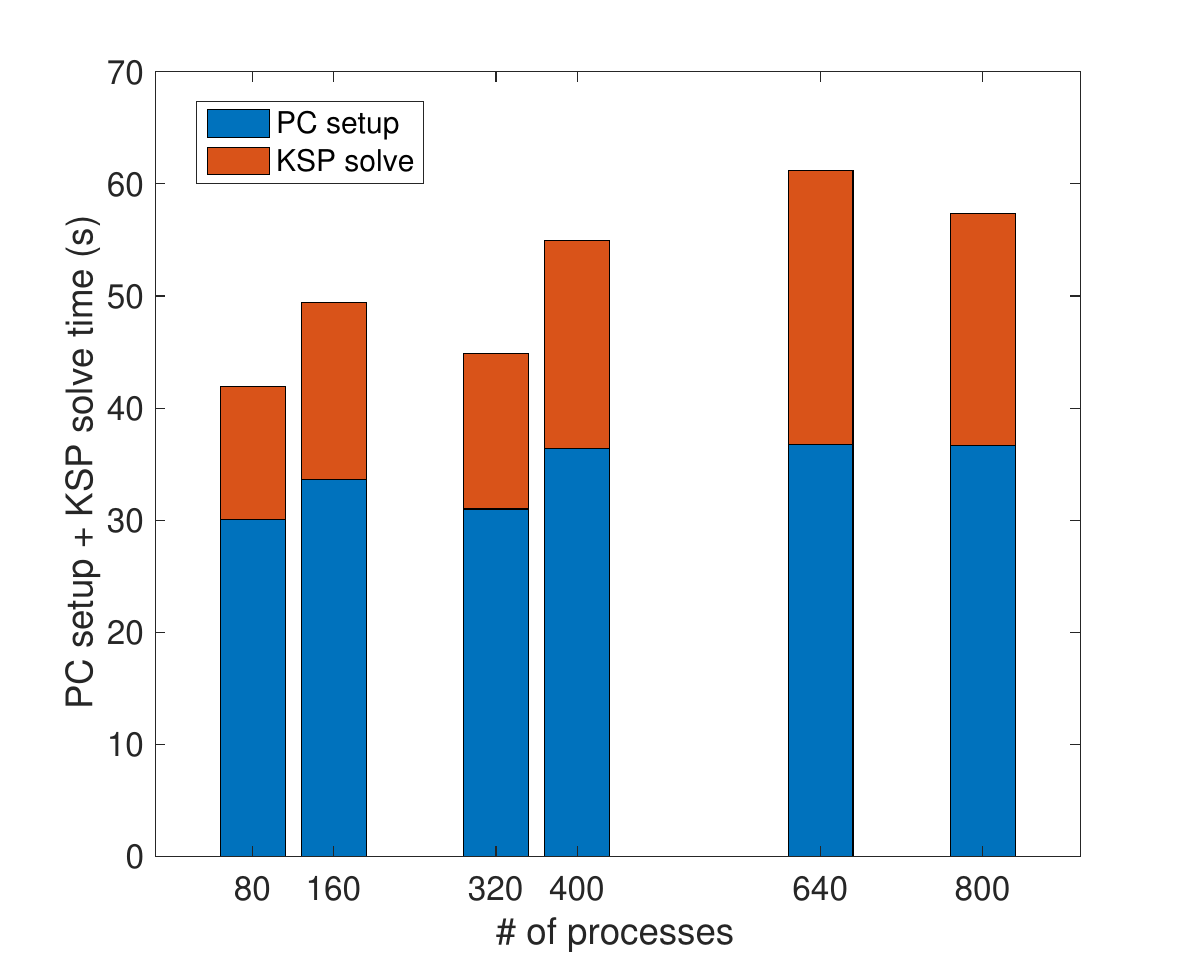} 
        \caption{2GV3}
    \end{subfigure}
    \centering
    \begin{subfigure}[b]{0.475\textwidth}
       \centering
        \includegraphics[width=\textwidth]{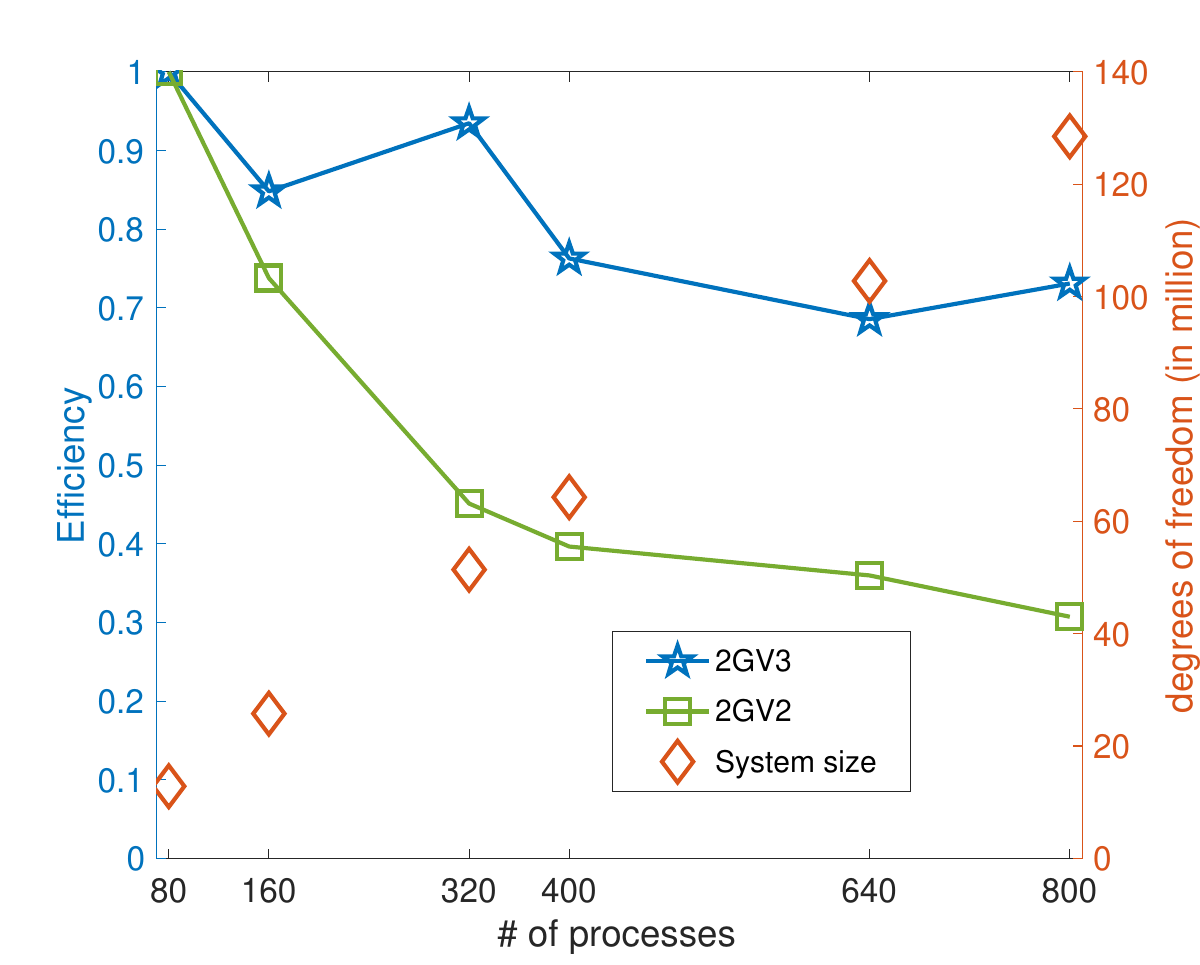} 
        \caption{Efficiency}
    \end{subfigure}
  \caption{Weak parallel scalability for the stochastic linear Poisson problem}\label{Fig:weakeff_LP}
\end{figure}

The weak parallel scaling for 2GV2 and 2GV3 is presented in Fig.~\ref{Fig:weakeff_LP}. Even though the PC setup time for 2GV2 does not increase with increasing global problem size and a fixed number of degrees of freedom per process in Fig.~\ref{Fig:weakeff_LP}a, the KSP solve time increases significantly. This can be understood from the poor performance of the one-level Schwarz preconditioner in the coarse grid. However, the PC setup and KSP solve time remain similar for an increasing number of processes in the case of 2GV3 (see Fig.~\ref{Fig:weakeff_LP}b) showing its improved performance due to AMG. This is also evident in the efficiency plot in Fig.~\ref{Fig:weakeff_LP}c, where 2GV2 shows a considerable reduction in efficiency up to $30 \%$ while 2GV3 maintains its efficiency close to $80 \%$. Even though 2GV2 shows comparable performance to 2GV3 for strong scaling, there is a considerable degradation in its performance for weak scaling.

\subsection{Scalability for Stochastic Linear Poisson Problem with Random Parameters}\label{sec.ss-lp}

In this section, the scalability of the preconditioner with respect to random parameters such as number of random variables and order of expansion is studied. A new type of PDE system with different coupling structure among PDEs emerges in the stochastic Galerkin method as the number of input random variables or the order of output expansion increases. This fact makes the weak scalability with respect to random parameters different from that of the scalability with respect to mesh size whereby the PDE system is solved with different discretization levels for a given geometry (as shown in Fig.~\ref{Fig:weakeff_LP}). \textcolor{ss}{Even though we solve different coupled system of PDEs for varying number of random variables or order of expansion, the scalability results are reported similar to the case of increasing mesh size for comparison. We note that the degradation of efficiency reported in these studies are not solely due to the performance of the solvers but also to the difference in the stochastic system solved with increasing number of random variable or order of expansion.}

\begin{figure}[htbp]
    \centering
    \begin{subfigure}[b]{0.475\textwidth}
       \centering
        \includegraphics[width=\textwidth]{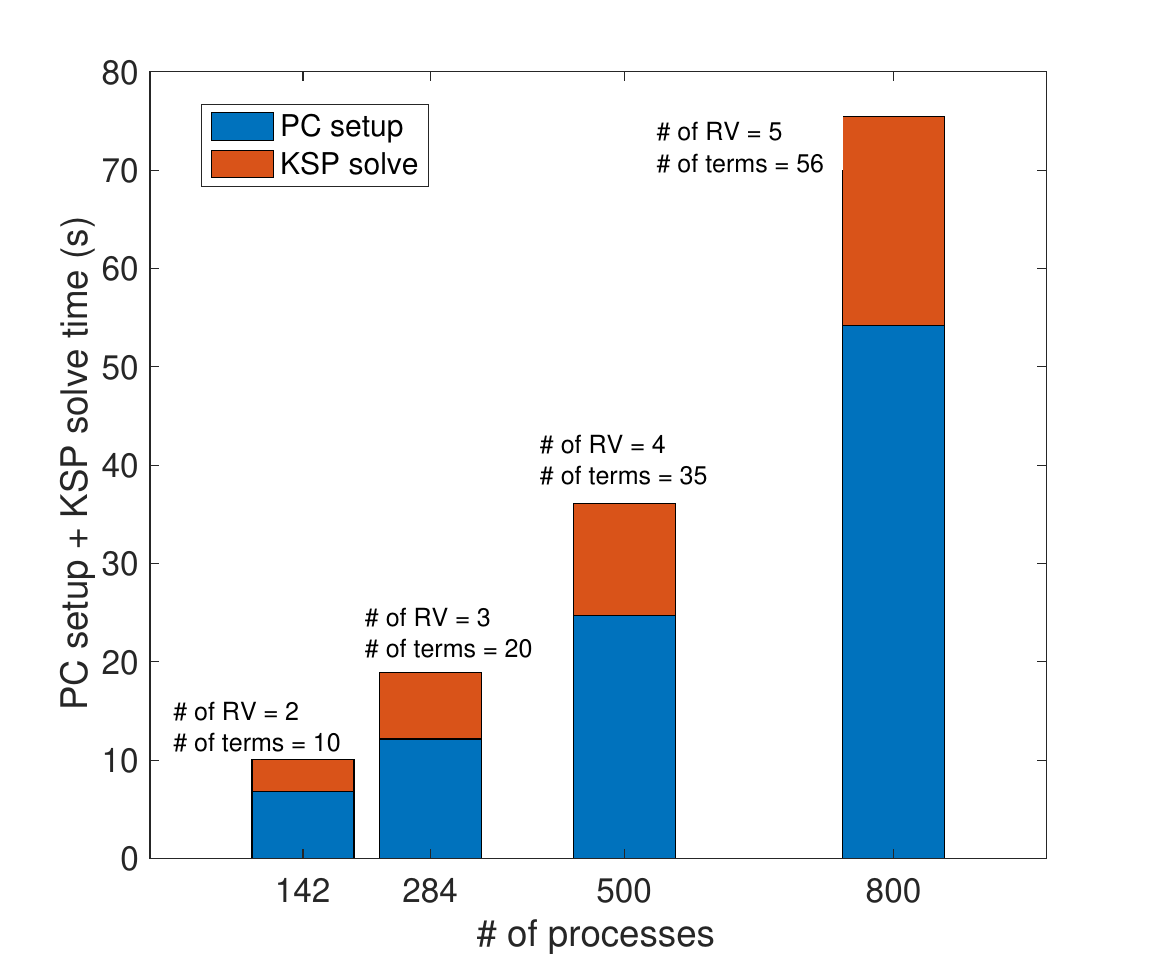} 
        \caption{PC setup and solve time}
    \end{subfigure}
    \centering
    \begin{subfigure}[b]{0.475\textwidth}
       \centering
        \includegraphics[width=\textwidth]{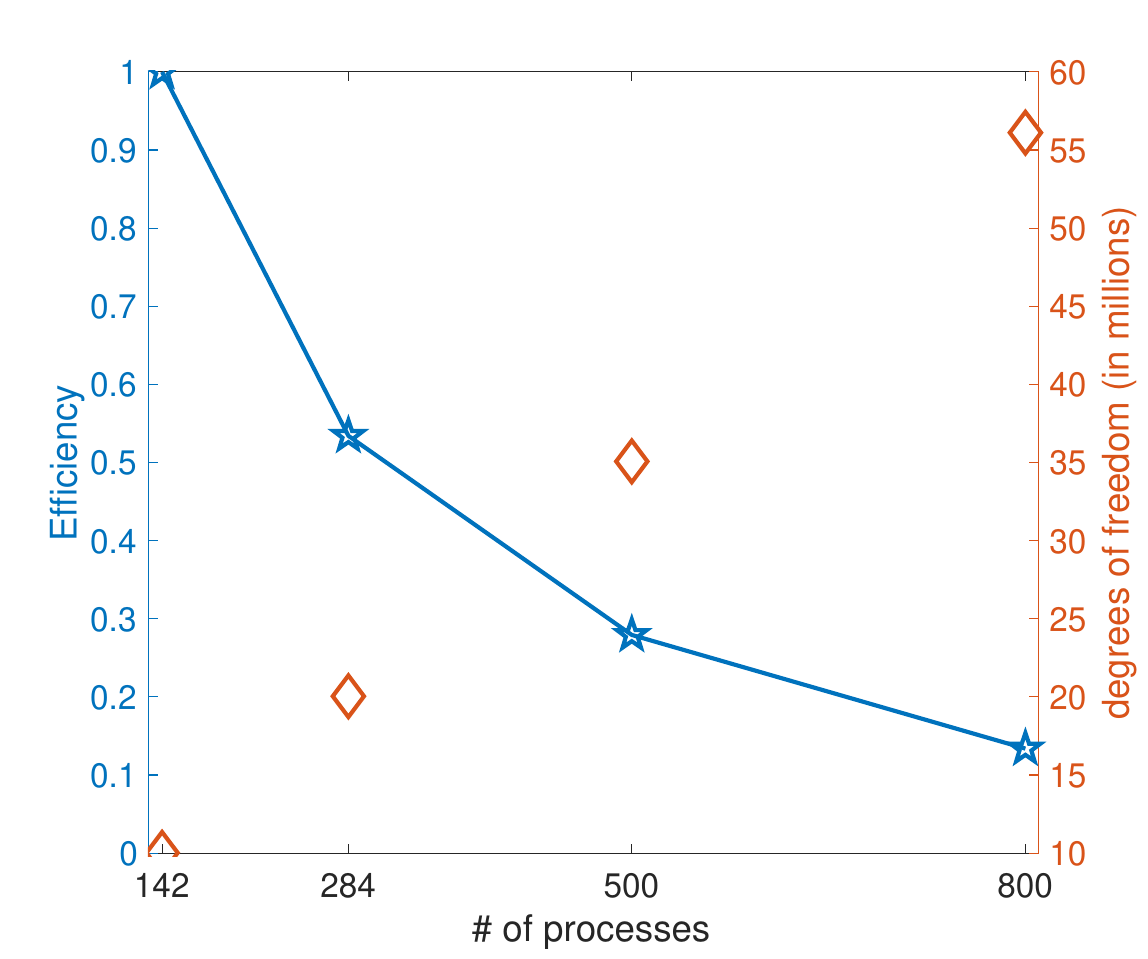}      \caption{Efficiency}
    \end{subfigure}
  \caption{Weak scalability with respect to the increasing number of random variables (RVs) (using $2^{nd}$ order input and $3^{rd}$ order output PCE) for the stochastic linear Poisson problem}\label{Fig:weakRV_LP}
\end{figure}

In this section, the scalability with respect to stochastic parameters is considered only for 2GV3 because of its superior performance with respect to 2GV2 as illustrated in the previous section. The finite element mesh used in this study contains $1{,}002{,}001$ vertices. For this fixed finite element mesh, the number of random variables or output order of expansion is increased while maintaining the problem size per core fixed. The time and efficiency plots with respect to increasing random variables for an increasing number of processes from $142$ to $800$ are shown in Fig.~\ref{Fig:weakRV_LP}. In these figures, the `$\#$ of terms' represents the number of terms in output PCE. \textcolor{ss}{It is observed that the PC setup time increases considerably in comparison to Krylov solve time} (see Fig.~\ref{Fig:weakRV_LP}a). This increasing setup and solve time is in part due to the increased condition number of the system matrix with increasing random variables (see \ref{app:condnumbersto}). Moreover, the strong stochastic coupling among output PCE coefficients increases the strength of off-diagonal blocks in the system matrix \cite{SG_GaussSeidel_Roger,SG_pellisetti,sousedik,sousedik_2}. This increase in time is reflected in the efficiency plot shown in Fig.~\ref{Fig:weakRV_LP}b where it decreases to $18\%$ for a $5$ random variable case having $56$ output PCE terms solved using $800$ processes. However, note that the computational requirements for a stochastic problem with a size of $56$ million necessitate the use of a parallel solver.

\begin{figure}[htbp]
    \centering
    \begin{subfigure}[b]{0.475\textwidth}
       \centering
        \includegraphics[width=\textwidth]{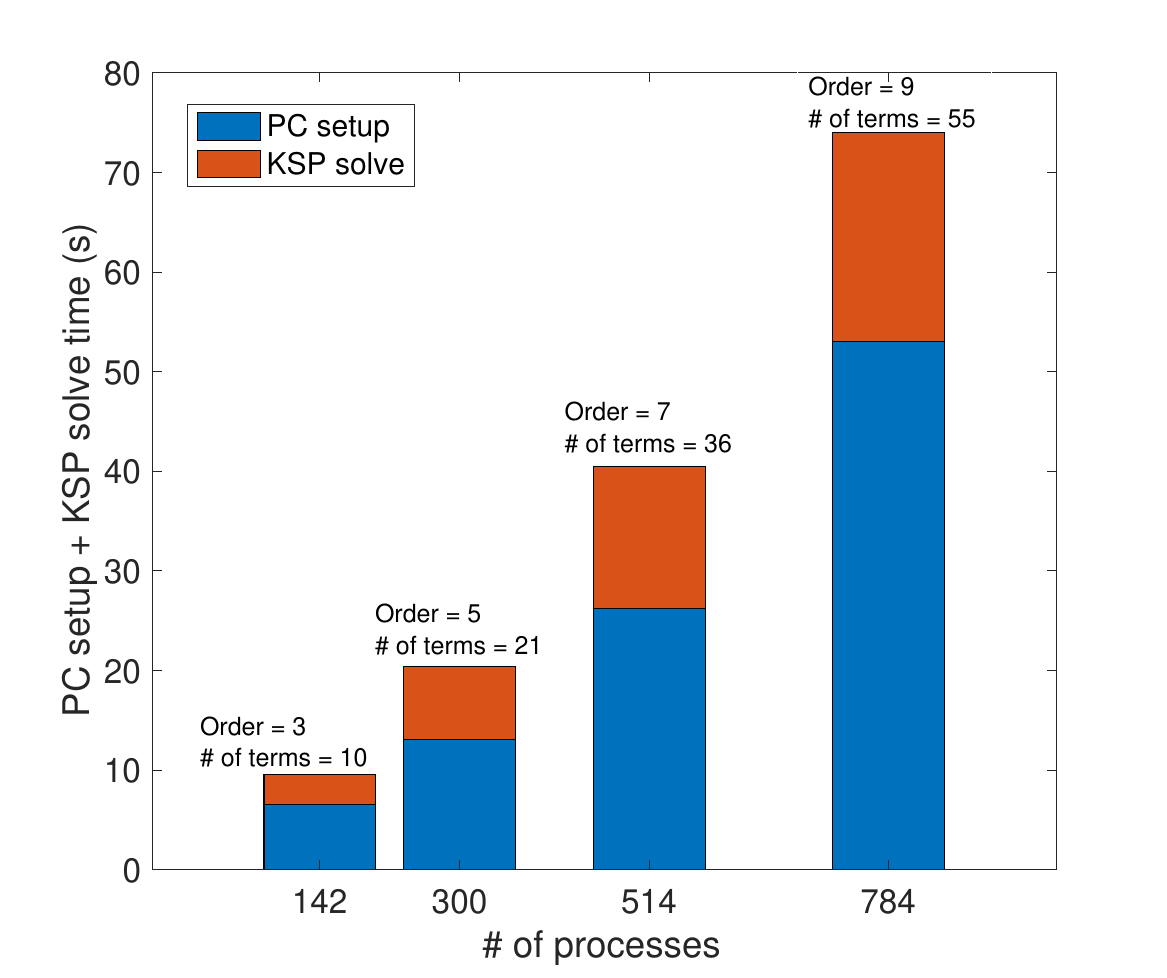} 
        \caption{PC setup and solve time}
    \end{subfigure}
    \centering
    \begin{subfigure}[b]{0.475\textwidth}
       \centering
        \includegraphics[width=\textwidth]{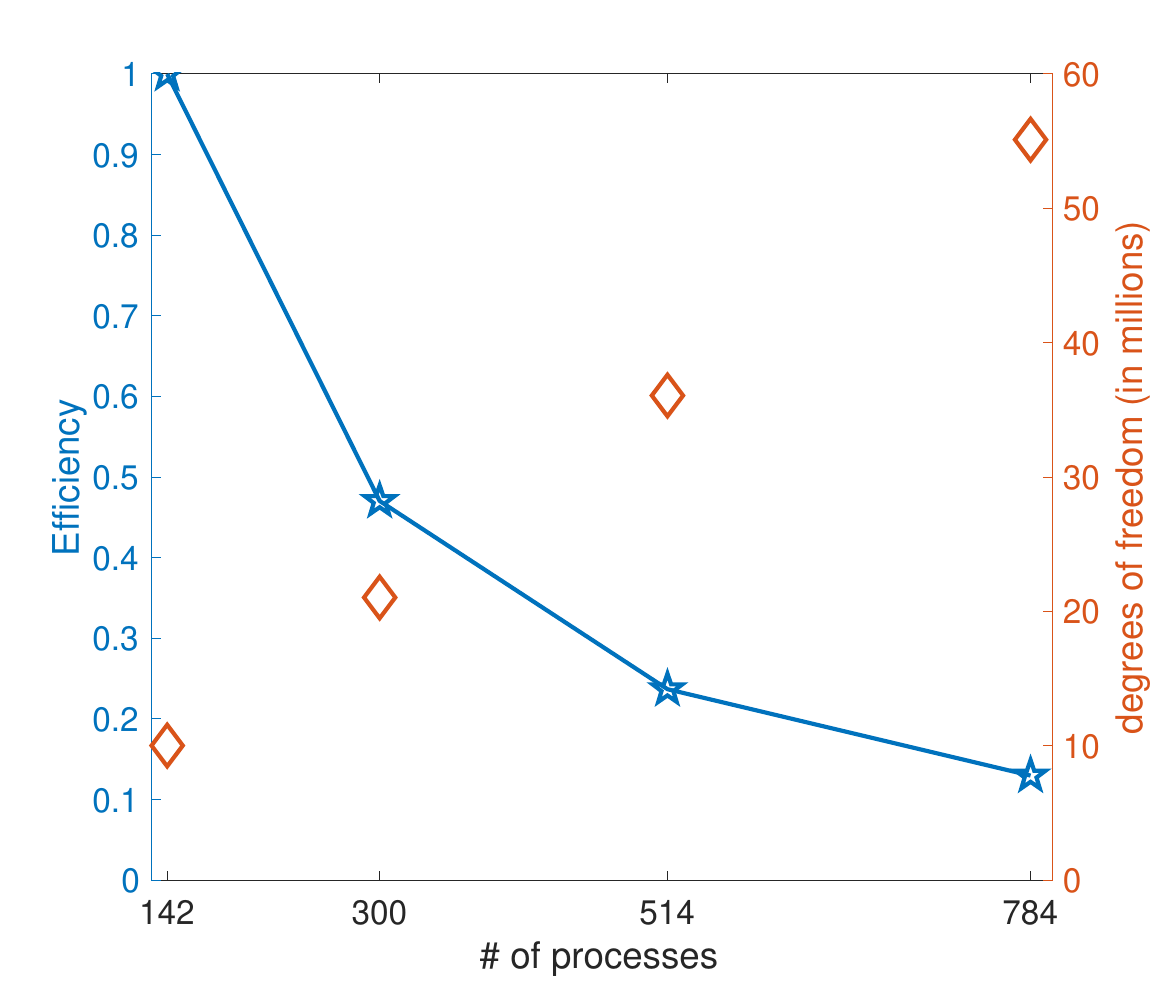}      \caption{Efficiency}
    \end{subfigure}
  \caption{Weak scalability with respect to increasing order of output PCE (using $2$ random variables and $2^{nd}$ order for input PCE) for the stochastic linear Poisson problem}\label{Fig:weakO_LP}
\end{figure}

The stochastic scalability with increasing order of expansions and fixed number of random variables is also shown in Fig.~\ref{Fig:weakO_LP}. As observed for the previous case of increasing random variables, increasing the order of expansion also increases the PC setup and KSP solve times. \textcolor{ss}{This may be due to the stronger coupling between the solution PCE coefficients with increasing order of expansion \cite{knio_book}}. We note that this decrease in efficiency (see Fig.~\ref{Fig:weakO_LP}b) can be tackled in several ways such as improving the coarse basis functions as in \cite{gander_ORAS,geneo_victorita} after suitable adaptations to the stochastic setting.

\subsection{Verification for Stochastic Nonlinear Poisson Problem}

\begin{figure}[htbp]
    \centering
    \begin{subfigure}[b]{0.475\textwidth}
       \centering
        \includegraphics[width=\textwidth]{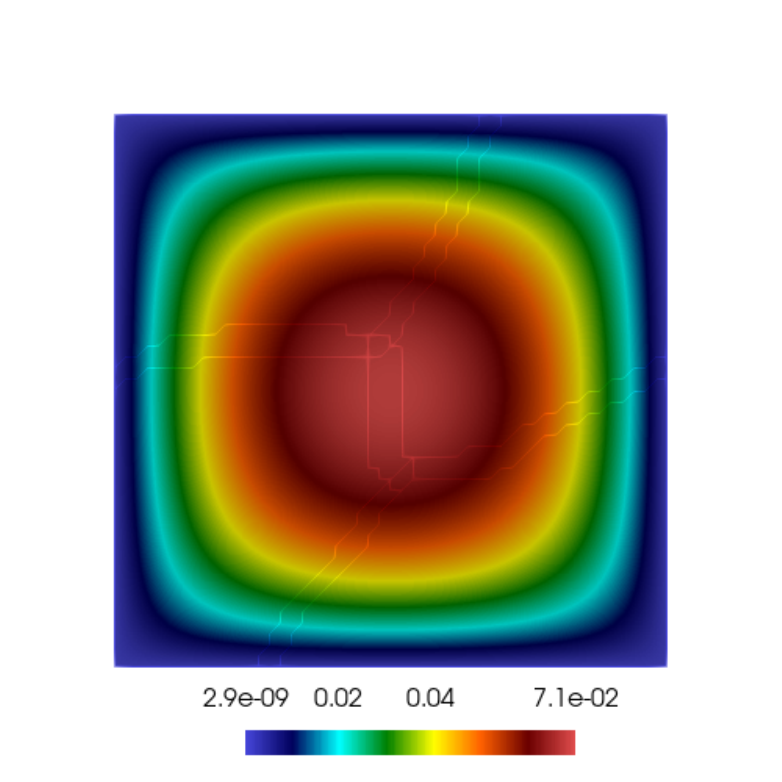} 
        \caption{Mean (SG)}
    \end{subfigure}
    \begin{subfigure}[b]{0.475\textwidth}
      		 \centering
        \includegraphics[width=\textwidth]{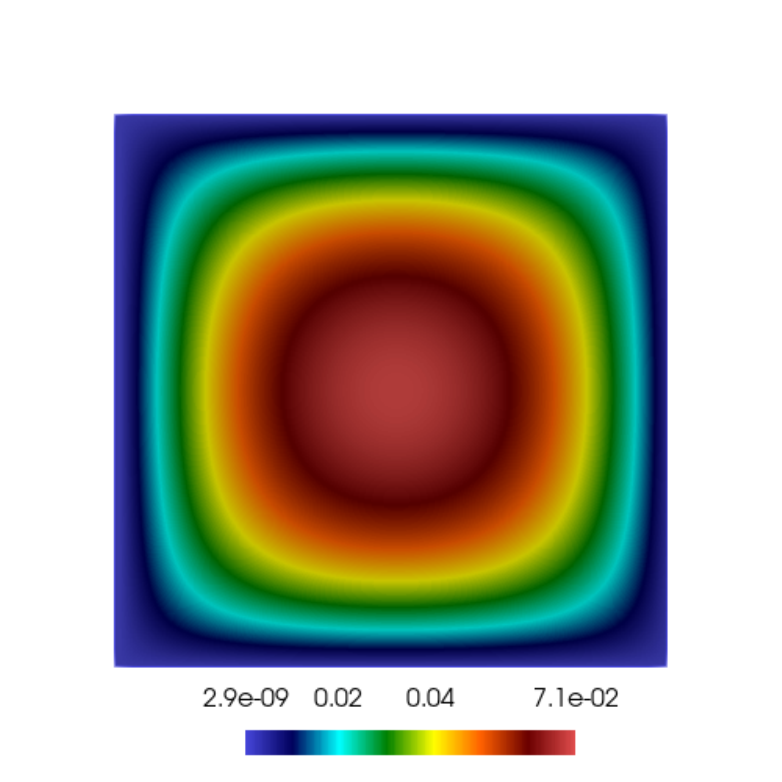} 
        \caption{Mean (MCS)}
    \end{subfigure} 
    \centering
    \begin{subfigure}[b]{0.475\textwidth}
       \centering
        \includegraphics[width=\textwidth]{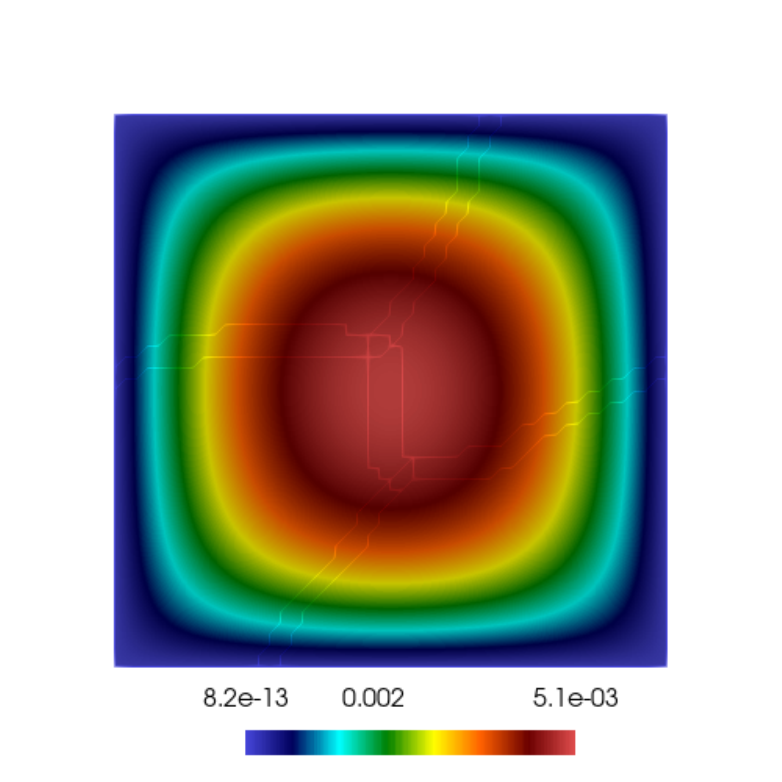} 
        \caption{Standard deviation (SG)}
    \end{subfigure}
    \begin{subfigure}[b]{0.475\textwidth}
      		 \centering
        \includegraphics[width=\textwidth]{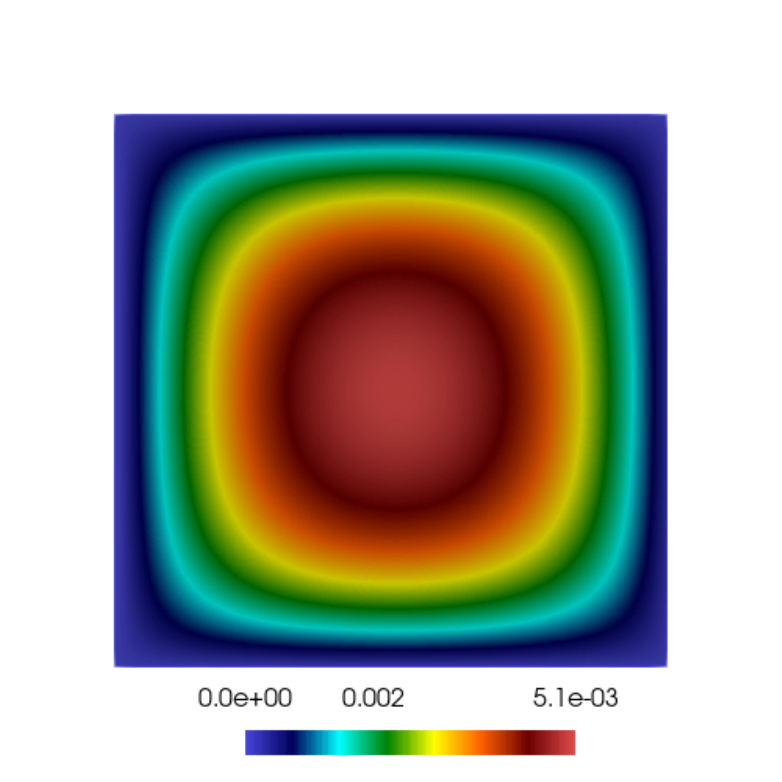} 
        \caption{Standard deviation (MCS)}
    \end{subfigure}
  \caption{Mean and standard deviation of the intrusive stochastic Galerkin (SG) method and MCS for the nonlinear Poisson problem}\label{Fig:mcs_verify_NLpoisson}
\end{figure}

\begin{figure}[htbp]
    \centering
\begin{subfigure}[b]{0.475\textwidth}
       \centering
        \includegraphics[width=\textwidth]{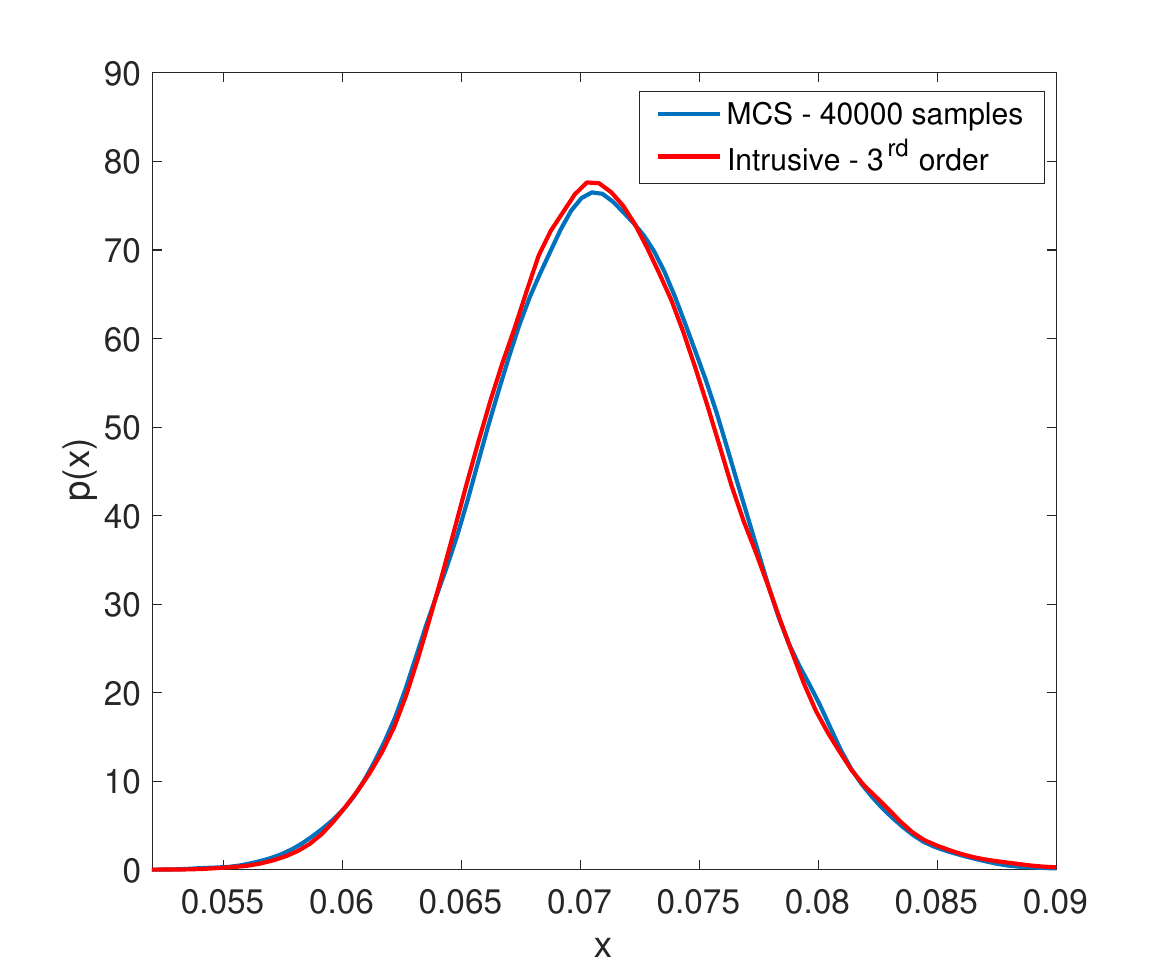} 
        \caption{pdf at the center of domain $(x, y) = (0.5, 0.5)$}
    \end{subfigure}
    \begin{subfigure}[b]{0.475\textwidth}
      		 \centering
        \includegraphics[width=\textwidth]{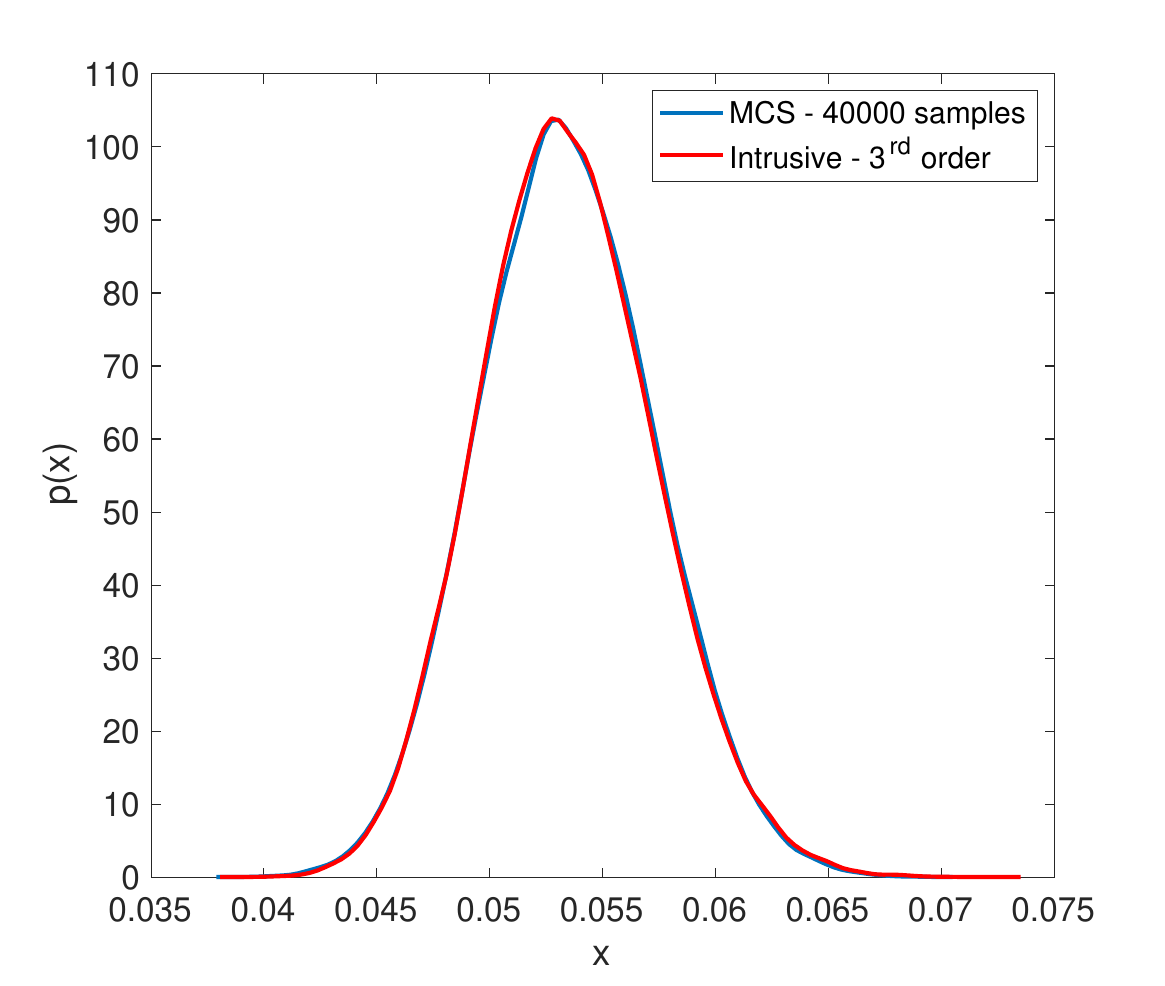} 
        \caption{pdf at $(x, y)  = (0.3, 0.7)$}
    \end{subfigure}
\caption{pdf at two points on the domain using MCS and intrusive stochastic Galerkin (SG) method for the nonlinear Poisson problem}\label{Fig:mcs_verify_NLpoisson_pdf}
\end{figure}

The verification for the nonlinear Poisson problem is carried out by comparing the results of the intrusive stochastic Galerkin method to the MCS. The square domain with four overlapping subdomains is discretized with $10{,}201$ vertices. The input and output PCE uses the same number of parameters as explained for the linear Poisson problem in subsection \ref{sec:verify_LP}. In this case, $40{,}000$ samples are used in MCS (converged) to compute the response statistics. The mean and standard deviation of the stochastic Galerkin and MCS are shown in Fig.~\ref{Fig:mcs_verify_NLpoisson}. The pdf of the solution at two different points shown in Fig.~\ref{Fig:mcs_verify_NLpoisson_pdf} illustrates the close agreement between both approaches.

\subsection{Scalability for Stochastic Nonlinear Poisson Problem with Mesh Size}

The strong and weak scalabilities of the stochastic nonlinear Poisson problem are demonstrated in this section. A PCE with $3$ random variables having $2^{nd}$ order for input and $3^{rd}$ order for output is used unless stated otherwise. The Picard iteration tolerance $\epsilon = 10^{-6}$ is used defined as:
\begin{equation}\label{Eq.tolerance}
 \frac{|| u^{k+1} - u^k ||_2}{|| u^k||_2} \leq \epsilon 
\end{equation}
where $k+1$ is the current iteration and $k$ is the previous iteration. 

\begin{figure}[htbp]
    \centering
      \begin{subfigure}[b]{0.475\textwidth}
       \centering
        \includegraphics[width=\textwidth]{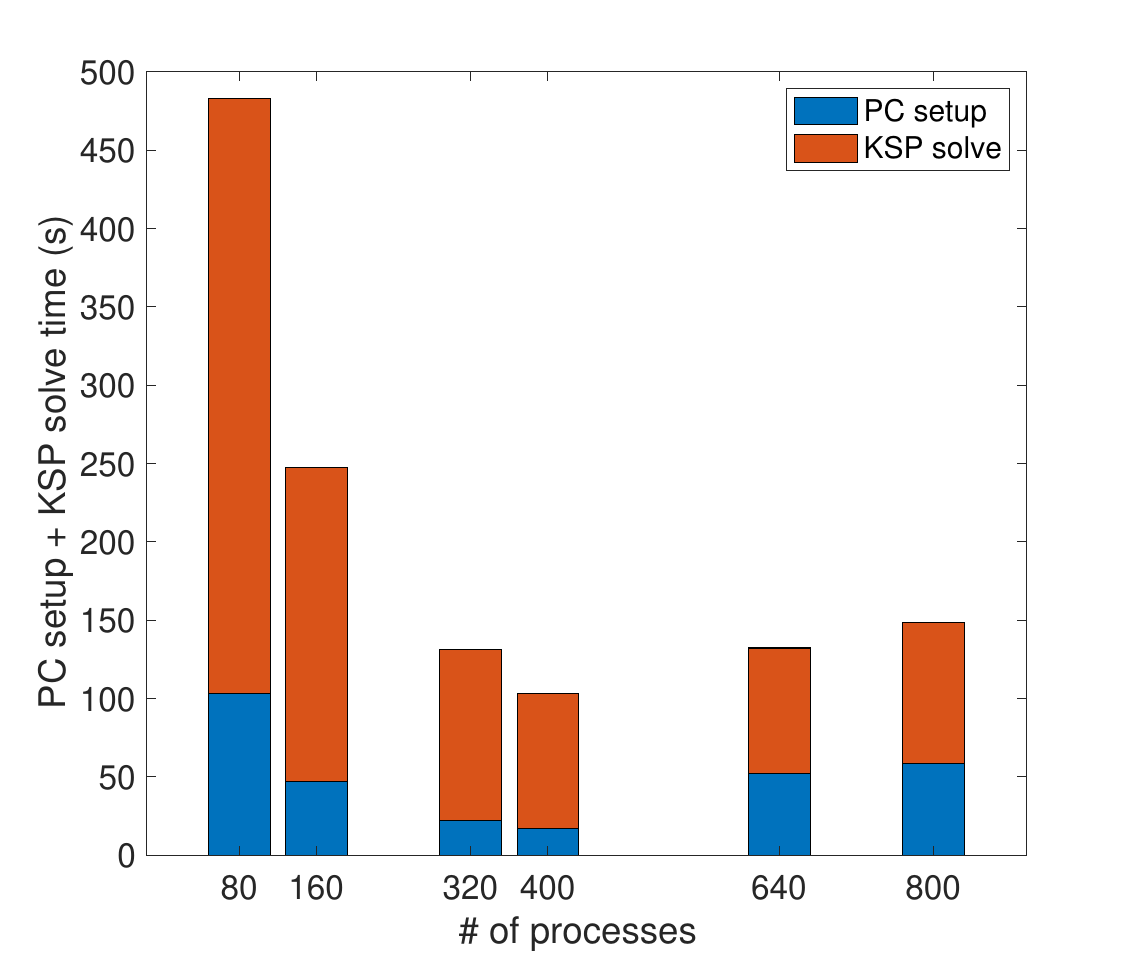} 
        \caption{2GV2}
    \end{subfigure}
    \begin{subfigure}[b]{0.475\textwidth}
       \centering
        \includegraphics[width=\textwidth]{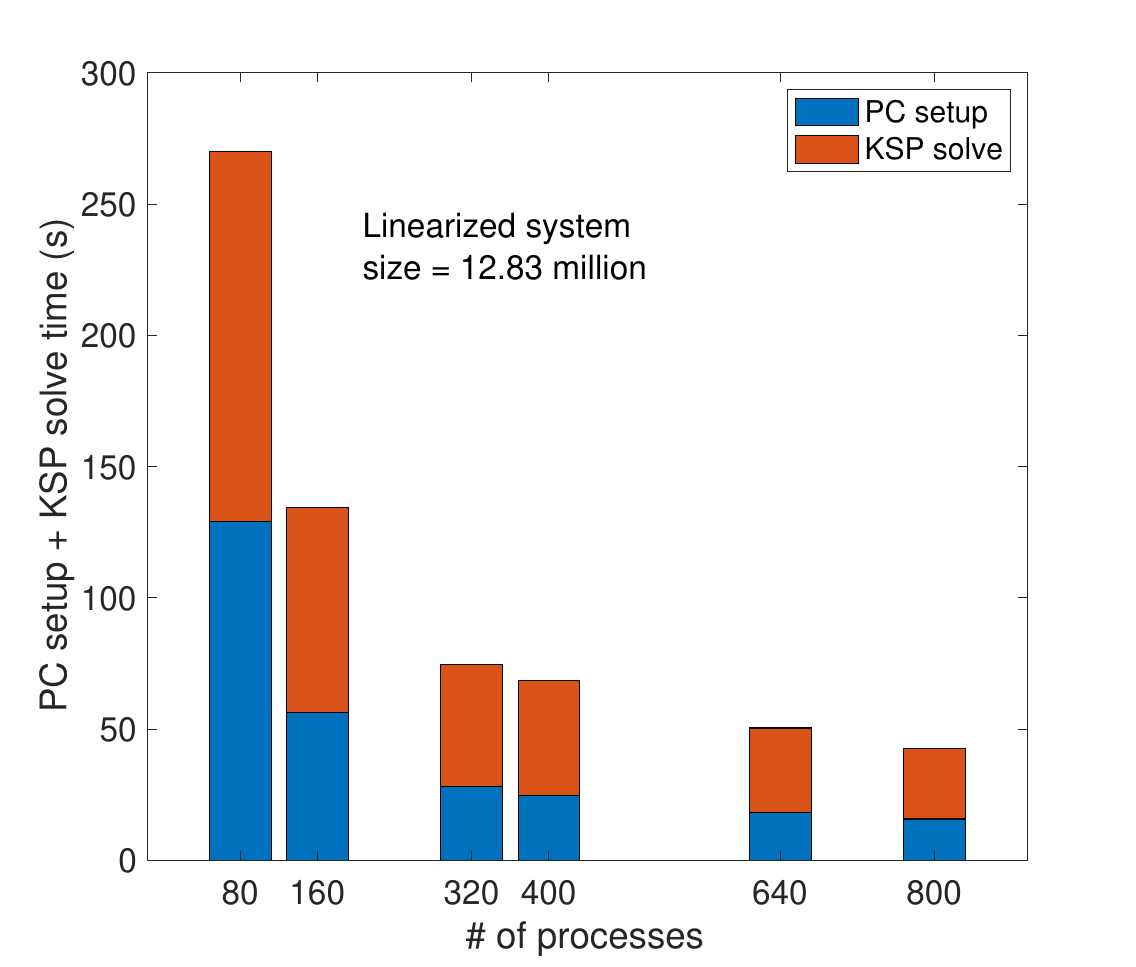} 
        \caption{2GV3}
    \end{subfigure}
    \begin{subfigure}[b]{0.475\textwidth}
      		 \centering
        \includegraphics[width=\textwidth]{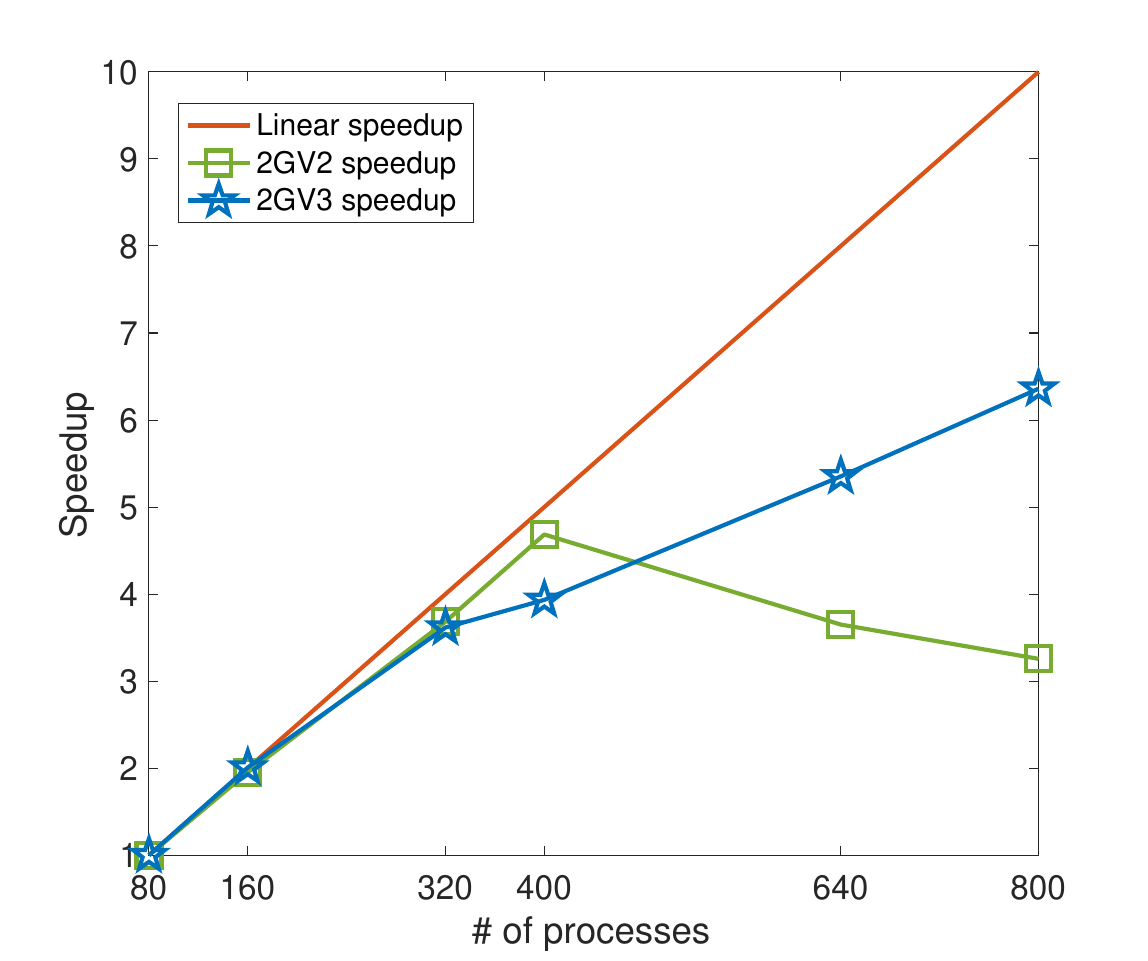} 
        \caption{Speedup}
    \end{subfigure} 
    \centering
    \begin{subfigure}[b]{0.475\textwidth}
       \centering
        \includegraphics[width=\textwidth]{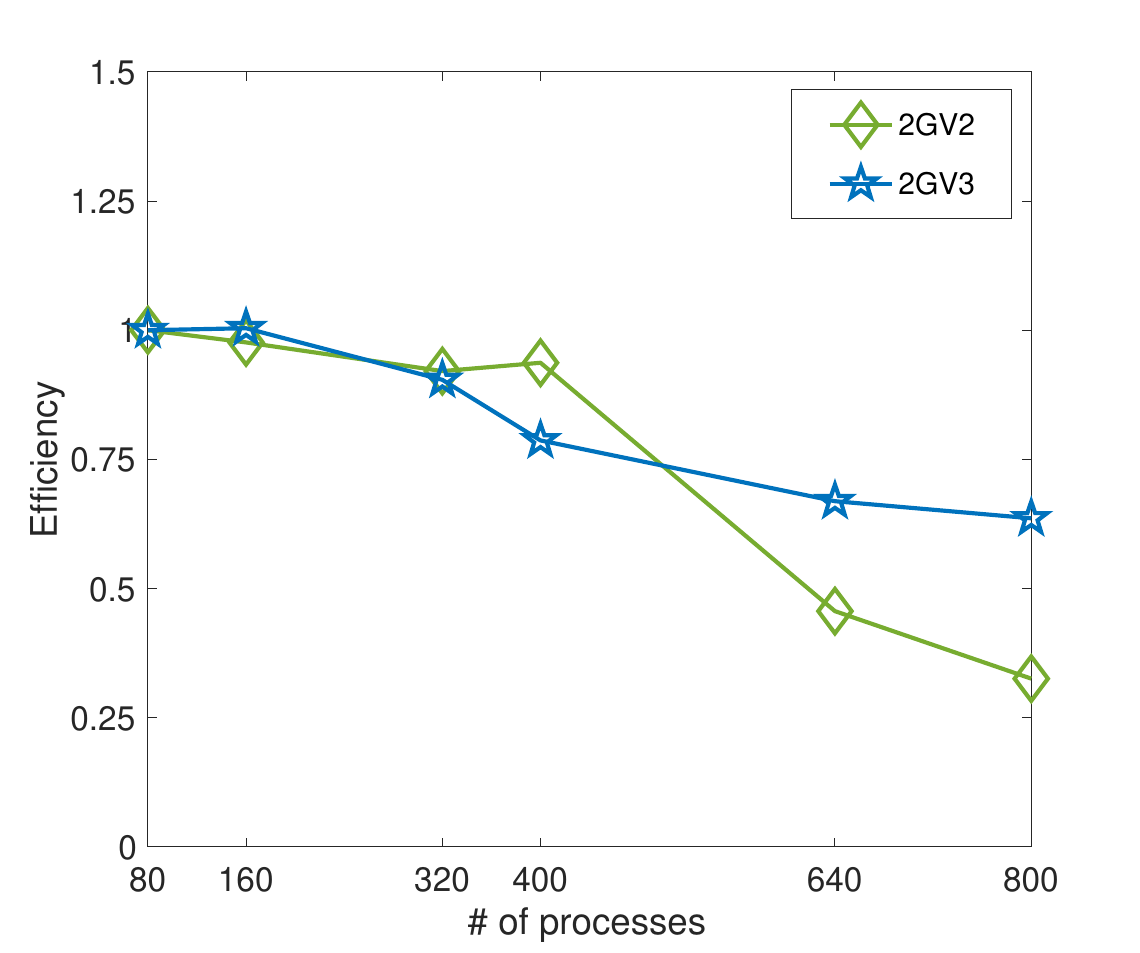} 
        \caption{Efficiency}
    \end{subfigure}
  \caption{Strong parallel scalability for the stochastic nonlinear Poisson problem}\label{Fig:NLPss-strong}
\end{figure}

The strong scalability results for both 2GV2 and 2GV3 is shown in Table \ref{tab:ss_nlp}. The GMRES iteration counts and Picard iteration counts for 2GV2 and 2GV3 remain constant with an increasing number of processes. However, the total time to solution decreases in both cases with slightly better timings for 2GV3. It is important to note that there is a considerable increase in the time to solution for stochastic nonlinear Poisson problems compared to stochastic linear Poisson problems. This is due to the increased cost of handling higher-order tensors and updating the system matrix at each Picard iteration. Similar to the linear Poisson problem, a detailed comparison using PC setup time and KSP solve time along with the speedup and efficiency plots for strong scaling are shown in Fig.~\ref{Fig:NLPss-strong}.
The PC setup and KSP solve time show a sudden increase after $400$ processes for 2GV2 while there is a monotonic decrease in these values for 2GV3 (see Fig.~\ref{Fig:NLPss-strong}a and \ref{Fig:NLPss-strong}b). This shows the stable performance of 2GV3 for strong scaling. This increase in time for 2GV2 is reflected in the speedup and efficiency plots in Fig.~\ref{Fig:NLPss-strong}c and Fig.~\ref{Fig:NLPss-strong}d. This is perhaps due to the communication overhead created by an increasing number of processes. The 2GV3 on the other hand, maintains a decrease in PC setup and KSP solve time and demonstrates better efficiency than 2GV2.

\begin{figure}[htbp]
    \centering
        \centering
        \begin{subfigure}[b]{0.475\textwidth}
       \centering
        \includegraphics[width=\textwidth]{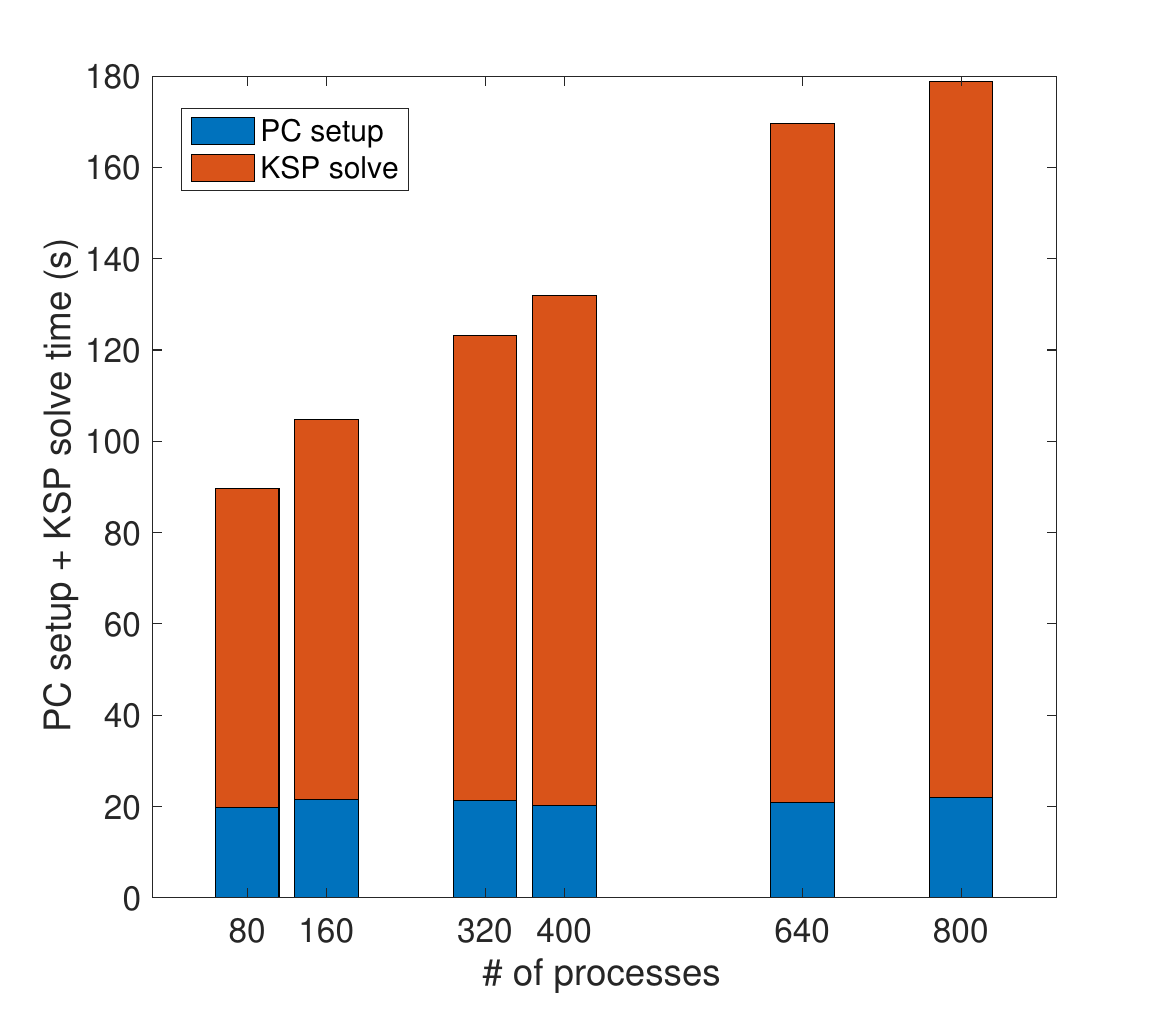} 
        \caption{2GV2}
    \end{subfigure}
        \begin{subfigure}[b]{0.475\textwidth}
       \centering
        \includegraphics[width=\textwidth]{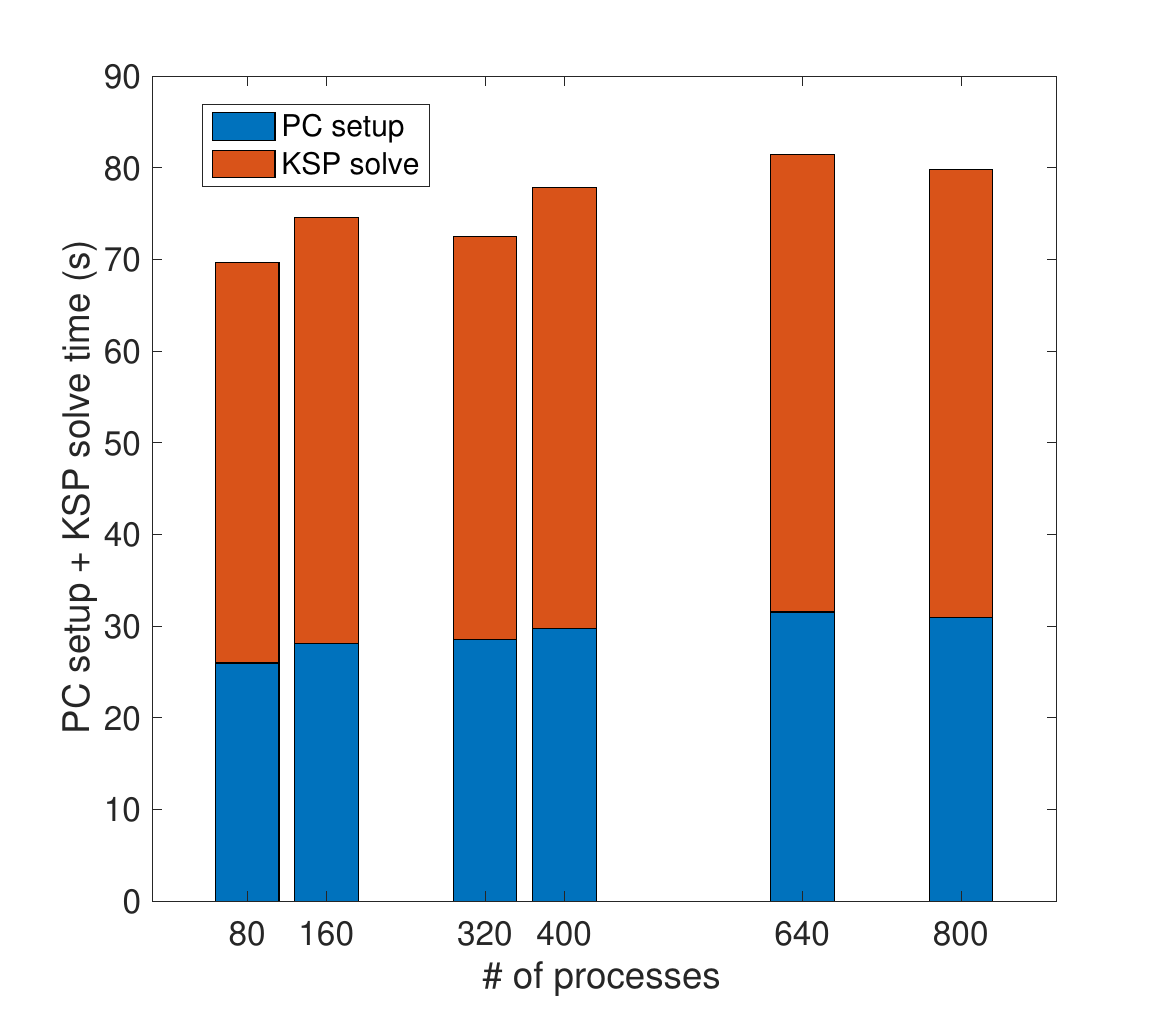} 
        \caption{2GV3}
    \end{subfigure}
    \begin{subfigure}[b]{0.475\textwidth}
      		 \centering
        \includegraphics[width=\textwidth]{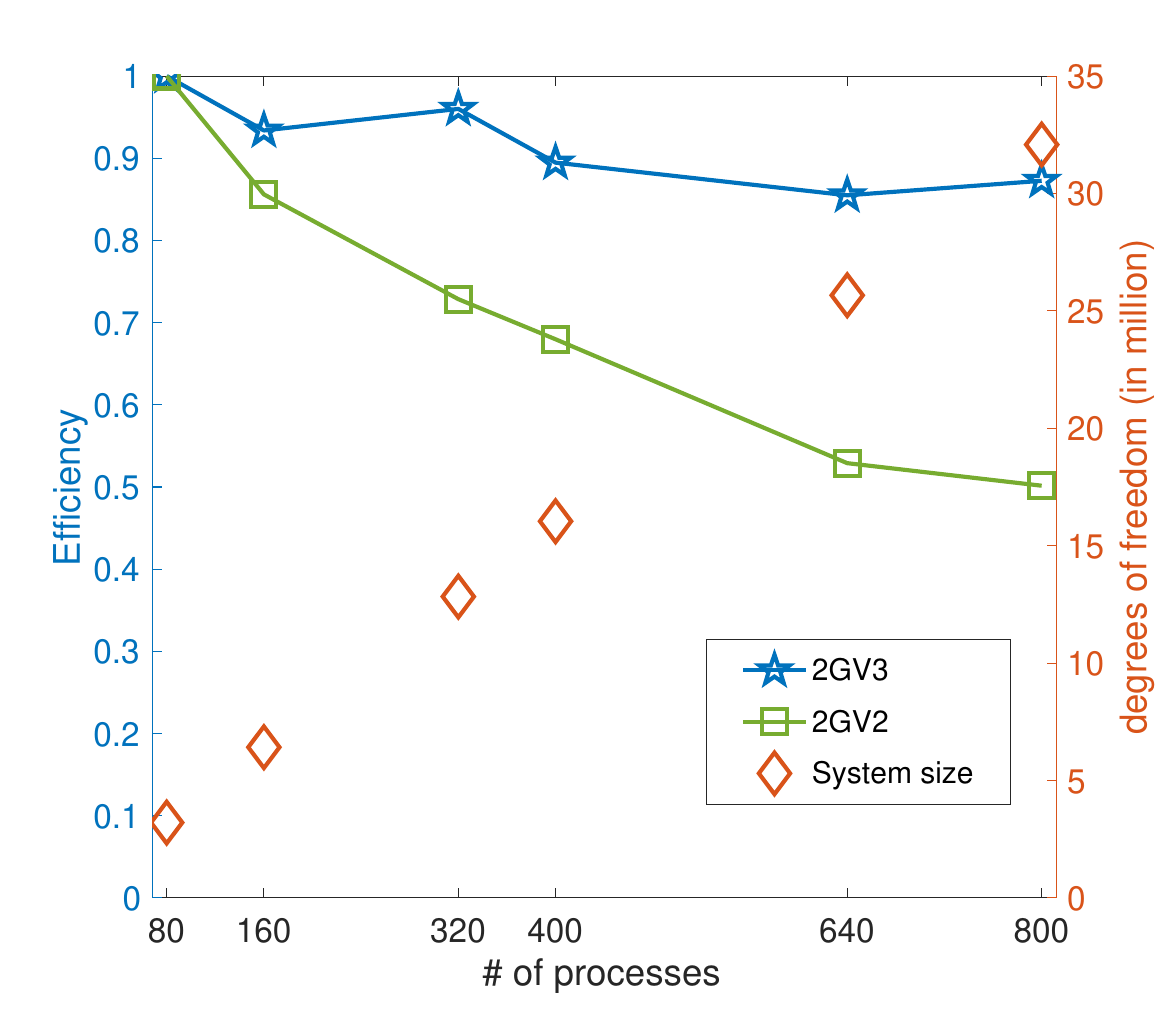} 
        \caption{Efficiency}
    \end{subfigure} 
  \caption{Weak parallel scalability for the stochastic nonlinear Poisson problem}\label{Fig:ss_NLpoisson_weak}
\end{figure}

A comparison of solvers with respect to weak scaling (increasing global problem size while fixing problem size per core constant) is also shown in Table \ref{tab:weak_nlp} for $80$ to $800$ processes. The number of GMRES iterations for 2GV2 increases with increasing global problem size while 2GV3 has smaller and constant iteration counts. However, for both solvers, Picard iteration counts remain the same. The total time to solution for 2GV3 remains almost the same but for 2GV2, the time increases. The PC setup and KSP solve time plots in Fig.~\ref{Fig:ss_NLpoisson_weak} show superior weak parallel scalability of 2GV3 compared to 2GV2. Similar to the linear Poisson problem, the KSP solve time for 2GV2 increases significantly while it remains the same for 2GV3. The speedup and efficiency for 2GV3 are better than 2GV2 in Fig.~\ref{Fig:ss_NLpoisson_weak}c.

\begin{table}[htbp]
    \centering
    \begin{tabular}{|c|c|c|c|c|c|c|}
    \hline
       \multirow{2}{*}{Number of processes} &  \multicolumn{2}{c|}{\specialcell{GMRES \\iteration count}} &  \multicolumn{2}{c|}{\specialcell{Picard \\iteration count}} & \multicolumn{2}{c|}{Total time (s) }\\
      \cline{2-7}
        & 2GV2 & 2GV3 & 2GV2 &  2GV3 & 2GV2 & 2GV3  \\
      \hline
       \hline
      $80$ & $10$  &  $7$ &  $5$ & $5$  & $5{,}809$ &$5{,}696$  \\ 
      $160$ & $10$ & $8$ & $5$ & $5$ & $3{,}146$ &  $3{,}029$ \\ 
      $320$ & $10$ & $8$ & $5$ & $5$ & $1{,}684$ & $1{,}611 $ \\ 
      $400$ & $10$ & $8$  & $5$ & $5$  & $1{,}314$ & $1{,}298 $ \\ 
      $640$ &$10$ & $8$  & $5$ & $5$  & $974$ & $862$ \\ 
      $800$ & $10$ & $8$  & $5$ & $5$  & $840$ & $707$ \\ 
      \hline
    \end{tabular}
    \caption{Strong scalability for the stochastic nonlinear Poisson problem with a total problem size of 12.83 million dof.}
    \label{tab:ss_nlp}
\end{table}

\begin{table}[htbp]
    \centering
    \begin{tabular}{|c|c|c|c|c|c|c|}
    \hline
      \specialcell{Number of processes \\ (linearized system size in million))}  &  \multicolumn{2}{c|}{\specialcell{GMRES \\ iteration count} } & \multicolumn{2}{c|} {\specialcell{Picard \\ iteration count} } & \multicolumn{2}{c|}{Total time (s)} \\
       \cline{2-7}
        & 2GV2 & 2GV3 & 2GV2 & 2GV3 & 2GV2 & 2GV3  \\
      \hline
       \hline
      $80$ ($3.22$) & $8$ & $8$ & $5$ & $5$ & $1{,}587$ & $1{,}559 $ \\ 
      $160$ ($6.43$) & $9 $ &  $8$ &  $5$ & $5$ &  $1{,}634$  & $1{,}570 $ \\ 
      $320$ ($12.83$) & $10 $  & $8$ & $5$ & $5$ &$1{,}622$ & $1{,}543 $ \\ 
      $400$ ($16.02$) &$11 $  & $8$ & $5$ & $5$ & $1{,}612$& $1{,}576 $ \\ 
      $640$ ($25.67$) &$14 $  & $8$ & $5$ & $5$ &$1{,}691$ & $1{,}600 $ \\ 
      $800$ ($32.11$) &$15 $  & $7$ & $5$ & $5$ & $1{,}707$ & $1{,}573 $ \\ 
      \hline
    \end{tabular}
    \caption{Weak scalability for the stochastic nonlinear Poisson problem}
    \label{tab:weak_nlp}
\end{table}

\subsection{Scalability for Stochastic Nonlinear Poisson Problem with Random Parameters}

The scalability studies with respect to random parameters differ from the scalability with respect to mesh size as alluded in subsection \ref{sec.ss-lp}. 
The increase in number of PDEs and the associated coupling among them with increasing number of random parameters influences the scalability.

\begin{figure}[htbp]
    \centering
    \begin{subfigure}[b]{0.475\textwidth}
       \centering
        \includegraphics[width=\textwidth]{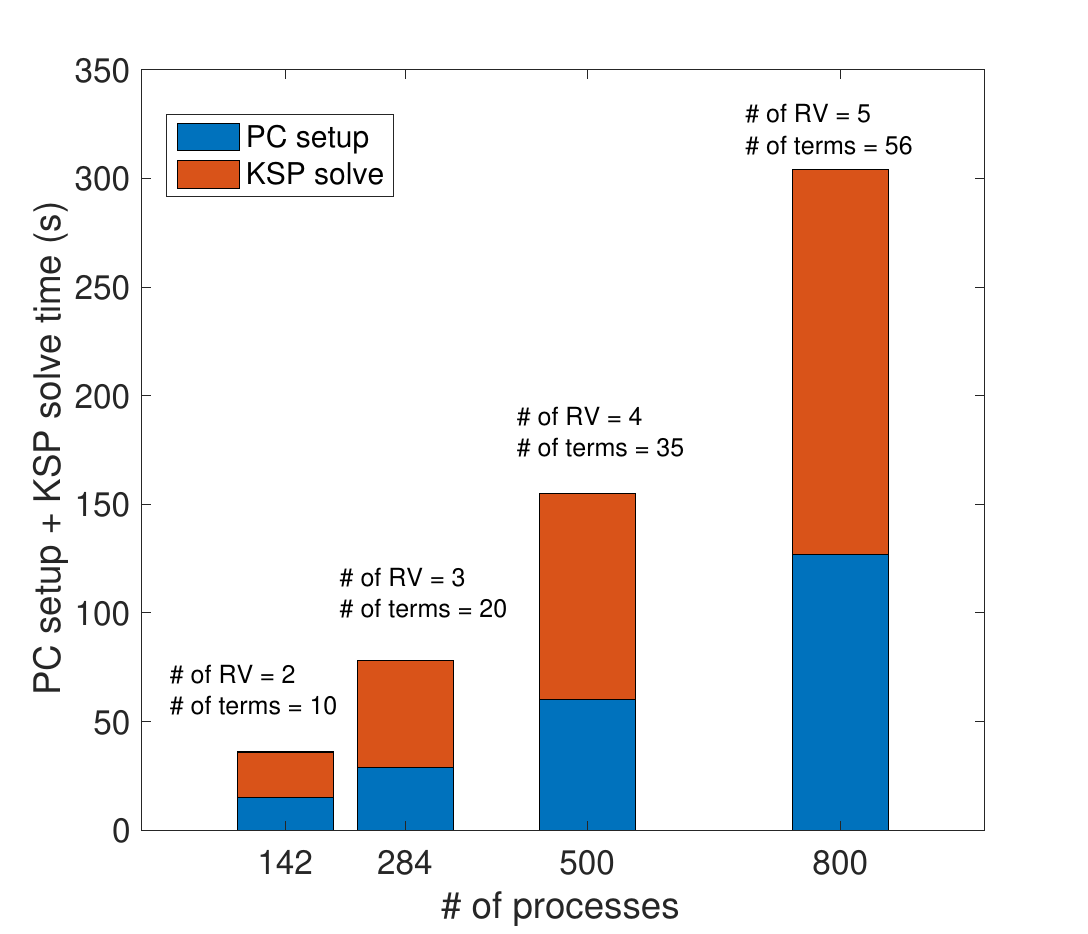} 
        \caption{PC setup and solve time}
    \end{subfigure}
    \centering
    \begin{subfigure}[b]{0.475\textwidth}
       \centering
        \includegraphics[width=\textwidth]{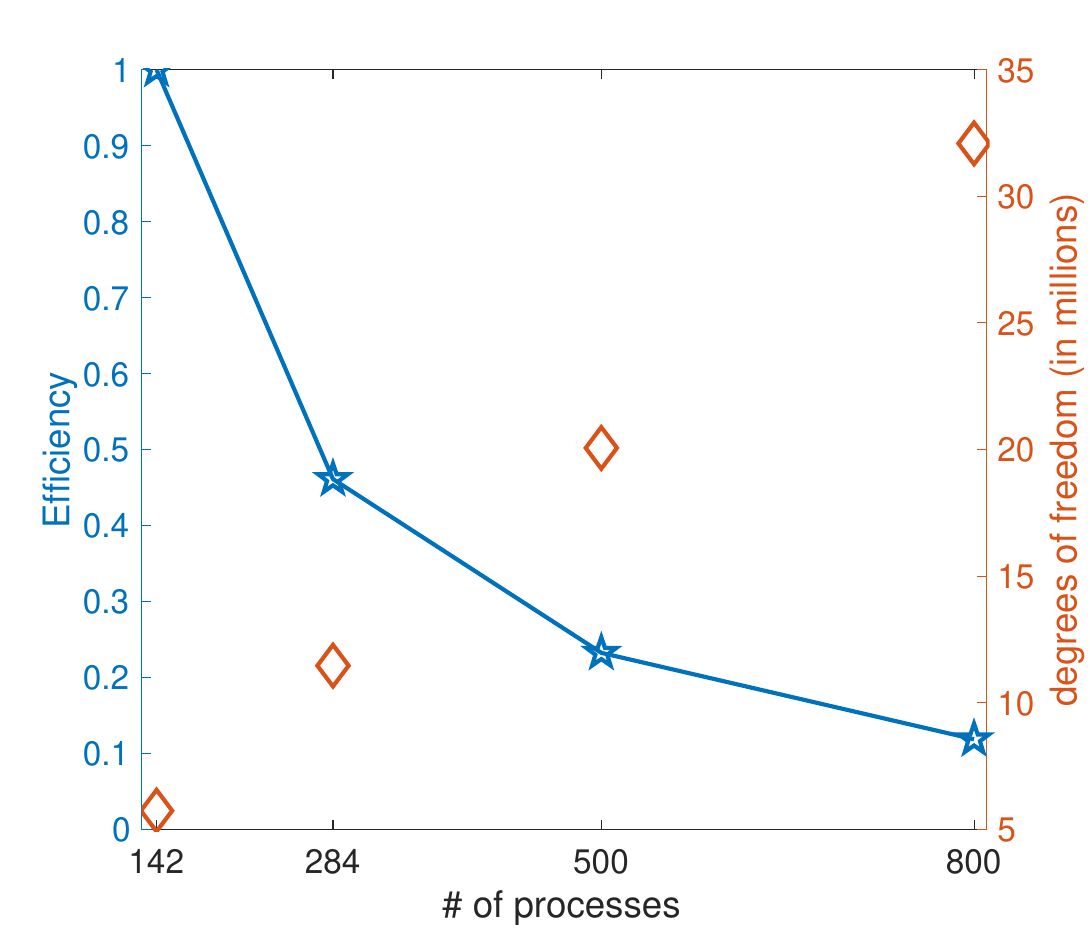}     \caption{Efficiency}
    \end{subfigure}
  \caption{Weak scalability with respect to the increasing number of random variables (RVs) (using $2^{nd}$ order for input and $3^{rd}$ order for output PCE) for the stochastic nonlinear Poisson problem}\label{Fig:NLPss_RV}
\end{figure}

\begin{figure}[htbp]
    \centering
    \begin{subfigure}[b]{0.475\textwidth}
       \centering
        \includegraphics[width=\textwidth]{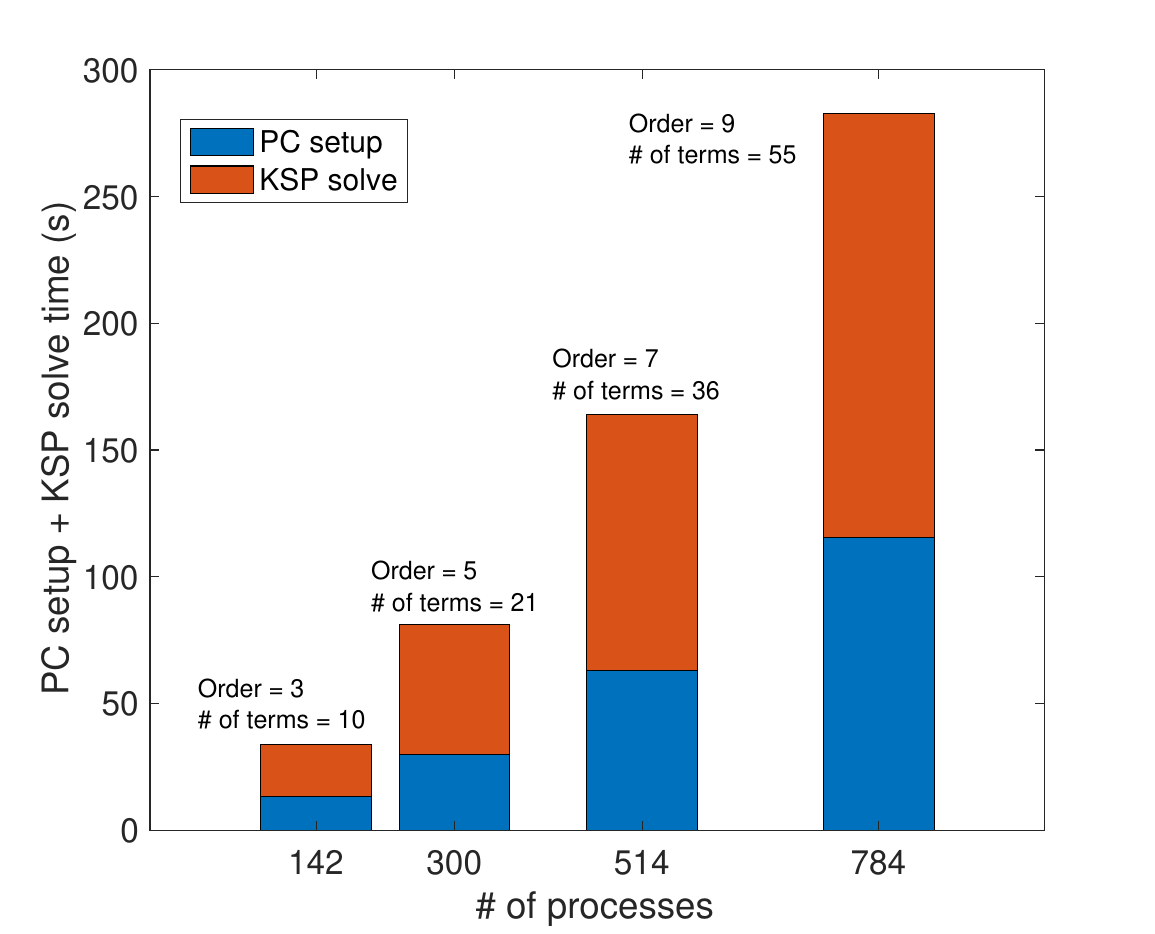} 
        \caption{PC setup and solve time}
    \end{subfigure}
    \centering
    \begin{subfigure}[b]{0.475\textwidth}
       \centering
        \includegraphics[width=\textwidth]{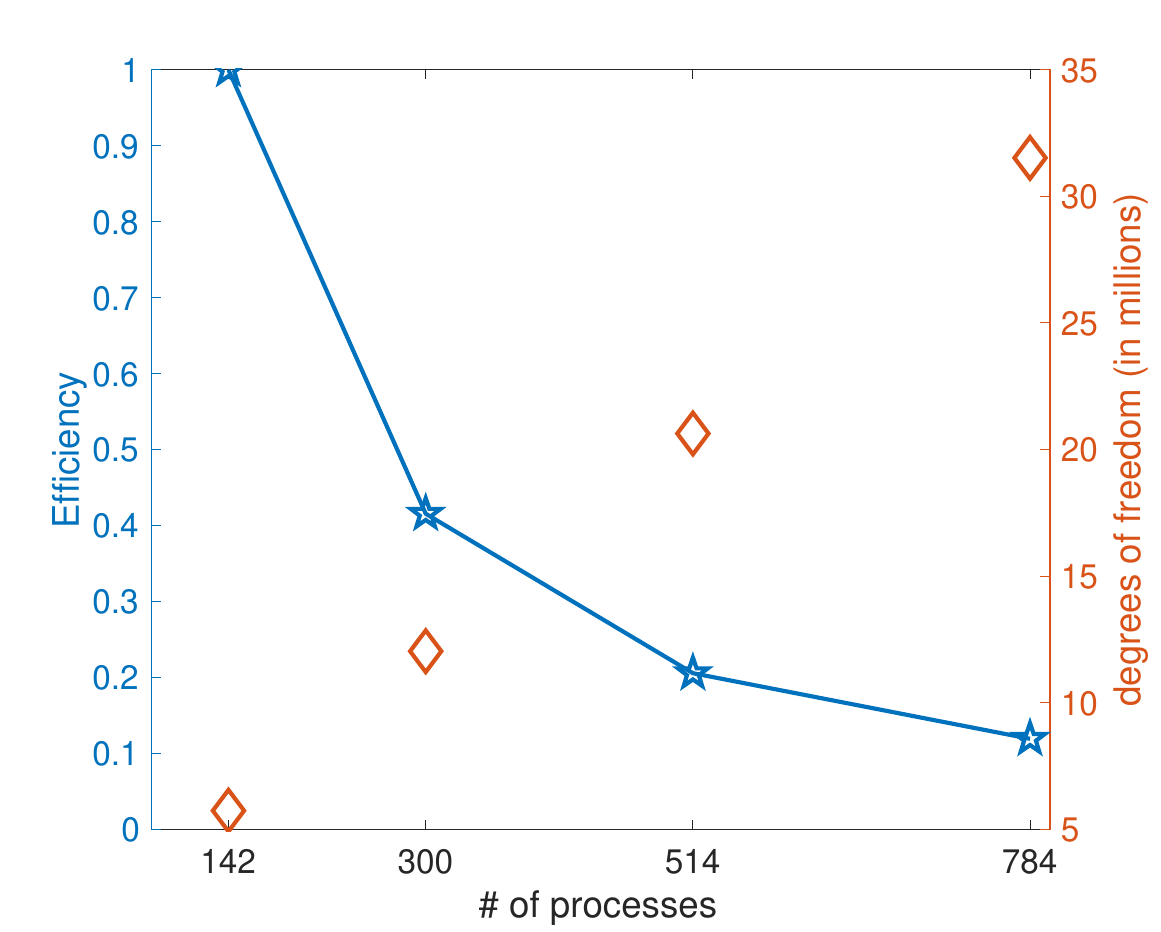}     \caption{Efficiency}
    \end{subfigure}
  \caption{Weak scalability with respect to increasing order of expansion (using $2$ random variables and $2^{nd}$ order for input PCE) for the stochastic nonlinear Poisson problem}\label{Fig:NLPss_order}
\end{figure}

The stochastic scalability of the 2GV3 preconditioner with respect to the increasing number of random variables and order of expansion is shown in Fig.~\ref{Fig:NLPss_RV} and Fig.~\ref{Fig:NLPss_order}. \textcolor{ss}{The Krylov solve time increases rapidly and has a larger proportion in comparison to the PC setup time.} The efficiency of the 2GV3 solver is seen to be decreasing in both cases. This could be due to the increasing strength of off-diagonal blocks and the worsening of the condition number of the system (see \ref{app:condnumbersto}) \cite{SG_GaussSeidel_Roger,SG_pellisetti,sousedik,sousedik_2}. As explained in subsection \ref{sec.ss-lp} for the linear Poisson problem, even though the solver has decreased efficiency, the use of DD-based (or any other parallel scalable) solvers is essential to tackle the computational requirements of the problem.
Moreover, improvements to the solver can be pursued along different directions such as improving the AMG coarse preconditioner by optimizing the parameters \cite{boomeramg,xu_AMG}, \textcolor{ss}{finding appropriate preconditioners for matrices written as Kronecker products} \cite{subber_schwarz}, applying a matrix-free approach to reduce the computational cost \cite{matrix-free} etc.

\section{Optimizing the Size of Coarse Grid}

For any DD algorithm, the coarse grid correction plays a crucial role in improving the convergence of the solver which is also the bottleneck of the algorithms. The ratio of the fine grid to the coarse grid can be varied to find an optimum value which reduces the degradation of the convergence of the solver while maintaining minimum computational cost. In all the previous numerical experiments, this ratio has been fixed to $4$ which is a fairly large size of coarse grid demanding higher computational cost.
This section studies the impact of this ratio for a fixed problem size in deterministic and stochastic settings. 

Table \ref{tab:ratio_deter} shows the average number of GMRES iterations (for both fine and coarse problems) and solve time (PC setup and KSP solve time) with $200$ cores for various sizes of coarse grids for a deterministic nonlinear Poisson problem. The total problem size is fixed at $49$ million. The coarse grid size is reduced from $12$ million to $78{,}961$ which ranges the fine grid to coarse grid ratio from $\approx 4$ to $\approx 620$. The number of outer GMRES iterations increases with decreasing coarse grid size while the time to solution increases. Since decreasing the coarse grid size increases the time to solution and number of outer GMRES iterations for convergence (see Table \ref{tab:ratio_deter}), one should use a smaller ratio of fine to coarse grid size.

\begin{table}[htbp]
    \centering
    \begin{tabular}{|c|c|c|c|c|}
    \hline
      \specialcell{Ratio of fine grid to \\ coarse grid}  &  \specialcell{Outer GMRES \\ iteration count  } & \specialcell{Coarse GMRES \\ iteration count  } & \specialcell{Picard \\ iteration count} & \multicolumn{1}{c|}{Solve time (s)} \\
      \hline
       \hline
     $3.99$ & $7$ & $5$ & $6$ &  $ 80 $ \\ 
      $15.9$ &  $9$ & $5$ &$6$ & $ 81 $ \\ 
      $63.7$ &  $11$ & $5$ &$6$ & $ 88 $ \\ 
      $99.4$ & $12$ & $5$ & $6$ & $ 93 $ \\ 
      $397.8$ &  $14$ & $5$ &$6$ & $ 108 $ \\ 
      $620.7$ &  $15$ & $5$ &$6$ & $ 116 $ \\  
      \hline
    \end{tabular}
    \caption{Convergence with respect to the ratio of fine grid size (fixed) and coarse grid size (varied) for a deterministic nonlinear Poisson problem}
    \label{tab:ratio_deter}
\end{table}

A similar experiment in the stochastic setting varying the coarse grid size with a fixed fine grid size of $12.832$ million solved using $80$ processes is shown in Table \ref{tab:ratio_stoc}. In this study, a $3^{rd}$ order output expansion with $3$ input random variables is considered, leading to $20$ terms for the output PCE. The coarse grid size is reduced from $3.22$ million to $21780$. The number of GMRES iterations increases with decreasing coarse grid size as in the deterministic case. However, the time to solution is found to be reduced only up to a ratio of $63.7$ after which it increases. Compared to the deterministic problem, savings in computational time can be achieved by
decreasing the fine-to-coarse grid ratio to an optimal value of $\approx 16$. However, this optimal value may change with an increasing number of random variables and order of expansion which can increase the condition number of the system matrix \cite{sousedik_2}. Similarly, the current experiment is done on a square domain with structured mesh while for a complex domain and unstructured mesh, the optimal value may differ.

\begin{table}[htbp]
    \centering
    \begin{tabular}{|c|c|c|c|c|}
    \hline
      \specialcell{Ratio of fine grid to \\ coarse grid} &  \specialcell{Outer GMRES \\ iteration count } & \specialcell{Coarse GMRES \\ iteration count}& \specialcell{Picard \\ iteration count}  & \multicolumn{1}{c|}{Solve time (s)} \\
      \hline
       \hline
    
       $3.99$ & $7$ & $5$ & $5$ & $ 260 $ \\ 
      $15.9$ &  $11$ & $5$ & $5$ & $ 243 $ \\ 
      $63.7$ &  $15$ & $5$ &$5$ & $ 258 $ \\ 
      $99.4$ &  $16$ & $5$ &$5$ & $ 271 $ \\ 
      $397.8$ &  $18$ &$5$ & $5$  & $ 288 $ \\ 
      $620.7$ & $5$ & $20$ & $5$ & $ 306 $ \\    
      \hline
    \end{tabular}
    \caption{Convergence with respect to the ratio of fine grid size (fixed) and coarse grid size (varied) for a stochastic nonlinear Poisson problem}
    \label{tab:ratio_stoc}
\end{table}

\section{Conclusion}
The stochastic Galerkin method is used for the uncertainty quantification of linear and nonlinear PDEs with random coefficients. The nonlinear mapping among input to output parameters (even for a linear stochastic PDE) in conjunction with the nonlinearity in the PDE can have compounding effects on the output uncertainty. To capture this uncertainty accurately, a large number of random variables and order of expansions may be necessary. Moreover, the repeated assembly of coefficient matrices of stochastic systems is required for these nonlinear problems solved using Picard/Newton iterations. For high-resolution numerical models of nonlinear PDEs, the excessive memory requirements and computational cost associated with the stochastic Galerkin method can be handled using DD-based solvers. The nonlinearity in the PDEs leads to a non-symmetric coefficient matrix which is solved using a GMRES iterative solver equipped with efficient parallel preconditioners. To this end, a two-grid Schwarz preconditioner is proposed exhibiting strong and weak scaling as illustrated using numerical experiments for linear and nonlinear time-independent stochastic PDEs. The use of algebraic multigrid as the preconditioner for the coarse grid correction in 2GV3 significantly improves the scalabilities of the solver from its counterpart (2GV2) due to the multiple levels of error reduction it offers. \textcolor{ss}{However, there is a degradation in performance of the solvers with increasing number of random variables or order of expansions. We note that the efficiency metric used for increasing mesh size may not be suitable for its counterpart (with increasing random parameters) with the stochastic problems differing significantly in their coupling structure and possibly leading to ill-conditioned matrices \cite{SG_GaussSeidel_Roger,SG_pellisetti,sousedik,sousedik_2}. To overcome these challenges, developing solvers utilizing matrix-free methods and efficient coarse corrections for stochastic systems will be pursued.}

 \bibliographystyle{elsarticle-num} 
 \bibliography{references}

\appendix

\section{Overlapping Schwarz Preconditioners for Deterministic Problems}\label{sec:overlapdd_deter}

This appendix investigates the performance of variants of (overlapping) Schwarz preconditioners having two-grid applied to linear and nonlinear Poisson problems in the deterministic setting. The two-grid Schwarz solver variants discussed in section \ref{sec.two-grid} is used here by removing the uncertainty in the input and output (see \cite{sudhi_MBE} for a complete description of the solver in a deterministic setting). The implementational details are same as explained in in section \ref{sec.two-grid}.

The boundary value problem of diffusion of a substance with the diffusion coefficient $q(u)$ on a unit square domain having vertices $(0,0), (0,1), (1,0)$ and $(1,1)$ can be written as \cite{fenics_book}:
\begin{align}\label{Eq.qunlc4}
   - \nabla \cdot \Big ( q(u) \nabla u(\mathbf{x}) \Big) &= f(\mathbf{x})\\
     u(\mathbf{x}) &= 0 \quad on \quad x = 0 \; (\text{left edge})\\
     u(\mathbf{x}) &= 1 \quad on \quad  x = 1 \; (\text{right edge})\\
     \frac{\partial u(\mathbf{x})}{\partial n}  &= 0 \quad on \quad y = 0\;(\text{bottom edge}) \; and\; y = 1 \; (\text{top edge})
\end{align}
where $q(u) = (1+\alpha u)^m $, $f = 0 $ and $m$ is the order of nonlinearity. The analytical solution for the above problem with $\alpha=1$ can be found as \cite{fenics_book},
\begin{equation}\label{Eq.nlpanalyticalsolc4}
    u(x,y) = \Big{(}(2^{m+1} - 1)x + 1 \Big{)}^{1/(m+1)} - 1.
\end{equation}

The relative error between the finite element solution and the analytical solution is computed as:

\begin{equation}\label{Eq.relerror}
  e = \frac{\parallel u_{truth} -  u_{numerical} \parallel_2}{\parallel u_{truth} \parallel_2}
\end{equation}

where $u_{truth}$ is the analytical solution and $u_{numerical}$ is the finite element solution.

\subsection{DD for Deterministic Linear Poisson Problem}

The linear Poisson problem can be considered as a special case of Eq.~(\ref{Eq.qunlc4}) with $m = 0$. We compute the analytical solution for this case using Eq.~(\ref{Eq.nlpanalyticalsolc4}) and apply the two-grid Schwarz preconditioners to solve the finite element system. 

\begin{figure}[!h]
    \centering
    \begin{subfigure}[b]{0.475\textwidth}
       \centering
        \includegraphics[width=\textwidth]{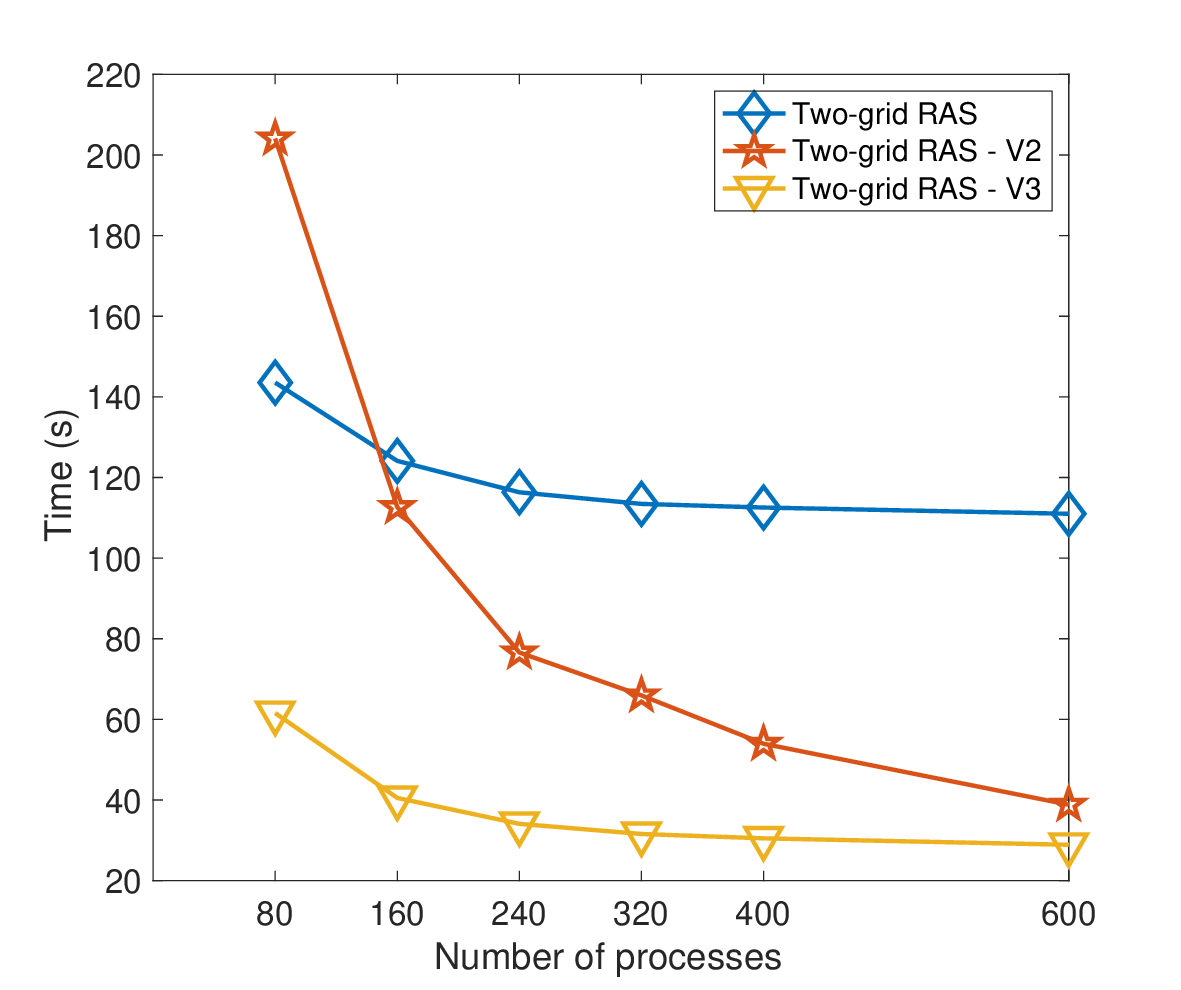} 
        \caption{Strong parallel scalability}
    \end{subfigure}
    \begin{subfigure}[b]{0.475\textwidth}
        \includegraphics[width=\textwidth]{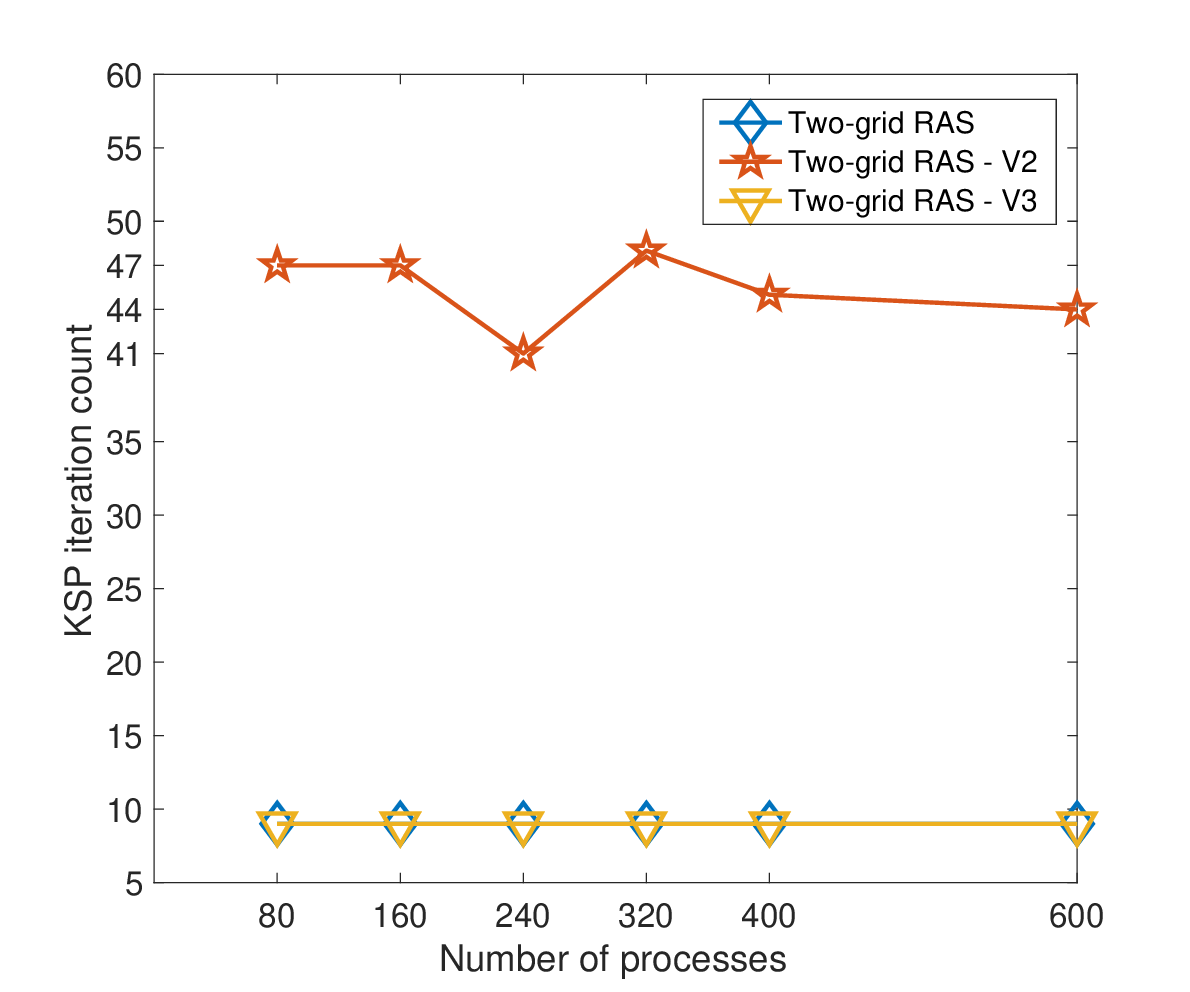} 
        \caption{Strong numerical scalability}
    \end{subfigure}
    \centering
  \caption{Strong scalability of two-grid RAS versions for deterministic linear Poisson problem}\label{Fig:Poisson_strong}
\end{figure}

\begin{figure}[!h]
    \centering
    \begin{subfigure}[b]{0.475\textwidth}
       \centering
        \includegraphics[width=\textwidth]{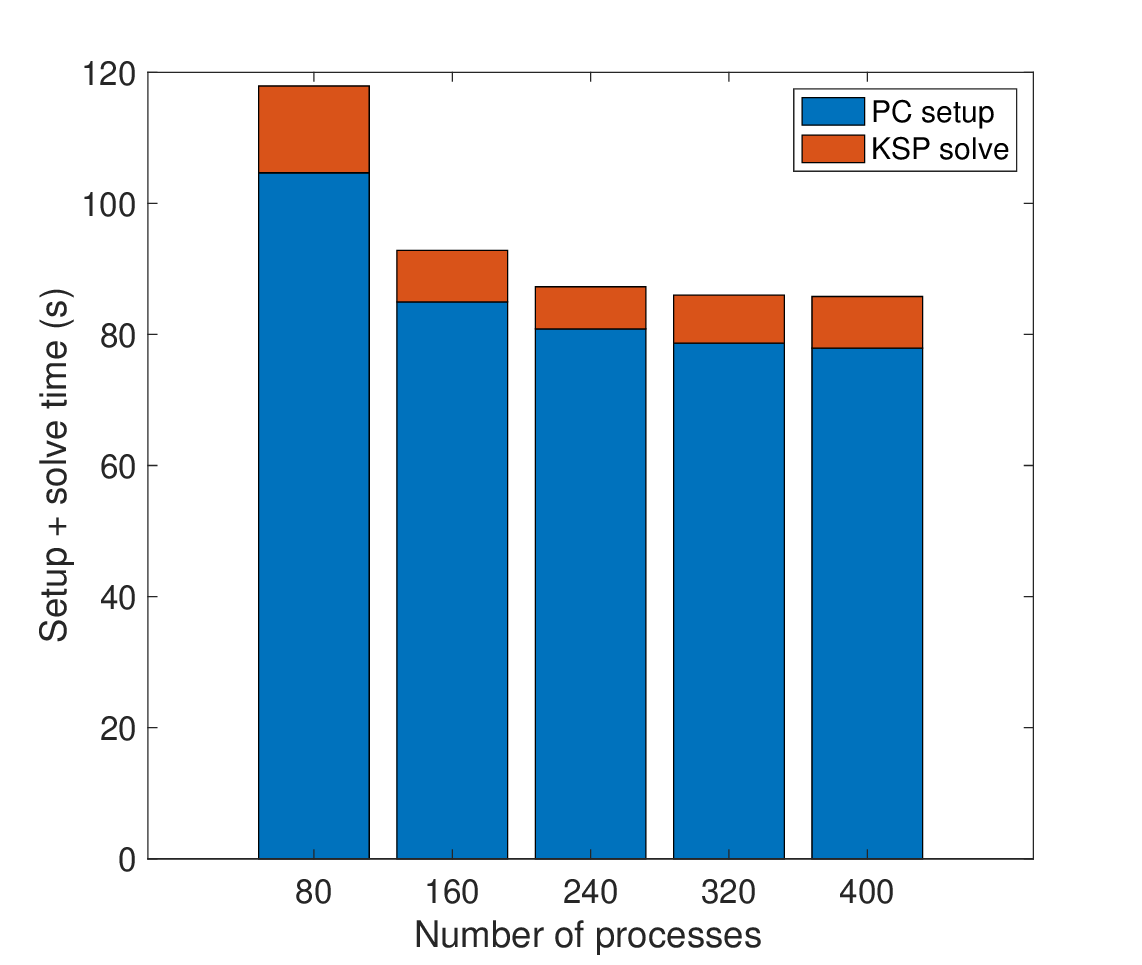} 
        \caption{Two-grid RAS}
    \end{subfigure}
     \centering
    \begin{subfigure}[b]{0.475\textwidth}
       \centering
        \includegraphics[width=\textwidth]{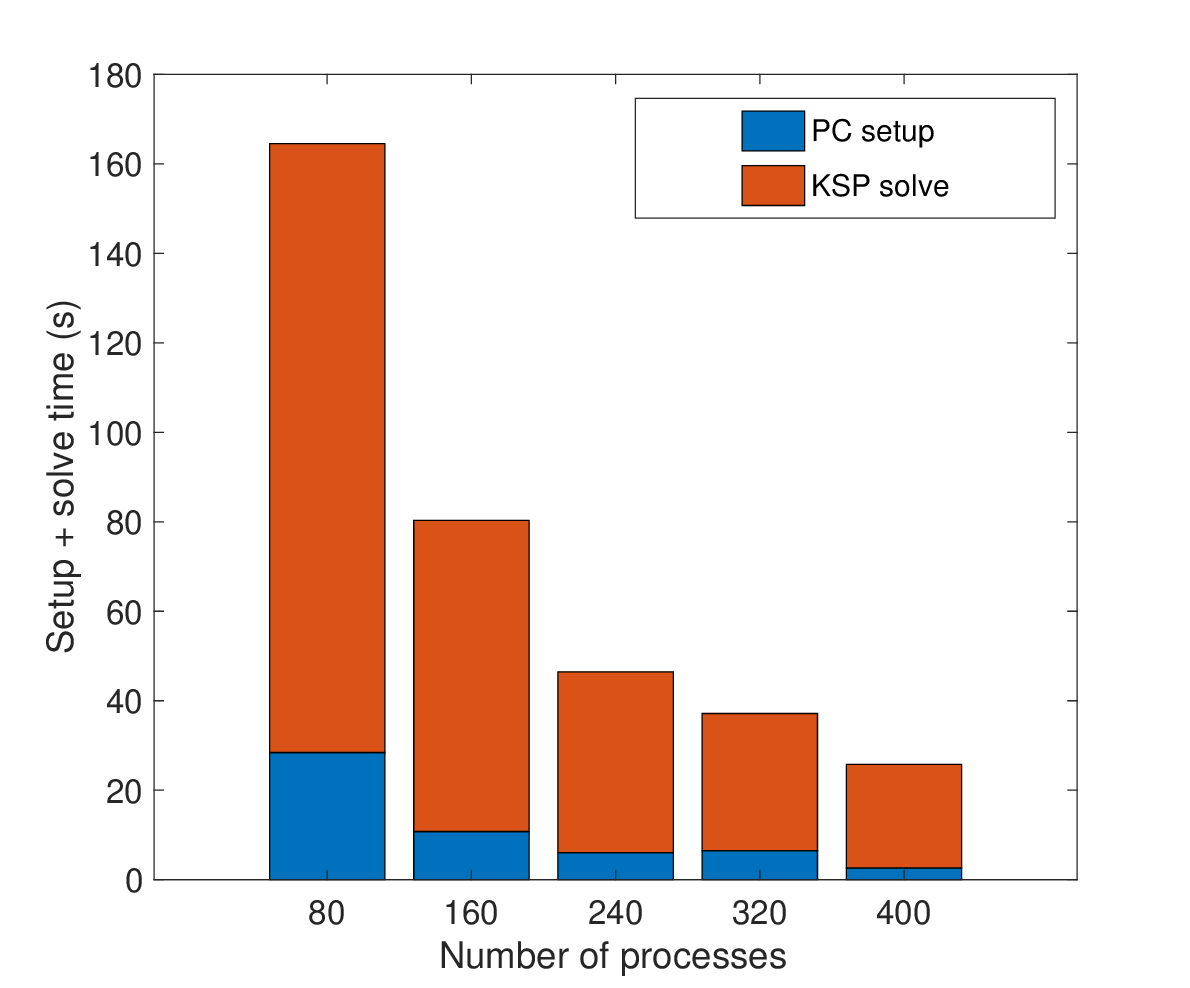} 
        \caption{Two-grid RAS-V2}
    \end{subfigure}
    \begin{subfigure}[b]{0.475\textwidth}
        \centering
        \includegraphics[width=\textwidth]{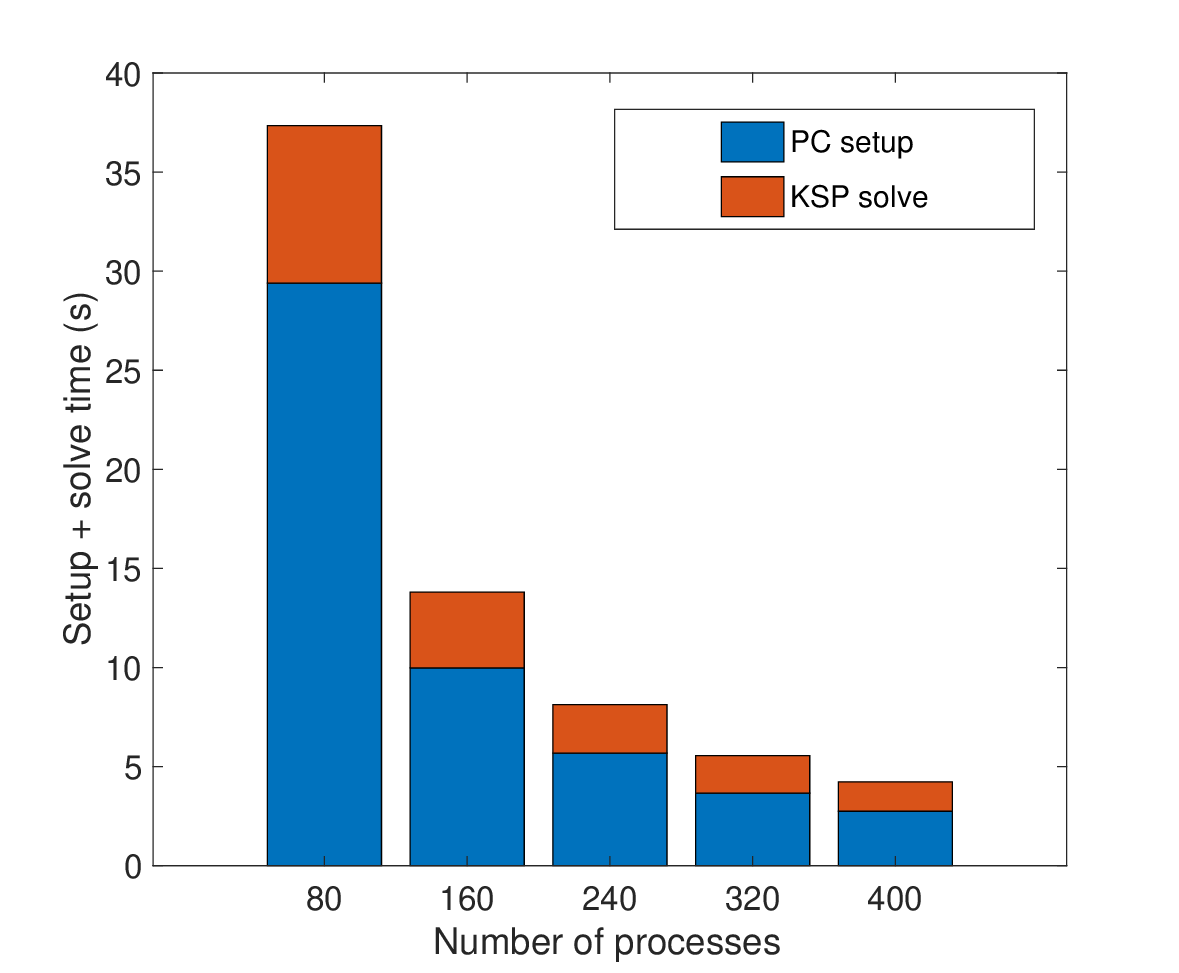} 
        \caption{Two-grid RAS-V3}
    \end{subfigure}
    
  \caption{PC setup and KSP solve time for variants of two-grid RAS for the deterministic linear Poisson problem}\label{Fig:Poisson_setup}
\end{figure}

Strong parallel and numerical scalability of these preconditioners with a linear system of size $32$ million are also shown in 
Fig.~\ref{Fig:Poisson_strong}. Two-grid RAS version stagnates very fast since the size of the coarse grid overwhelms the LU solver. Other variants show significant reductions in time up to $600$ processes. Two-grid RAS V3 has the smallest time compared to other versions. The large difference in iteration counts for both cases can be explained by the rapidly converged coarse grid in 2GV3 and the slow convergence of the coarse solver in 2GV2. This is also observed from the setup and solve times for different solvers in Fig.~\ref{Fig:Poisson_setup}. The two-grid RAS with LU factorization for coarse grid solve takes a significant amount of time to setup which is reflected in the PC setup time. The 2GV2 consumes more solve time than the setup of the preconditioner because of the poor one-level RAS preconditioner for the coarse grid solve. The 2GV3 has a large setup time due to AMG. As the problem size increases, the coarse grid size increases which necessitates a multilevel solver for coarse level (in this case AMG) for scalability. Among these solvers, 2GV3 which has an AMG preconditioner for coarse grid used in conjunction with the fine grid preconditioner shows promising results. This is possible due to the restricted additive Schwarz method used in the fine grid (acting as a smoother) which removes high-frequency components of the error and the multilevel error reduction strategy offered by AMG which reduces the low-frequency components in the coarse grid. 

\begin{figure}[!h]
    \centering
    \begin{subfigure}[b]{0.475\textwidth}
       \centering
        \includegraphics[width=\textwidth]{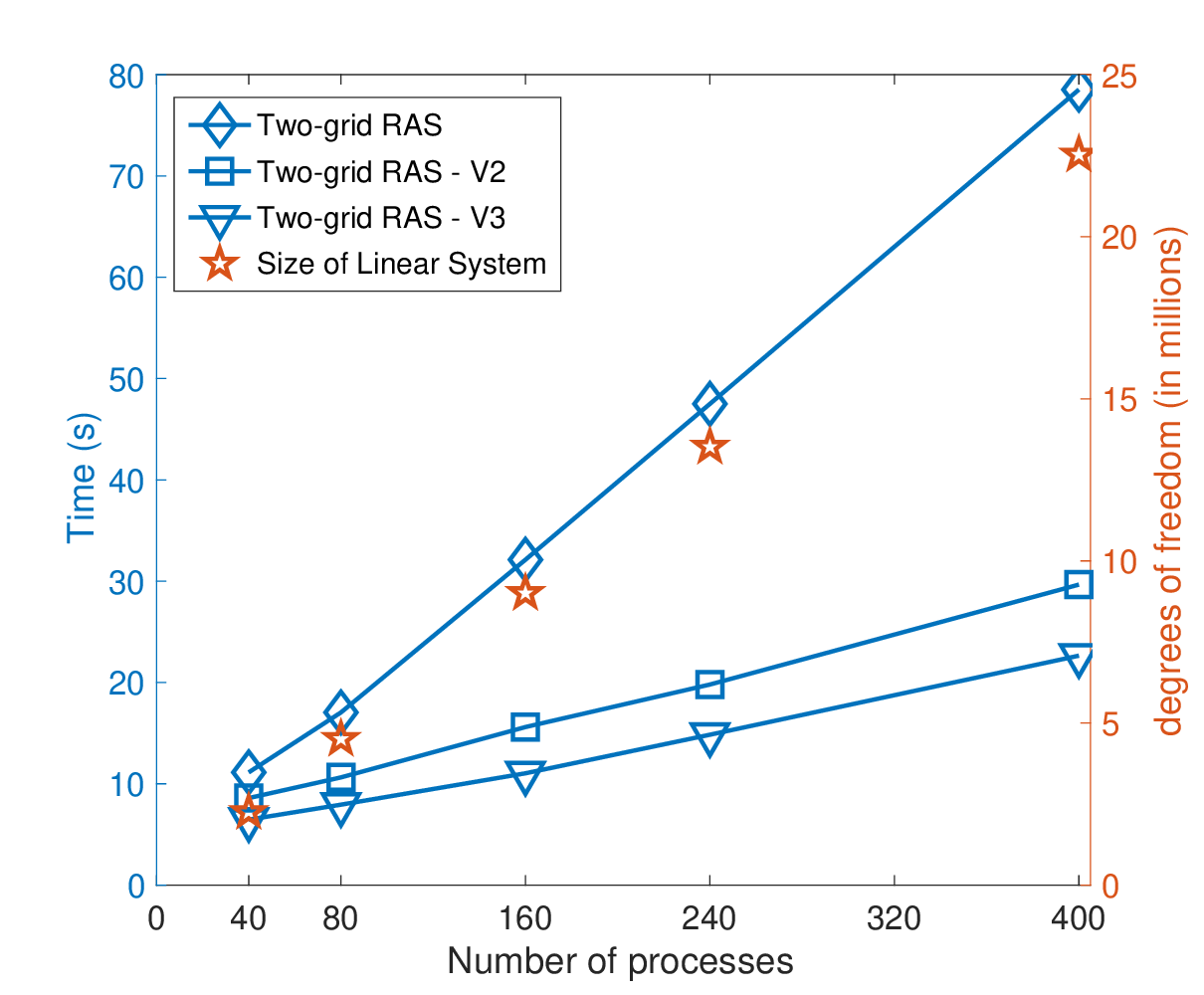} 
        \caption{Weak parallel scalability}
    \end{subfigure}
    \begin{subfigure}[b]{0.475\textwidth}
      		 \centering
        \includegraphics[width=\textwidth]{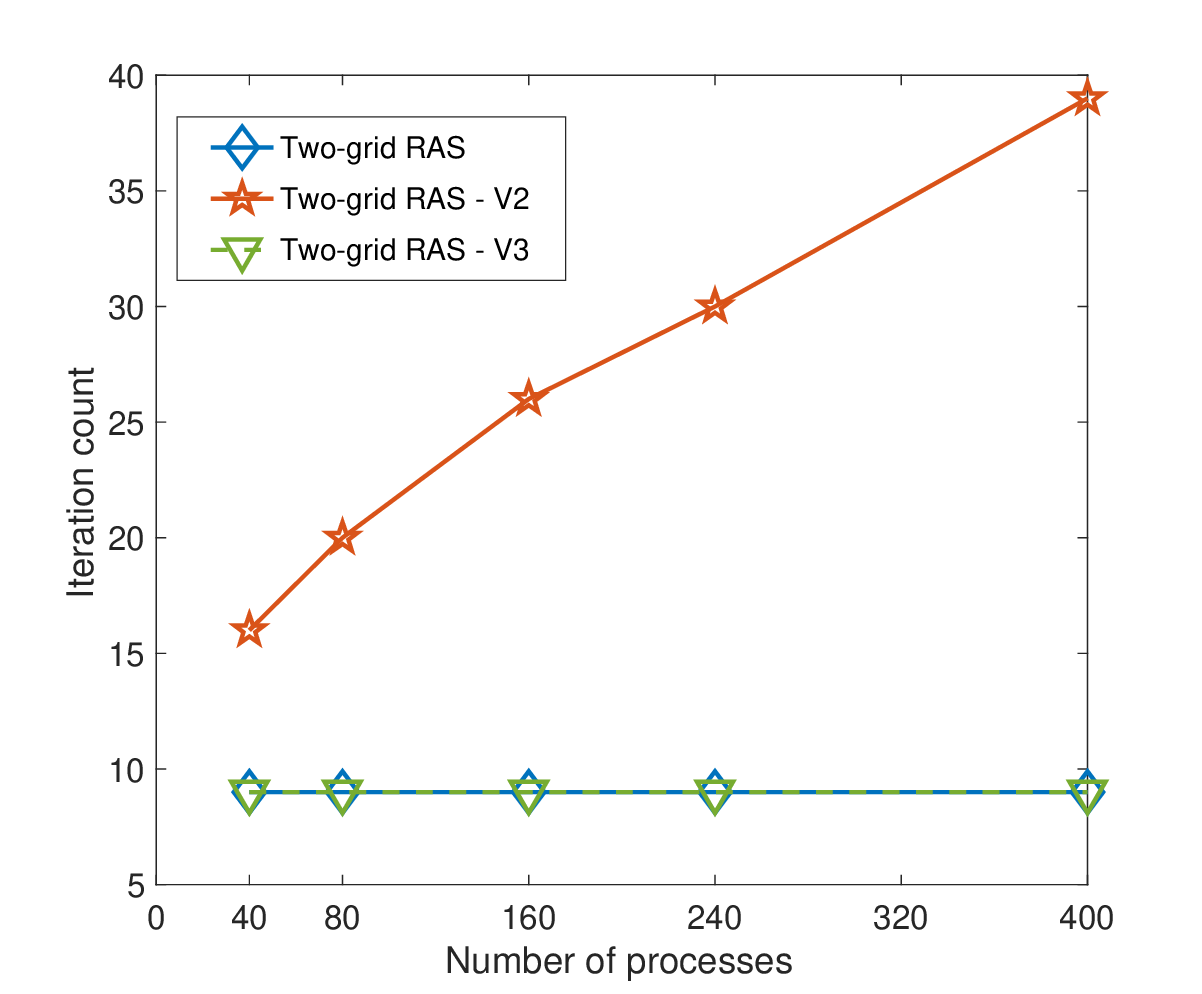} 
        \caption{Weak numerical scalability}
    \end{subfigure} 
    \centering
    \begin{subfigure}[b]{0.475\textwidth}
       \centering
        \includegraphics[width=\textwidth]{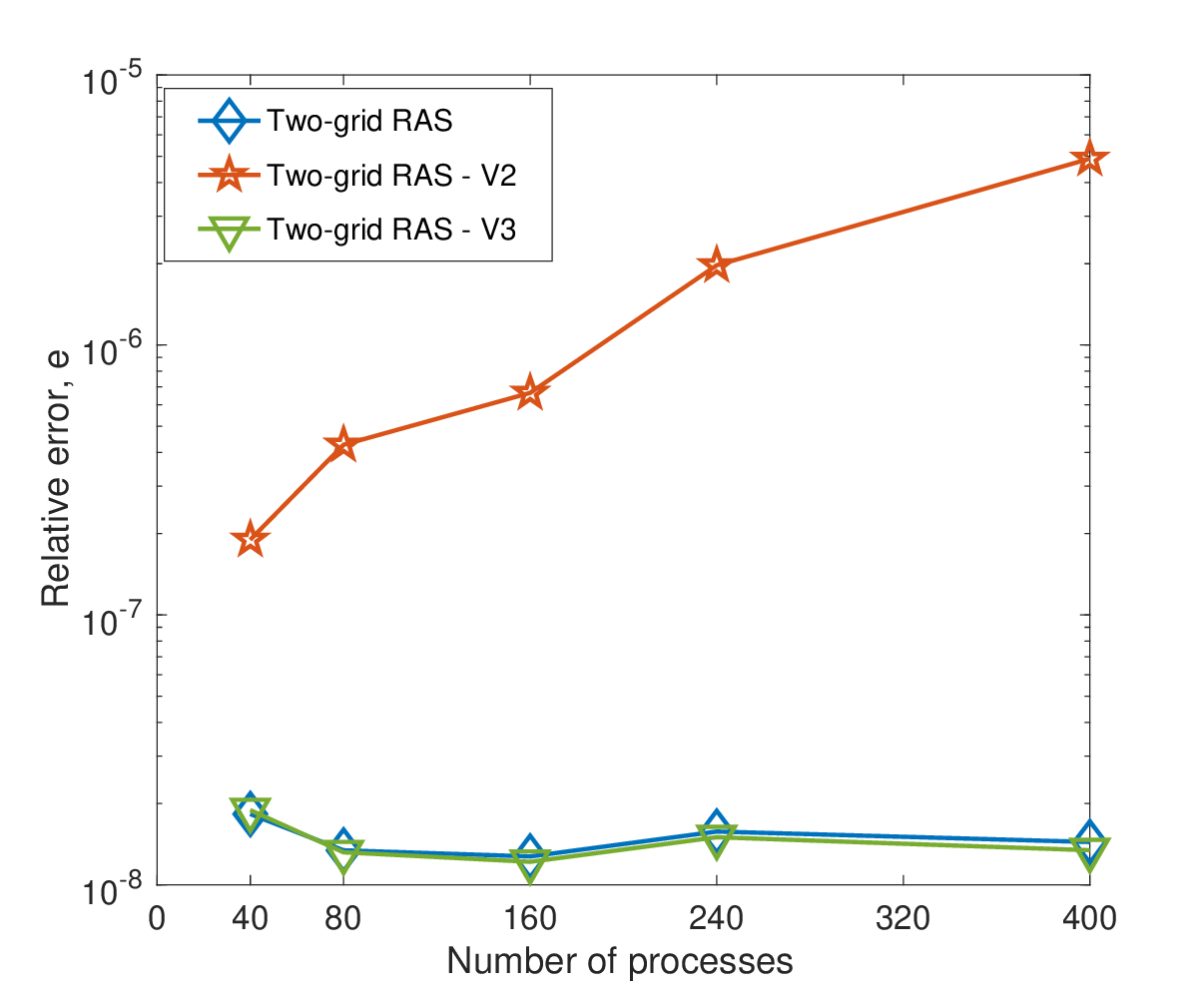} 
        \caption{Variation in relative error}
    \end{subfigure}
  \caption{Comparison of time, iteration count and relative error of two-grid RAS versions for deterministic linear Poisson problem}\label{Fig:2L_weak}
\end{figure}

A comparison of two-grid Schwarz preconditioners with respect to weak scalabilities is shown in Fig.~\ref{Fig:2L_weak}. As the size of the linear system increases, the coarse grid size also increases which results in higher solution time for the two-grid RAS method as shown in
Fig.~\ref{Fig:2L_weak}a. This time is reduced by solving the coarse grid by an iterative approach with one level preconditioner as in the two-grid RAS-V2. However, it has a higher number of iterations since the coarse grid convergence rate is slow as evident from Fig.~\ref{Fig:2L_weak}b. By comparing the relative error (computed using Eq.~(\ref{Eq.relerror})), it can be observed that the two-grid RAS-V2 can obtain solutions much faster than two-grid RAS with comparable accuracy. The two-grid RAS-V3 performs best of all the three methods with the lowest computational time, iteration counts and comparable accuracy to two-grid RAS as evident from 
Fig.~\ref{Fig:2L_weak}a, \ref{Fig:2L_weak}b and \ref{Fig:2L_weak}c. Comparisons with other versions such as block Jacobi and additive Schwarz approach for the coarse grid are also conducted but not shown since AMG outperforms all. For further experiments, we focus only on 2GV2 and 2GV3 and neglect two-grid RAS (with LU factorization for coarse grid solve) due to its poor scalability characteristics.

\subsection{DD for Deterministic Nonlinear Poisson Problem}

A nonlinear Poisson problem with a quadratic nonlinearity ($m = 1$ in Eq.~(\ref{Eq.qunlc4}) ) and forcing term $f = 0$ is solved using Picard iterations and various two-grid RAS preconditioners in this section.

\begin{figure}[!h]
    \centering
    \begin{subfigure}[b]{0.475\textwidth}
       \centering
        \includegraphics[width=\textwidth]{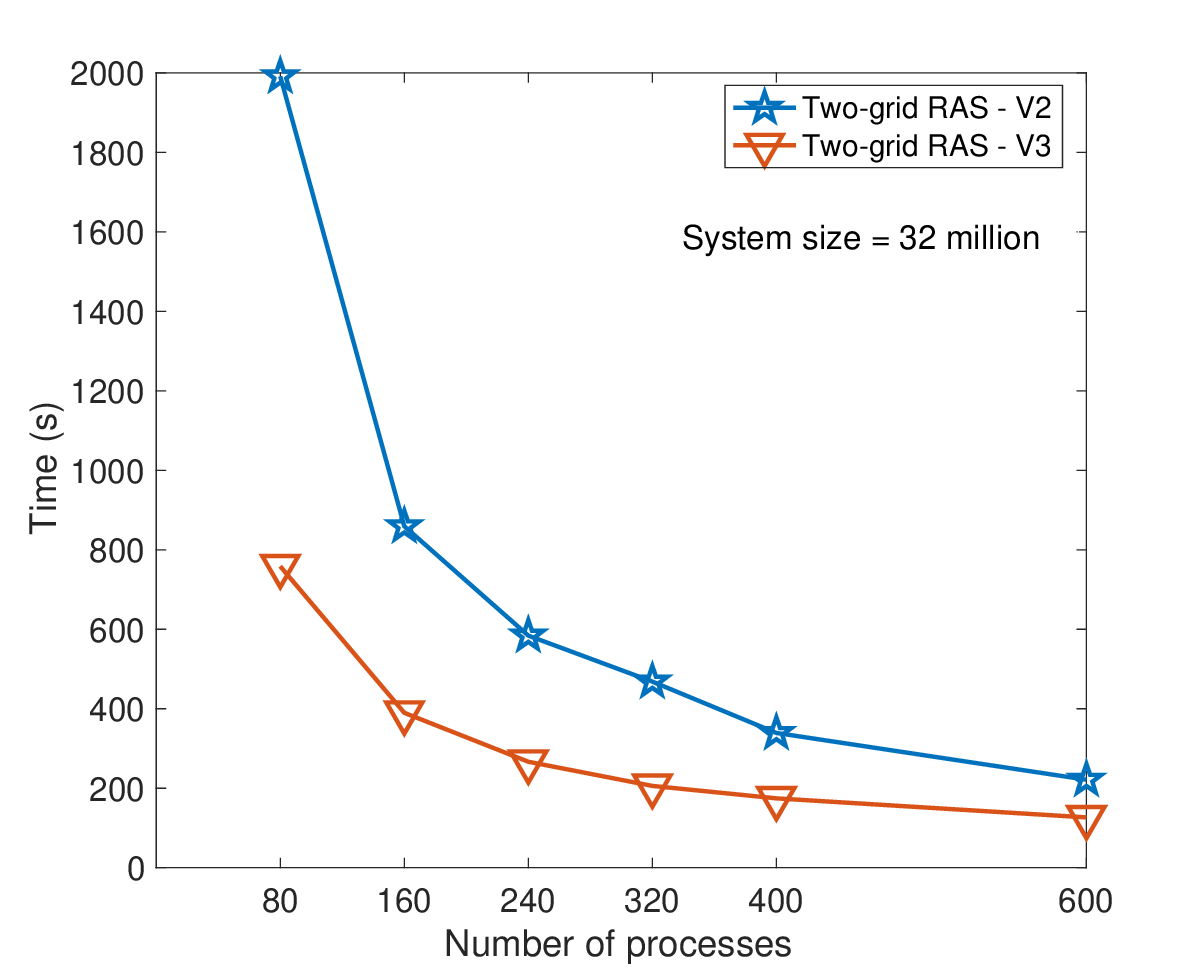} 
        \caption{Strong parallel scalability}
    \end{subfigure}
    \begin{subfigure}[b]{0.475\textwidth}
      		 \centering
        \includegraphics[width=\textwidth]{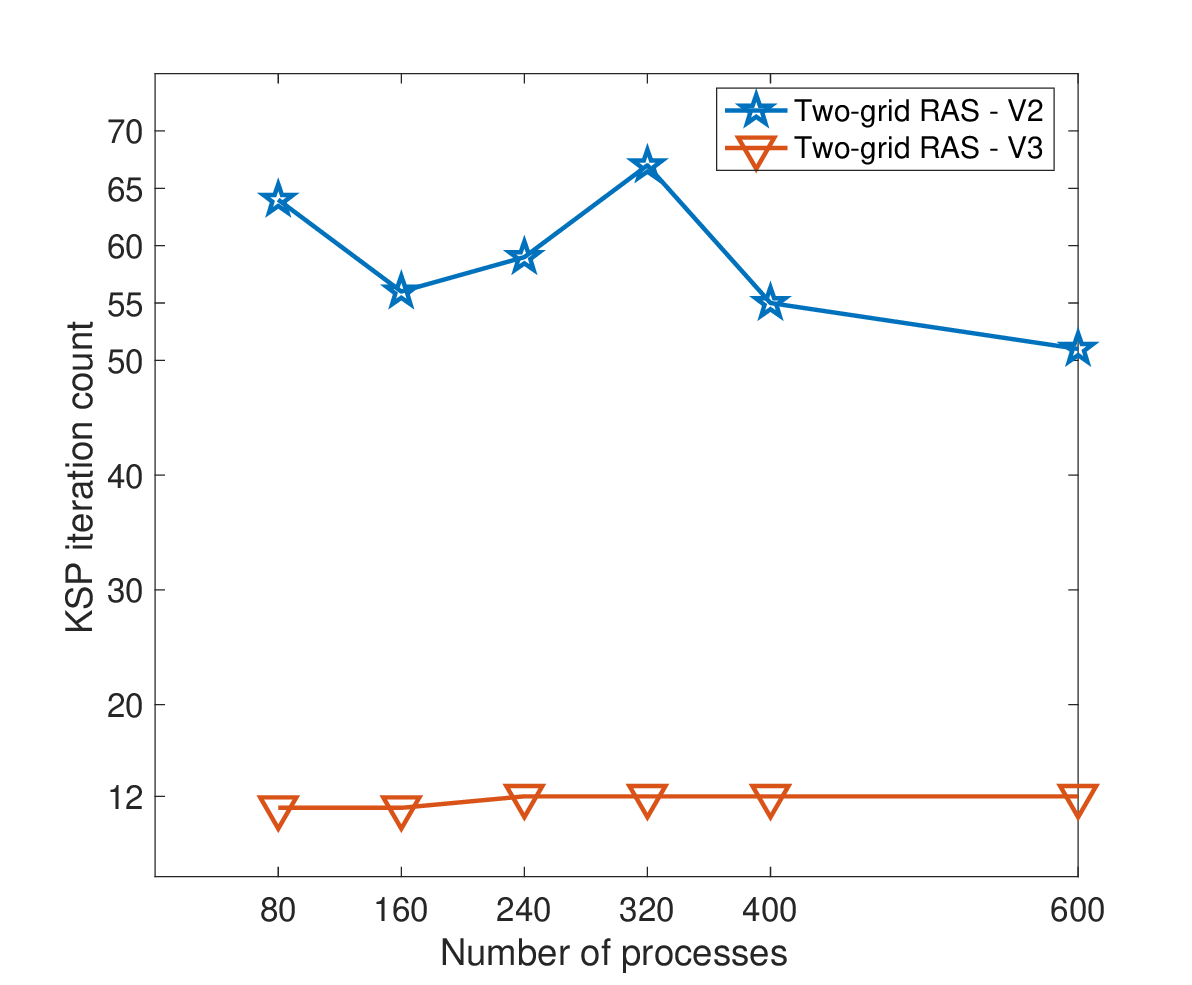} 
        \caption{Strong numerical scalability}
    \end{subfigure} 
    \centering
    \begin{subfigure}[b]{0.475\textwidth}
       \centering
        \includegraphics[width=\textwidth]{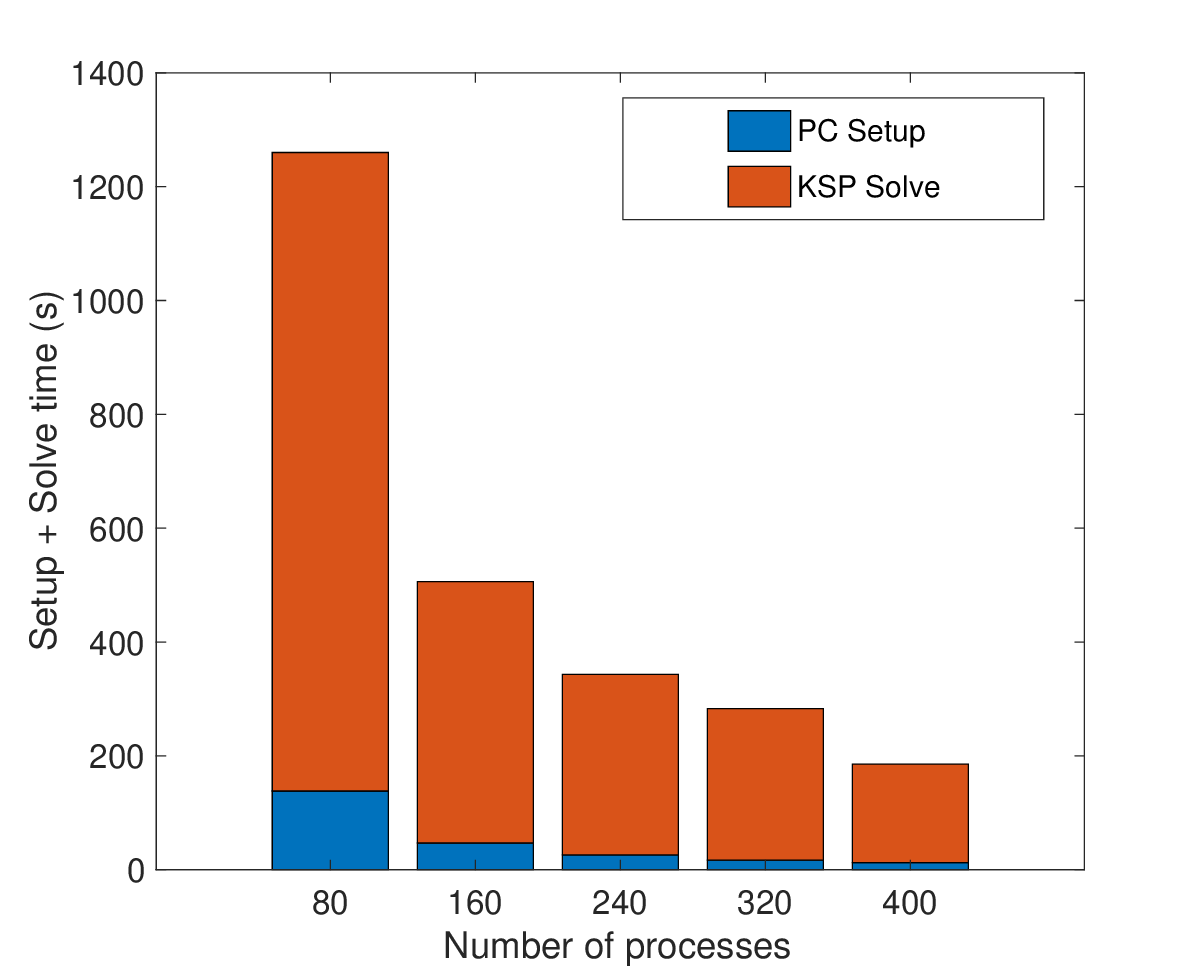} 
        \caption{Two-grid RAS-V2}
    \end{subfigure}
    \begin{subfigure}[b]{0.475\textwidth}
      		 \centering
        \includegraphics[width=\textwidth]{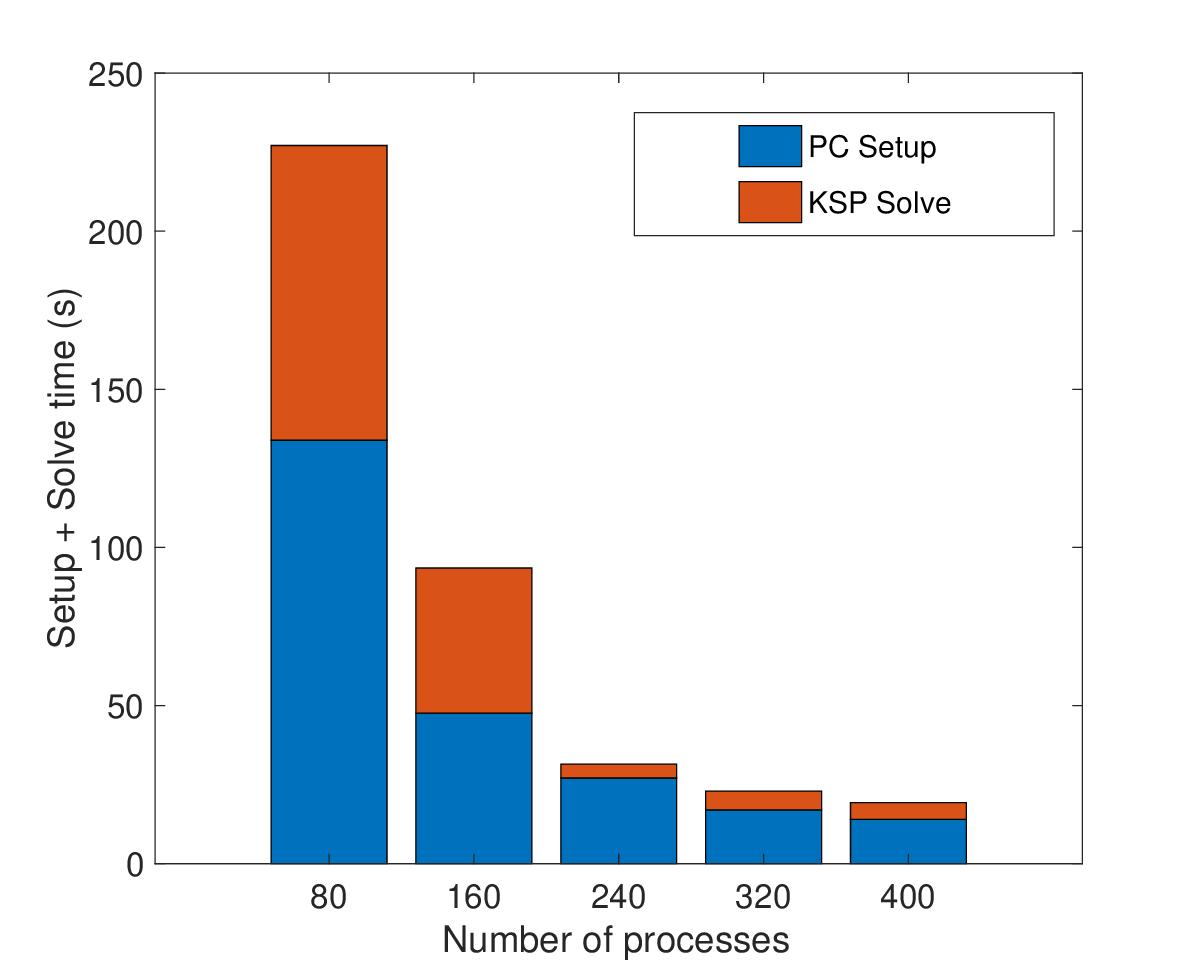} 
        \caption{Two-grid RAS-V3}
    \end{subfigure} 
  \caption{Strong scalability of two-grid RAS versions for deterministic nonlinear Poisson problem}\label{Fig:NLpoisson_strong}
\end{figure}

The strong scalability for 2GV2 and 2GV3 is shown in Fig.~\ref{Fig:NLpoisson_strong}. Even though both solvers have comparable time to solution for $600$ processes, the difference is significant for $80$ processes. The KSP (GMRES) iteration counts (mean for all Picard iterations) for 2GV2 are much higher than 2GV3 (see Fig.~\ref{Fig:NLpoisson_strong}b). This is reflected in the KSP solve time for 2GV2 in Fig.~\ref{Fig:NLpoisson_strong}c. For 2GV3, a higher PC setup time is observed as expected due to the cost of setup for AMG while KSP solve time is much lower than 2GV2 (see Fig.~\ref{Fig:NLpoisson_strong}d).

\begin{figure}[!h]
    \centering
    \begin{subfigure}[b]{0.475\textwidth}
       \centering
        \includegraphics[width=\textwidth]{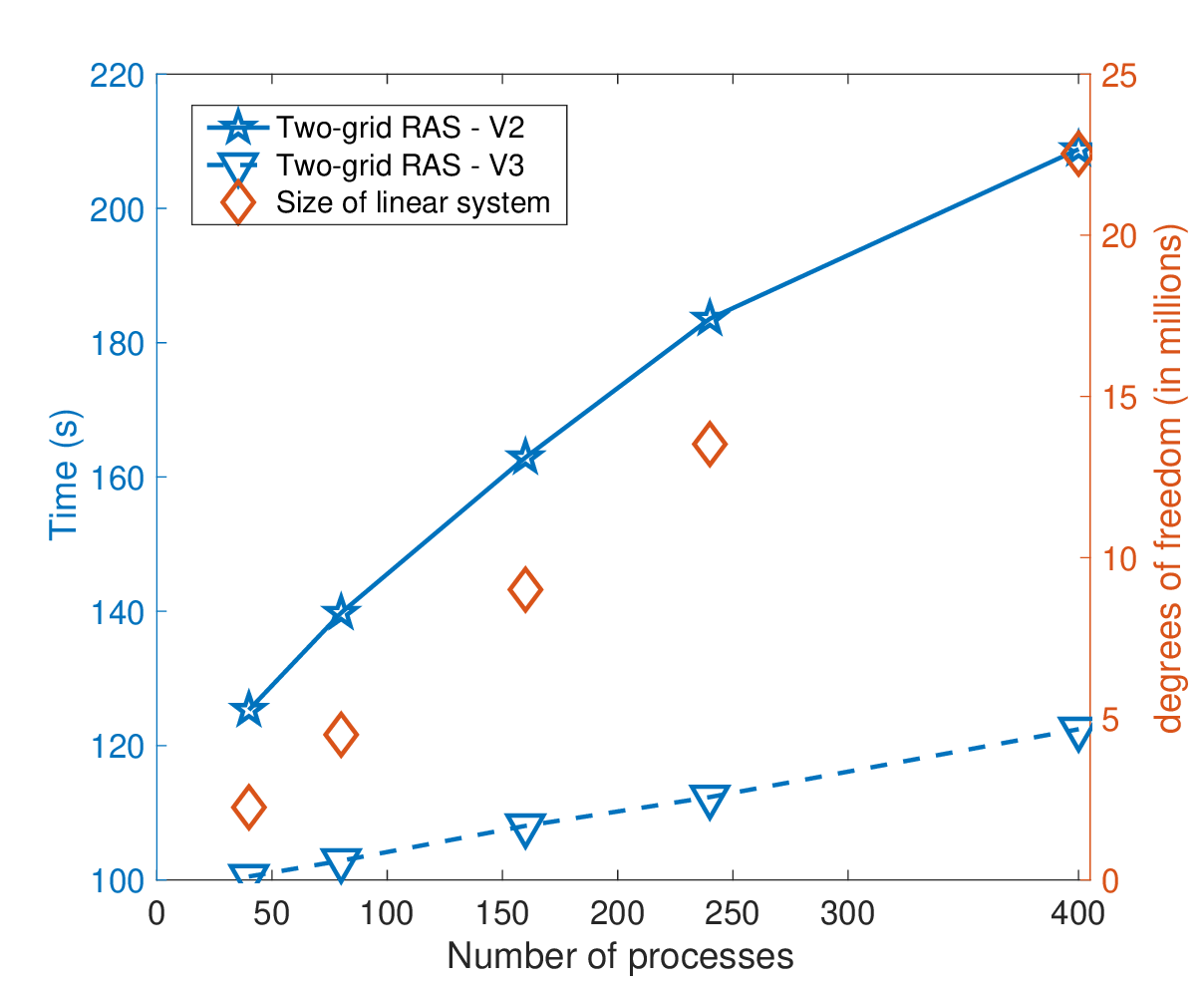} 
        \caption{Weak parallel scalability}
    \end{subfigure}
    \begin{subfigure}[b]{0.475\textwidth}
      		 \centering
        \includegraphics[width=\textwidth]{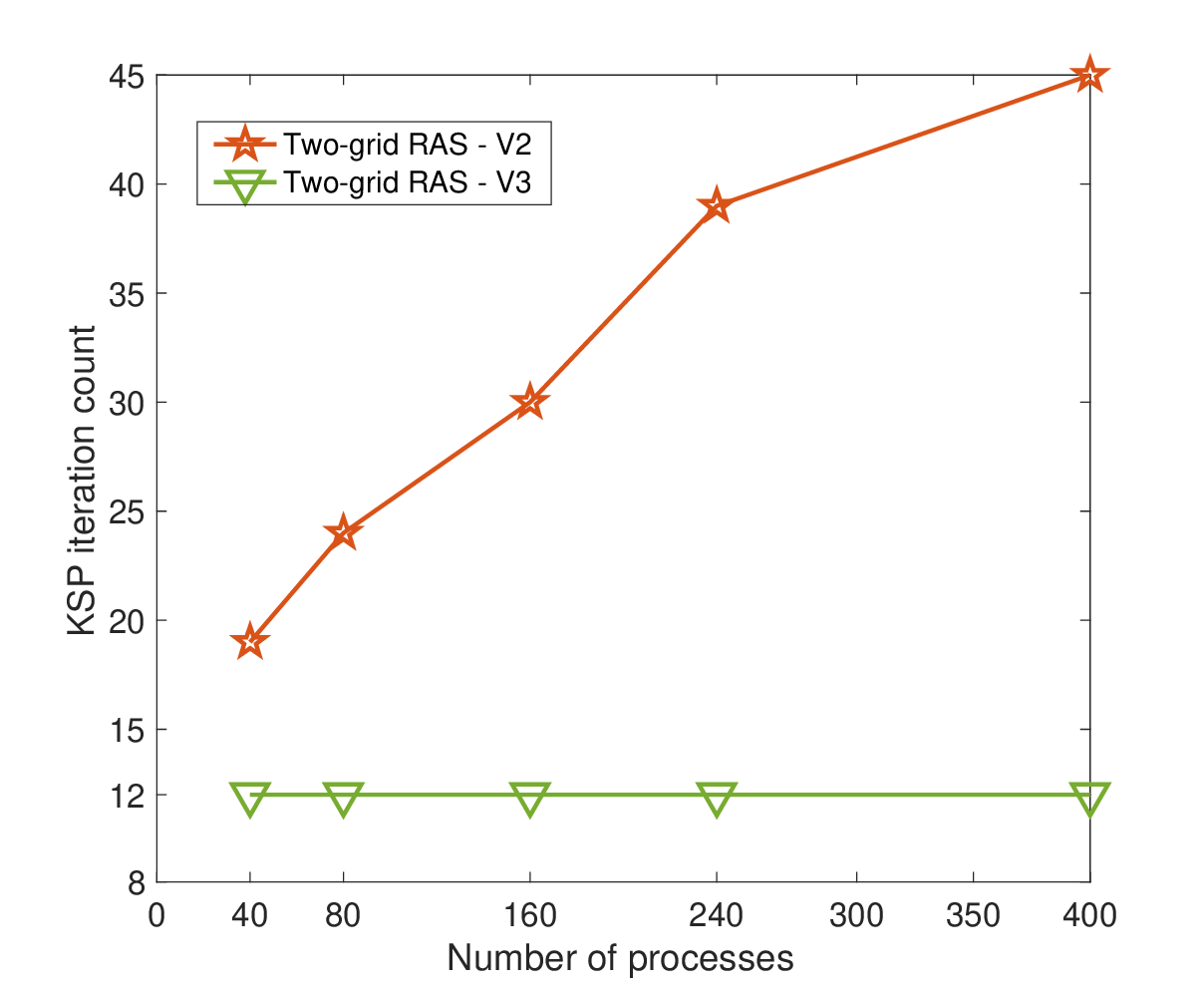} 
        \caption{Weak numerical scalability}
    \end{subfigure} 
    \centering
    \begin{subfigure}[b]{0.475\textwidth}
       \centering
        \includegraphics[width=\textwidth]{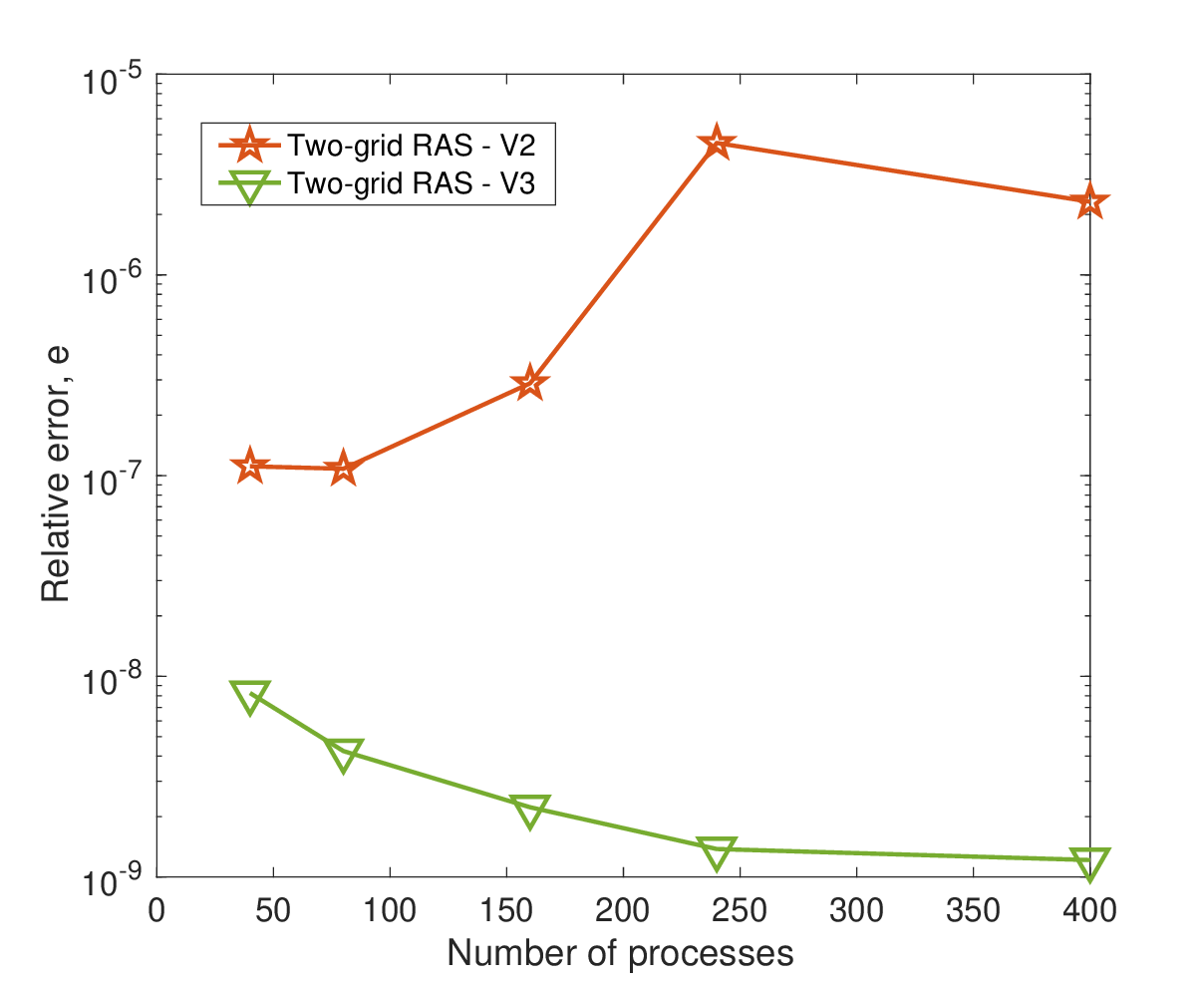} 
        \caption{Variation in relative error}
    \end{subfigure}
  \caption{Comparison of time, iteration count and error of two-grid RAS versions for deterministic nonlinear Poisson problem}\label{Fig:NLpoisson_weak}
\end{figure}

Fig.~\ref{Fig:NLpoisson_weak} shows the comparison of two-grid RAS-V2 and two-grid RAS-V3 for weak scaling experiments. The 2GV2 method has a rapid increase in the time and iteration number with increasing global problem size. This is due to the slow convergence of the one-level RAS preconditioner for the coarse grid solve which influences the iteration count (see Fig.~\ref{Fig:NLpoisson_weak}b) and time to solution. However, when this is replaced by an AMG preconditioner, significant improvement can be observed in computational time and iteration counts. The difference between 2GV2 and 2GV3 is significantly larger than the linear Poisson problem. The relative error between both methods is also shown in Fig.~\ref{Fig:NLpoisson_weak}c which shows the better accuracy of 2GV3 than 2GV2.

\begin{figure}[!h]
    \centering
    \begin{subfigure}[b]{0.475\textwidth}
       \centering
        \includegraphics[width=\textwidth]{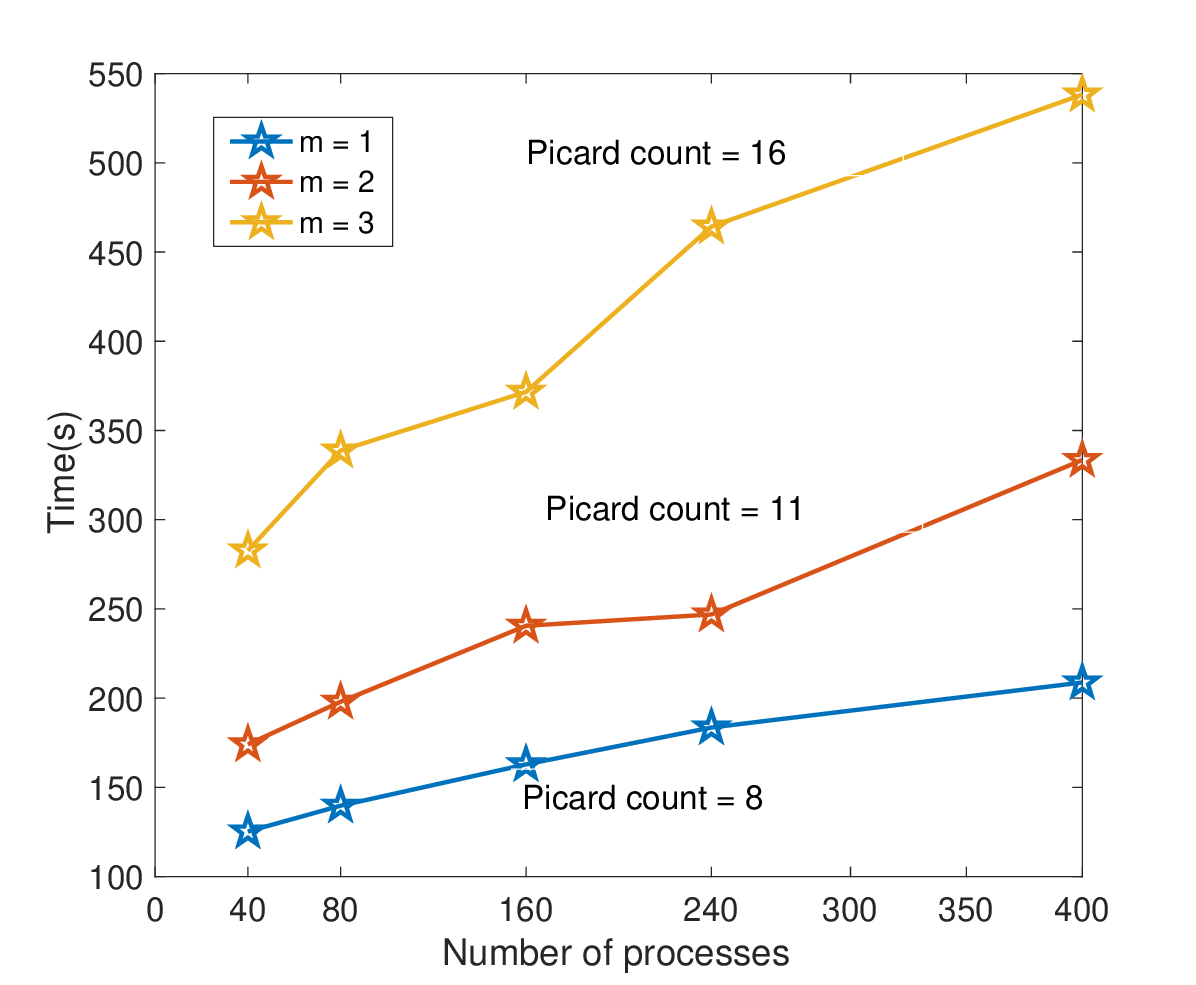} 
        \caption{Two-grid RAS-V2}
    \end{subfigure}
    \begin{subfigure}[b]{0.475\textwidth}
      		 \centering
        \includegraphics[width=\textwidth]{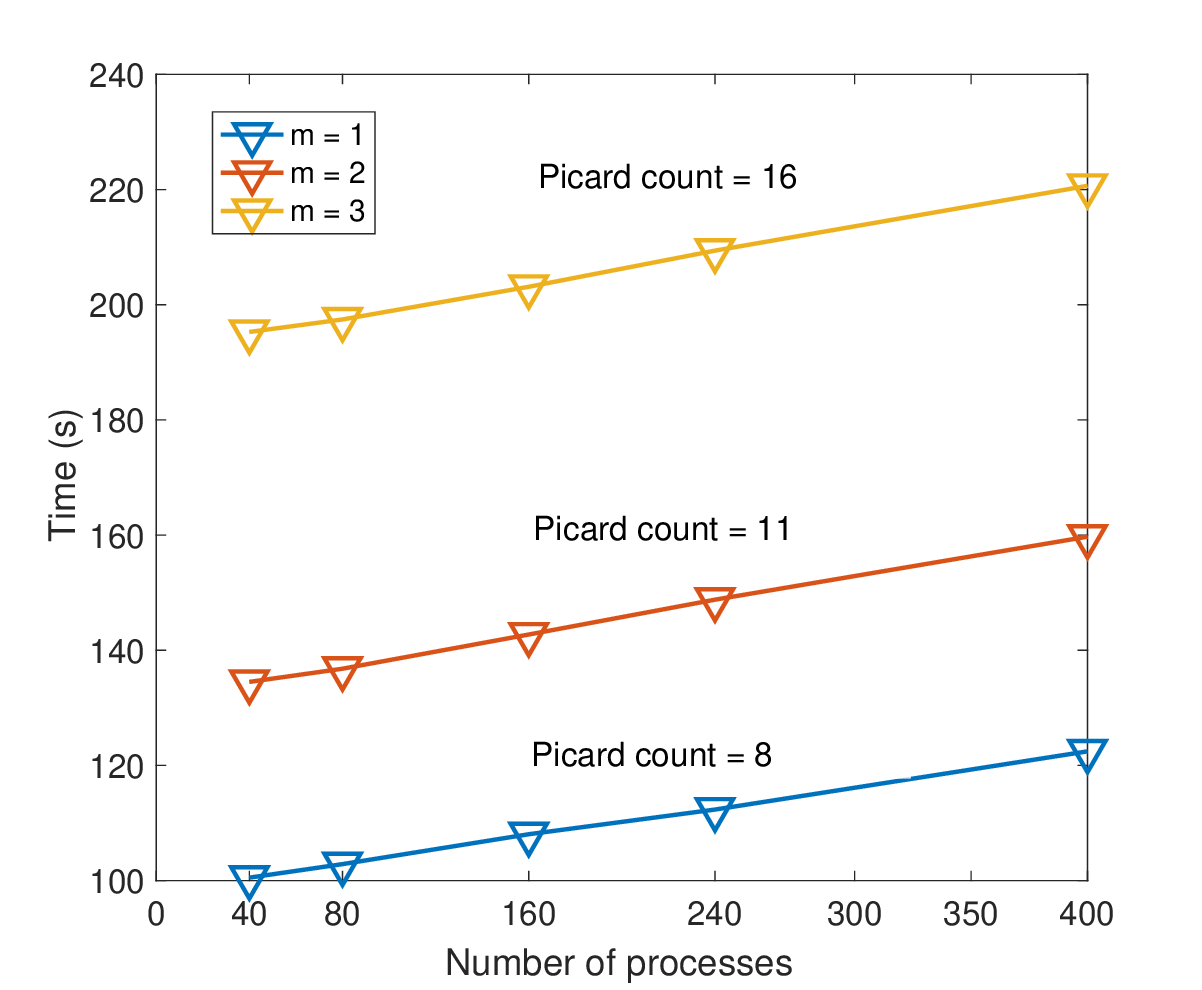} 
        \caption{Two-grid RAS-V3}
    \end{subfigure} 
    \centering
    \begin{subfigure}[b]{0.475\textwidth}
       \centering
        \includegraphics[width=\textwidth]{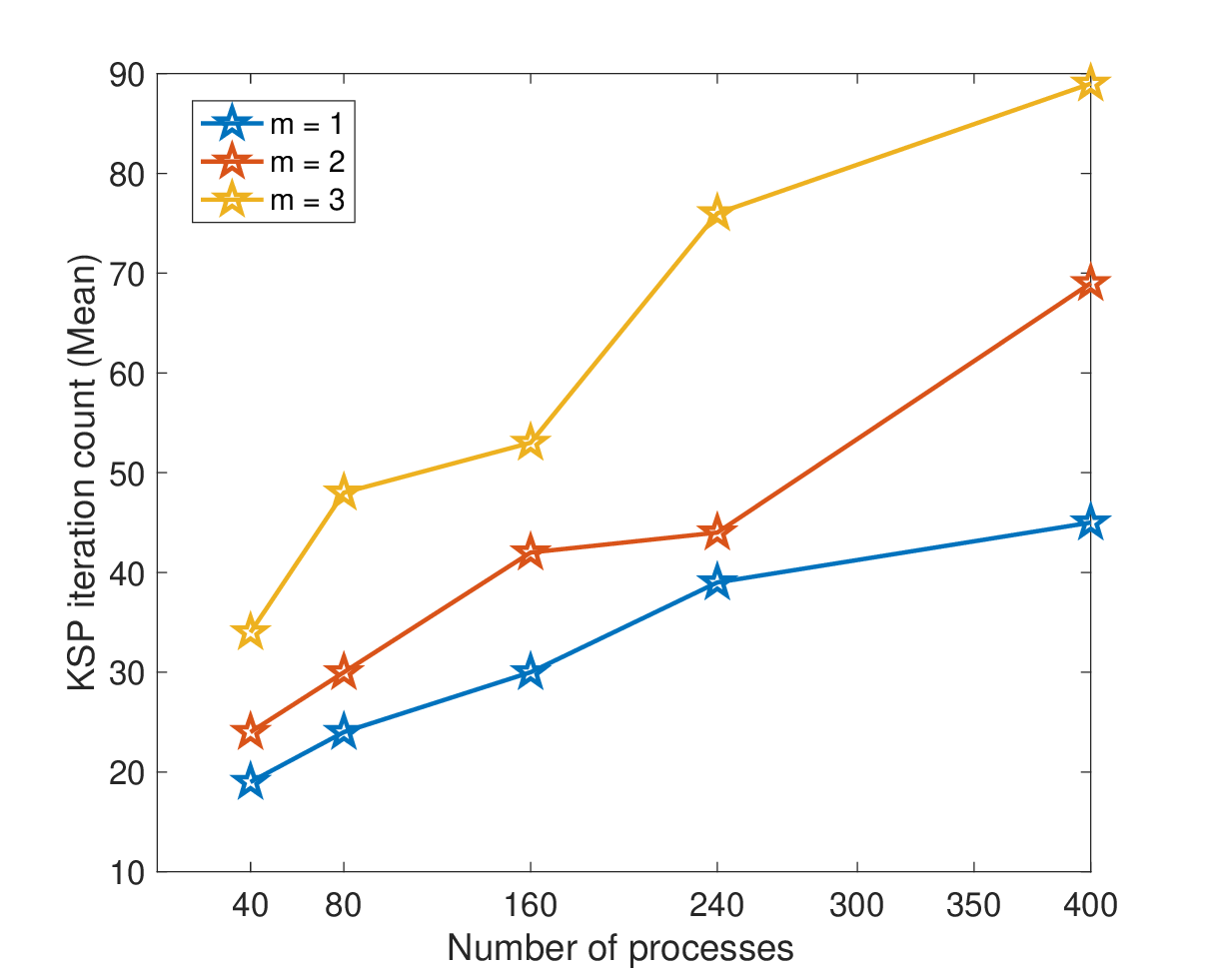} 
        \caption{Two-grid RAS-V2}
    \end{subfigure}
    \begin{subfigure}[b]{0.475\textwidth}
      		 \centering
        \includegraphics[width=\textwidth]{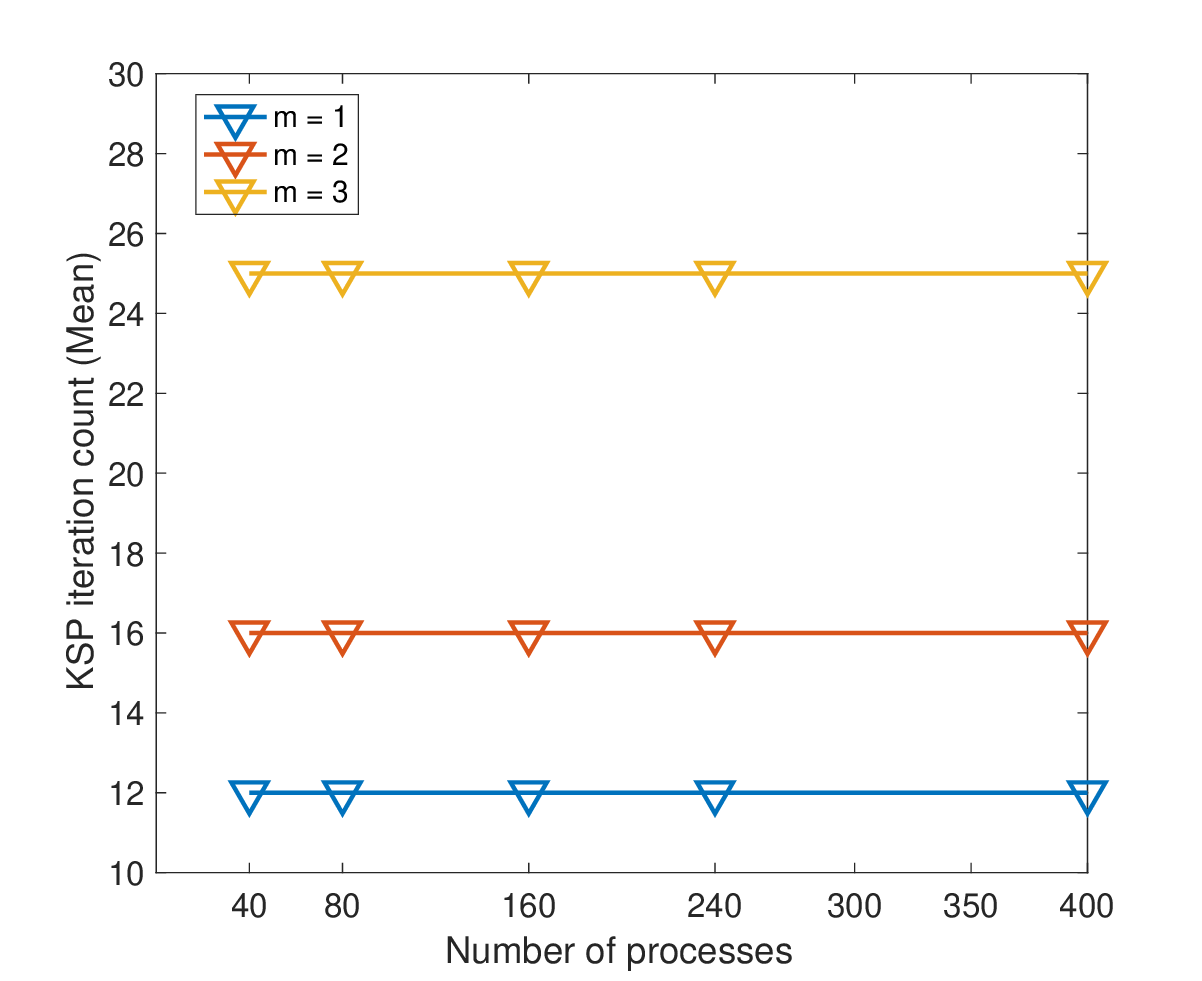} 
        \caption{Two-grid RAS-V3}
    \end{subfigure}
  \caption{Comparison of weak scalability of two-grid RAS versions with increasing order of nonlinearity for deterministic nonlinear Poisson problem}\label{Fig:NLpoisson_compare}
\end{figure}

Another comparison with increasing the order of nonlinearity for $m = 1,2,3$ in Eq.~(\ref{Eq.qunlc4}) (using the same system sizes as in weak scaling experiment) is shown in Fig.~\ref{Fig:NLpoisson_compare}. 2GV2 shows a non-monotonic increase in time with the order $m$, but 2GV3 has a linear increase in time (see Fig.~\ref{Fig:NLpoisson_compare}b). This increase in time can be attributed to the increase in Picard iteration counts with increasing order of nonlinearity $m$. Similarly, for KSP iteration counts 2GV2 shows a significant increase with the number of processes while 2GV3 has constant iteration counts as shown in Fig.~\ref{Fig:NLpoisson_compare}c and Fig.~\ref{Fig:NLpoisson_compare}d.

\subsection{Condition Number for a Stochastic System}\label{app:condnumbersto}

This section deviates from the previous discussions on overlapping Schwarz solvers in the deterministic setting to the stochastic setting. The stochastic Galerkin method transforms a stochastic PDE using polynomial chaos expansions for the random quantities along with Galerkin projection into a system of coupled deterministic PDEs. These system of PDEs (upon numerical discretization) has a much larger size than that of the corresponding deterministic system. In this section, a comparison of the condition number of a deterministic Poisson problem and that of a stochastic problem with increasing number of random variables is demonstrated. \textcolor{ss}{Note, similar studies showing the condition number of the stochastic system matrix with respect to increasing random variables and order of expansion is shown in \cite{sousedik_2}.}

A Poisson problem in a square domain discretized with $29$ vertices is considered. A smaller grid is chosen to allow faster computation of condition number of the stochastic system. The Dirichlet boundary condition on all sides of the square domain is enforced using the penalty approach (with a penalty term of $1\times10^{7}$). The stochastic system having $2$ to $9$ random variables are constructed using a $2^{nd}$ order input and $3^{rd}$ order output PCE. The condition number of the coefficient matrix for the deterministic problem is computed as $2.478\times 10^{6}$. The ratio of the condition number of the stochastic system matrix to the deterministic system is reported below. 

\begin{table}[!htbp]
    \centering
    \begin{tabular}{|c|c|c|c|c|c|}
    \hline
      \specialcell{Number of random \\ variables}  &  $2$ & $3$ & $5$ & $7$ & $9$  \\
       \hline
       \specialcell{Ratio of condition  \\ number} & $17.94 $ & $24$ & $35.75$ & $47.26$ & $58.66$\\ 
      \hline
    \end{tabular}
    \caption{Growth in condition number of the stochastic system with respect to the deterministic system of same mesh size}
    \label{tab:ss_cond}
\end{table}

A significant growth in the condition number of the stochastic system matrix from the corresponding deterministic system matrix is observed. The increase in number of random variables not only increases the system size but also increases the condition number of the coefficient matrix. An increase in the initial deterministic mesh size from $29$ to $1660$ vertices by finer discretization increases the condition number of the system by $9.57$ while a similar increase in the stochastic dimension ($5$ random variable to obtain a stochastic system size of $1624$ degrees of freedom) increases the condition number by $35.75$. This demonstrates that the coupling introduced by increase in degrees of freedom in the stochastic dimension is stronger than the increase in spatial direction.







\end{document}